\documentclass[
aps,prd,
nofootinbib,
superscriptaddress,
showpacs
]{revtex4}

\usepackage{amsmath}
\usepackage{amssymb}
\usepackage{bm}
\usepackage{color,graphicx}
\usepackage{slashed}

\usepackage{mciteplus}
\usepackage[dvipsnames]{xcolor}

\newcounter{comment}

\begin{document}

\title{Lorentz Invariance and QCD Equation of Motion Relations for Generalized Parton Distributions and the Dynamical Origin of Proton Orbital Angular Momentum}
%

\author{Abha Rajan}
\email{ar5xc@virginia.edu}
\affiliation{University of Virginia - Physics Department,
382 McCormick Rd., Charlottesville, Virginia 22904 - USA} 

\author{Michael Engelhardt}
\email{engel@nmsu.edu}
\affiliation{New Mexico State University - Department of Physics, Box 30001 MSC 3D, Las Cruces NM, 88003 - USA}

\author{Simonetta Liuti }
\email{sl4y@virginia.edu}
\affiliation{University of Virginia - Physics Department,
382 McCormick Rd., Charlottesville, Virginia 22904 - USA \\ and Laboratori Nazionali di Frascati, INFN, Frascati, Italy.}

\begin{abstract}
We derive new Lorentz Invariance and Equation of Motion Relations between twist-three Generalized Parton Distributions (GPDs) and moments in the parton transverse momentum, $k_T$, of twist-two Generalized Transverse Momentum-Dependent Distributions (GTMDs), as a function of the parton longitudinal momentum fraction $x$. Although GTMDs in principle define the observables for partonic orbital motion, experiments that can unambiguously detect them appear remote at present. The relations presented here provide a solution to this impasse in that, e.g., the orbital angular momentum density is connected to directly measurable twist-three GPDs.
Out of 16 possible Equation of Motion relations that can be written in the T-even sector, we focus on three helicity configurations that can be detected analyzing specific spin asymmetries: two correspond to longitudinal proton polarization and are associated with quark orbital angular momentum and spin-orbit correlations; the third, obtained for transverse proton polarization, is a generalization of the relation obeyed by the $g_2$ structure function. We also exhibit an additional relation connecting the off-forward extension of the Sivers function to an off-forward Qiu-Sterman term.
\end{abstract}

\maketitle
%
\baselineskip 3.0ex

\section{Introduction}
A fundamental way of characterizing the internal structure of the proton is through sum rules that express how global properties of the proton are composed from corresponding quark and gluon quantities. For example, one may ask what portion of a proton's momentum is carried by either quarks or gluons; or one may ask how the spin of the proton is composed from the spins and orbital angular momenta of its quark and gluon constituents. Elucidating this latter question, the so-called proton spin puzzle \cite{Jaffe:1989jz}, indeed counts among the prime endeavors of hadronic physics in the last decades.
These questions can be cast in field-theoretic language by considering proton matrix elements of the energy momentum tensor, $T_{q,g}$ ($q$ and $g$ denote the quark and gluon sectors), 
\begin{eqnarray}
\label{eq:EMT_P}
\langle p'\mid T^{0i}_{q,g} \mid p \rangle & = & A_{q,g} \, P^i \, \overline{U}(p') \gamma^o U(p)  \\
\label{eq:EMT_J}
 \epsilon^{ijk}\langle p'\mid \left(x^j T^{0k} - x^k T^{0j} \right)\mid p \rangle & = & A_{q,g} \, P^i \, \overline{U}(p') \gamma^o U(p) + B_{q,g} \, P^i \, \overline{U}(p') \frac{\sigma^{o \alpha}\Delta_\alpha}{2M^2} U(p) \ ,
\end{eqnarray}
where $p$ and $p'$ describe the incoming and outgoing proton states, and $A_{q,g}(t), B_{q,g}(t)$, $(t=\Delta^2= (p'-p)^2, P= (p'+p)/2)$ are the relevant gravitomagnetic form factors parametrizing the proton matrix elements (Refs.\cite{Ji:1994av,Ji:1996ek}, for reviews see Ref.\cite{Wakamatsu:2014zza,Liu:2015xha}).
These basic constructs of the theory can be accessed experimentally owing to 
the connection, through the operator product expansion (OPE), of the gravitomagnetic form factors to the Mellin moments of specific parton distributions parametrizing both the forward ($p=p'$) and off-forward ($p\neq p'$) quark and gluon correlation functions. One obtains the following sum rules for momentum and angular momentum, respectively,
\begin{eqnarray}
\label{eq:MomSR}
A_{q,g} & = & \int_0^1 dx x H_{q,g} \;\;\;\;\;   \Rightarrow \sum_{i=q,g} A_i  = \epsilon_q + \epsilon_g = 1 \\
\label{eq:JiSR}
 B_{q,g} &  = & \int_{0}^1 dx x (H_{q,g}+E_{q,g}) \Rightarrow \sum_{i=q,g} (A_i + B_i) = J_q + J_g = \frac{1}{2}.
\end{eqnarray}
Eq.(\ref{eq:JiSR}), the angular momentum sum rule, is also known as the Ji sum rule \cite{Ji:1996ek}. All of the distributions entering Eqs.(\ref{eq:MomSR}) and (\ref{eq:JiSR}) are observable in a wide class of experiments probing the deep inelastic structure of the proton. 
$H_{q,g}(x,\xi,t)$ and $E_{q,g}(x,\xi,t)$ are the Generalized Parton Distribution (GPD) functions which depend on the longitudinal momentum transfer between the initial and final proton, represented through the skewness parameter $\xi$, and the four-momentum transfer squared, $t$, $x$ being the light cone momentum fraction carried by the parton \cite{Belitsky:2005qn,Diehl:2003ny}. In particular, $H_{q}(x,0,0) \equiv q(x), H_g(x,0,0) \equiv g(x)$, where $q(x)$ and $g(x)$ are the unpolarized quark (antiquark) and gluon distributions, or the Parton Distributions Functions (PDFs). 
PDFs have been measured in decades of Deep Inelastic Scattering (DIS) experiments, with impressive accuracy and kinematical coverage, confirming to high precision the momentum sum rule, Eq.(\ref{eq:MomSR}). To verify the angular momentum sum rule it is necessary to extract the GPDs from experiment, in particular, $E_{q,g}$. Sufficiently accurate values for the GPDs have just fairly recently started to become available from exclusive deeply virtual scattering experiments, namely Deeply Virtual Compton Scattering (DVCS), Deeply Virtual Meson Production (DVMP) and related processes, conducted most recently at Jefferson Lab and COMPASS (see \cite{Kumericki:2016ehc} for a recent review). 

DVCS experimental measurements are necessarily more involved than ones for inclusive scattering, since they require the simultaneous detection of all products of reaction. The extraction of observables, the GPDs, from experiment is also more complex owing to the increased number of kinematical variables they depend on. An additional hurdle is present for the analysis of angular momentum in both identifying and giving a physical interpretation to the components of the sum rule (\ref{eq:JiSR}): while the momentum sum rule has an immediate dynamical interpretation in terms of the average longitudinal momentum carried by the different parton components, to obtain a dynamically transparent expression for the angular momentum sum rule one has to break it down into its spin and Orbital Angular Momentum (OAM) components,
while simultaneously preserving the gauge invariance of the theory. The decomposition can be performed within two different approaches,  by Jaffe and Manohar (JM) \cite{Jaffe:1989jz},
\begin{eqnarray}
\label{eq:JM_OAM}
&& \frac{1}{2}\Delta \Sigma_q + L_q^{JM} + \Delta G + L_g^{JM} = \frac{1}{2} \end{eqnarray}
and by Ji \cite{Ji:1996ek},
\begin{eqnarray}
&& \frac{1}{2}\Delta \Sigma_q + L_q^{Ji} + J_g^{Ji} = \frac{1}{2}\quad .
\label{eq:Ji_OAM}
\end{eqnarray}
Longitudinal OAM distributions have been identified with parton Wigner distributions weighted by the cross product of position and momentum in the transverse plane, $b_T\times k_T$ \cite{Lorce:2011kd,Lorce:2011ni}. Parton Wigner distributions can be related, through Fourier transformation, to specific Generalized Transverse Momentum-Dependent Parton Distributions (GTMDs), which are off-forward TMDs. 
The correlation defining OAM corresponds to the GTMD  $F_{14}$ (we follow the naming scheme of Ref.\cite{Meissner:2009ww}). In particular, the OAM distribution is described by the $x$-dependent $k_T^2$ moment of $F_{14}$. 

The OAM term differs in the JM and Ji approaches with regard to how the gauge invariance of the theory intervenes through the gauge link in the relevant parton correlator \cite{Hatta:2011ku}. The difference was recently explicated in the quark sector in Refs.\cite{Hatta:2012cs,Burkardt:2012sd}, where it was shown that JM OAM, $L_q^{JM}$, can be written as the sum of Ji's OAM, $L_q^{Ji}$, plus a matrix element including the gluon field. The latter was interpreted in the semi-classical picture of Ref.\cite{Burkardt:2012sd} as having the physical meaning of an integrated torque stemming from the chromodynamic force between the struck quark and the proton remnant interacting in the final state. 

To summarize, in both Ji's and JM's expressions, OAM is defined through an imbalance in the distribution of the number density of quarks in longitudinally polarized proton states, when the quark's displacement in the transverse plane is simultaneously orthogonal to its intrinsic transverse motion. 
%
%
JM's definition includes a quark re-interaction which could be, in principle, process-dependent. 
%
%
How can these two pictures of the proton's angular momentum coexist, and what are experimental measurements really probing?

The work presented here was motivated by the question of defining a way to test these ideas through observables that would enable direct access to OAM in experimental measurements. While $J_{q,g}$ measurements through GPDs are in progress, 
GTMDs, providing in principle the density distributions for OAM, remain experimentally elusive objects, since they require exclusive measurements of particles in the two distinct hadronic planes disentangling the $k_T$ and $b_T$ (or $\Delta_T$) directions  \cite{Bhattacharya:2017bvs,Hagiwara:2016kam,Liuti:2017uxp}.  
GTMDs can, however, be evaluated in ab initio calculations \cite{Engelhardt:2017miy}.  

In a previous publication \cite{Rajan:2016tlg}, we showed that the $x$-dependent $k_T^2$ moment of $F_{14}$ entering Eq.~(\ref{eq:Ji_OAM}) can be written in terms of a twist-three GPD, $\widetilde{E}_{2T}$ \cite{Meissner:2009ww}, as
\begin{equation}
\label{eq:LIR2_alt}
\int d^2 k_T  \, \frac{ k_T^2}{M^2} \, F_{14} =
- \int_x^1 dy \, \left( \widetilde{E}_{2T} + H + E \right) 
\end{equation}
Here, we present several extensions of this relation, and describe the details of the derivation comprehensively. In particular, we show that a more general relation holds,
\begin{equation}
\label{eq:LIR3_alt}
\int d^2 k_T  \, \frac{ k_T^2}{M^2} \, F_{14} =
- \int_x^1 dy \, \left( \widetilde{E}_{2T} + H + E + {\cal A}_{F_{14} } \right)
\end{equation}
where ${\cal A}_{F_{14} } (x)$ is a term containing the gauge link dependent, or quark-gluon-quark, components of the correlation function. For a straight gauge link, ${\cal A}_{F_{14} }(x)=0$, thus recovering the result displayed in Eq.~(\ref{eq:LIR2_alt}).  
These relations are specific generalized Lorentz Invariance Relations (LIR) connecting the $x$-dependent $k_T^2$ moments of GTMDs and GPDs. Just as in the forward case \cite{Goeke:2003az,Goeke:2005hb,Kanazawa:2015ajw}, generalized LIR are based upon the covariant decomposition of the fully unintegrated quark-quark correlation function in off-forward kinematics: the number of independent functions parametrizing the correlator is less than the total number of GTMDs and GPDs, thus inducing relations among the latter. 
Several LIRs have been found between forward twist-three PDFs and $k_T$ moments of TMDs. The most remarkable example of an LIR is perhaps 
the relation between the TMD $g_{1T}$ and the twist-three PDF $g_T$, leading to the Wandzura-Wilczek relation between the helicity distribution $g_1$ and $g_T = g_1+ g_2$ \cite{Wandzura:1977qf}. In the presence of a gauge link other than the straight one ({\it e.g.} a staple link), LIRs acquire an additional term that cannot be encoded in the available GTMD and GPD structures. As we show in the present paper, this term produces a correction to Eq.~(\ref{eq:LIR2_alt}), leading eventually to the Qiu-Sterman type term of Ref.~\cite{Burkardt:2012sd}.       
Furthermore, by combining Eqs.~(\ref{eq:LIR2_alt},\ref{eq:LIR3_alt}) with the quark field Equations of Motion (EoM), we can ascribe the difference between the integrated quark total angular momentum, $J_q$, and the spin, $S_q \equiv (1/2) \Delta \Sigma_q$, in Ji's description to the integral of the Wandzura-Wilczek component of the GPD combination $\widetilde{E}_{2T} + H + E$. 
We find that, at the unintegrated level, a quark-gluon-quark term is also present which integrates to zero consistently with Ji's sum rule.  
Our relation, therefore, allows one to connect the partonic sum rule originating from the dynamical definition of OAM -- through the unintegrated correlation function --  and the gravitomagnetic form factors which define the energy-momentum tensor  (Eq.~(\ref{eq:JiSR})). On the other hand, having access to relations at the unintegrated level allows us to extend the treatment to the JM case, where we obtain that the quark-gluon-quark contribution does not vanish upon integration. We show it to reproduce the Qiu-Sterman type term in \cite{Burkardt:2012sd}.  

In principle, 32 individual EoM relations can be constructed, associated with the 8 twist-two GTMDs in the vector and axial-vector sectors, which each feature independent real and imaginary components; an additional doubling of the number of relations is given by contracting the EoMs in the transverse plane either with the transverse momentum $k_T$ or with the transverse momentum transfer $\Delta_T$. 
However, we place a special focus in the present paper on just three further relations besides Eq.~(\ref{eq:LIR2_alt}) \cite{Rajan:2016tlg} that describe spin correlations stemming from a similar operator structure as for OAM,
\begin{eqnarray}
\int d^2k_T \, \frac{k_T^2}{M^2} \, G_{11}  & = & \int_x^1 dy \, \left(2\widetilde{H}_{2T}' + E_{2T}'
  + \widetilde{H}- {\cal A}_{G_{11} } \right)\\
 \frac{1}{2} \int d^2k_T \, \frac{k_T^2}{M^2} \, G_{12} &=& - \int_x^1 dy \left(H_{2T}' -\frac{\Delta_T^2}{4M^2}E_{2T}'-
\left(1+\frac{\Delta_T^2}{4M^2}\right)
\widetilde{H} + {\cal A}_{G_{12} } \right) \\
\int d^2 k_T \, \frac{k_T^2}{M^2}  \, F_{12}^{o} &  \equiv & - f_{1T}^{\perp (1)}  = -\left. {\cal M}_{F_{12}} \right|_{\Delta_{T} =0}
\label{pres_sivers}
\end{eqnarray}
The three additional relations presented here for the first time involve
the $k_T^2$ moments of the following GTMDs: $G_{11}$, which was observed to provide information on the longitudinal part of the quark spin-orbit interaction, or the projection of quark OAM along the quark spin \cite{Lorce:2011kd}; $G_{12}$, which corresponds to a transverse proton spin configuration and generalizes the TMD $g_{1T}$ leading to the original Wandzura-Wilczek relation \cite{Wandzura:1977qf,Accardi:2009au},
and, finally, the naive T-odd part of $F_{12}$ which corresponds to the off-forward generalization of the Sivers function, $f_{1T}^\perp$ \cite{Sivers:1989cc}, which we relate to a generalized Qiu-Sterman term represented by ${\cal M}_{F_{12}}$ in Eq.~(\ref{pres_sivers}).    
For $G_{11}$, in particular, by using the EoM we find a relation whose integral in $x$ is consistent with the sum rule found in \cite{Kiptily:2002nx} and revisited in \cite{Lorce:2014mxa}. However, our derivation,
valid for arbitrary gauge link structure, allows for a new term representing final state interactions. Furthermore, we stress the importance of the term proportional to the quark mass which appears in this relation as being generated from quark transverse spin contributions.

This paper is organized as follows. In Section \ref{sec2} we define the general framework: the correlation functions, the gauge link structure, the parametrization of the correlation functions which ensues, and the helicity amplitudes; in Section \ref{sec3} we give a detailed derivation of the EoM relations, including explicit quark-gluon-quark terms; in Section \ref{sec4} we derive the LIRs for both OAM and spin-orbit correlations. We discuss their Mellin moments to order $n=3$; in Section \ref{sec5} we discuss the relations for transverse proton spin configurations and their connection to the forward limit; finally, in Section \ref{sec6} we give our conclusions and outlook.

\section{Formal framework and definitions}
\label{sec2}
We base our treatment on the
complete parametrization of the quark-quark correlation functions in the proton up to twist four given in Ref.\cite{Meissner:2009ww}. By applying time reversal invariance, charge conjugation, parity and hermiticity one finds that, at twist two, there are three independent PDFs: $f_1$, $g_1$, in the chiral even sector, and the chiral odd $h_1$; eight GPDs (four chiral even and four chiral odd); eight TMDs, and sixteen GTMDs. At twist three, one has many more functions due to both the presence of additional couplings (scalar, and pseudoscalar), and to the larger number of kinematical terms in the correlation function parametrizations for the vector, axial vector and pseudoscalar couplings. Each one of the PDFs, TMDs, and GPDs corresponds to specific quark-proton helicity amplitude combinations that can be extracted from various hard inclusive, semi-inclusive and deeply virtual exclusive  processes, respectively, and that represent specific polarization configurations, or spin correlations, of partons inside the proton. 

It is important to distinguish between different types of twist-three objects that will be dealt with in this paper.
{\it Canonical} twist three effects, describing quark-gluon correlations in the nucleon, appear in the OPE as coefficients of the inverse power terms in a large characteristic scale of the process, {\it e.g.} ${\cal O}(M/Q)$, $M$ being a nonperturbative mass scale. 
A different class of twist three effects, {\it geometrical}
twist three, arises from the quark field components which are not dynamically independent solutions of the equations of motion, and that can be expressed, through the equations of motion, as composites of the quark and gluon fields. These are also suppressed by inverse powers of $Q$ (the classification of parton distributions given above concerns this type of twist three objects).  The order of canonical and geometrical twist does not match beyond order two: 
contributions  with the same power in $M/Q$, or same dynamical twist, can be written in terms of matrix elements of operators with different canonical twist.
The Wandzura-Wilczek (WW) \cite{Wandzura:1977qf} relations between matrix elements of operators of different dynamical and same canonical twist encode this mismatch, 
as first exemplified for the polarized distribution functions $g_1$ and $g_2$. 

A complete set of relations between twist two TMDs and twist three PDFs was presented and discussed for various correlation functions in Refs.\cite{Goeke:2003az,Kanazawa:2015ajw}. These relations are based upon the Lorentz invariant decomposition of the fully unintegrated correlation function with the two quark fields located at different space-time positions, and they necessarily involve parton transverse momentum and off-shellness both through the $k_T$-moments of twist-two TMDs (where $k_T $ denotes the quark transverse momentum), and the twist-three PDFs.
The different kinds of twist three functions were renamed: {\it intrinsic} for geometric, {\it dynamic} for canonical, {\it i.e.}, when an extra gluon field operator is directly involved in the definition, and {\it kinematic} which are related to $k_T$-moments of TMDs.

These distinctions are useful to keep in mind as we extend both the Lorentz Invariance Relations (LIRs) and the Equation of Motion relations (EoMs) to off-forward kinematics involving GTMDs and their kinematic twist-three constructs, intrinsic twist-three GPDs, and off-forward dynamical twist-three terms.

Already the construction of the aforementioned relations between TMDs and PDFs, once taken beyond a purely formal level, encounters obstacles rooted in divergences of the $k_T $-integrations connecting TMDs to collinear objects such as the PDFs. These divergences must be separated off to ultimately contribute to the scale evolution of the collinear quantities. Our treatment similarly relates GTMDs to GPDs through $k_T $-integrations, and thus inherits these issues in complete analogy. In the present paper, we do not present any further developments on this topic beyond what is given in the literature on the connection between ordinary TMDs and PDFs. In general, the precise connection of GTMDs to GPDs still requires further specification. The relations we derive can also be read purely at the GTMD level, before identifying $k_T $-integrals of GTMDs with GPDs. In that form, all components of our relations can be regularized on an identical footing, before identifying their collinear limits. To the extent that our relations derive from symmetries (such as Lorentz invariance), any regularization that respects these symmetries can be expected to leave the relations we derive intact. At appropriate places in our treatment, we will indicate points at which modifications of our results must be countenanced owing to issues of regularization; an example is the standard deformation of TMD gauge links off the light cone, associated with the introduction of a Collins-Soper evolution parameter. This procedure applies likewise to a proper definition of GTMDs. We will also refrain from writing explicitly the soft factors that are required \cite{Echevarria:2016mrc} to regulate divergences associated with the gauge connections contained in the bilocal operators defining TMDs and GTMDs.

\subsection{Kinematics and correlators}
The completely unintegrated off forward quark-quark correlation function is defined as the matrix element between proton states with momenta and helicities $p, \Lambda$ and $p',\Lambda'$,
\begin{eqnarray}
\label{eq:unintcorr}
W_{\Lambda' \Lambda}^\Gamma(P,k,\Delta; {\cal U}) & = & \displaystyle\frac{1}{2}\int \frac{d^4 z} {(2 \pi)^4} e^{i k\cdot z}
 \left. \langle p', \Lambda' \mid 
 \bar{\psi} \left(-\frac{z}{2}\right) \Gamma {\cal U} \psi\left(\frac{z}{2}\right)   \mid p, \Lambda \rangle \right.,
\end{eqnarray} 
where the gauge link structure ${\cal U} $ connecting the quark operators at positions $-z/2$ and $z/2$ is discussed in detail in the next section, $\Gamma$ is a Dirac structure, $\Gamma = {\bf 1}, \gamma^5, \gamma^\mu,\gamma^\mu\gamma^5,i\sigma_{\mu\nu}$, and the choice of four-momenta is defined with $P=(p+p')/2$ along the z-axis,
$\Delta = p' -p$ as in Ref.\cite{Meissner:2009ww},
\begin{eqnarray}
\label{kinematics}
P & \equiv & \left(P^+, \frac{\Delta^2_T + 4M^2}{8(1-\xi^2)P^+},0\right)\stackrel{\xi=0}{=}\left(P^+, \frac{\Delta^2_T + 4M^2}{8P^+},0\right) \\
\Delta & \equiv & \left(-2\xi P^+, \frac{\xi(\Delta^2_T +4M^2)}{4(1-\xi^2)P^+}, \Delta_{T} \right) \stackrel{\xi=0}{=} \left(0, 0, \Delta_{T} \right)\\
k & \equiv & \left(xP^+, k^-,k_{T} \right)
\end{eqnarray}
where the initial and final quark momenta are $k-\Delta/2$ and $k+\Delta/2$, respectively. Four-vectors $w^{\mu } $ are represented in terms of light-cone components, $w^\mu \equiv(w^+,w^-,w_T)$; $\xi = -\Delta^+/2P^+$ is the skewness parameter, $\Delta_T \equiv(\Delta^1,\Delta^2)$, $k_T\equiv(k^1,k^2)$, and the four-momentum transfer squared is $\Delta^2 \equiv t$; we displayed the kinematics also specifically for the $\xi=0$ case, which is the case on which we will focus in this study.

\subsection{Gauge link structures}
\label{sec:upath}
\begin{figure}[h]
\includegraphics[width=10cm]{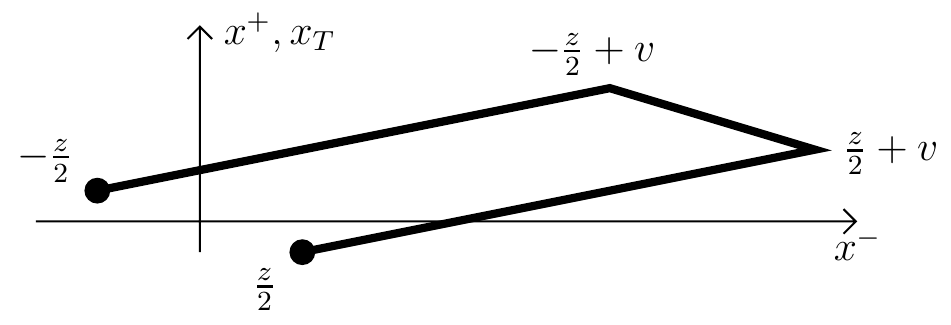}
\caption{Staple-shaped gauge link path connecting quark operators located at $-z/2$ and $z/2$. The legs of the staple are described by the four-vector $v$. GTMDs are defined at separation $z^{+} =0$; the vector $z$ thus deviates from the $x^{-} $ axis by a transverse component $z_T $, i.e., $z=(0,z^{-},z_T )$. On the other hand, $v$ in general is taken to deviate from the $x^{-} $ axis by a plus component $v^{+} $ in order to regulate rapidity divergences occurring if $v$ is taken to point purely in the minus direction; i.e., $v=(v^{+} ,v^{-} ,0)$. Note that, in the two-dimensional projection displayed, the $x^{+} $ and the $x_T $ axes fall on top of one another; they are nevertheless of course distinct axes. The separation $z$ is Fourier conjugate to the quark momentum $k$. Integrating over transverse momentum $k_T $ sets $z_T =0$, i.e., the quark operator positions then fall on the $x^{-} $ axis. Nevertheless, for $v^{+} \neq 0$, the path then still retains its staple shape. Only in the $v^{+} =0$ limit (staple legs become horizontal in figure) does the staple path collapse onto the $x^{-} $ axis upon $k_T $ integration, leading to a bona fide GPD limit in which all parts of the staple link cancel, except for a residual straight link directly connecting $-z/2$ to $z/2$. One can alternatively define GTMDs with a straight gauge link from the outset; in terms of the vectors defined in the figure, this simply corresponds to the limit $v=0$.}
\label{fig:staple}
\end{figure}

To ensure gauge invariance, the quark bilocal operator (\ref{eq:unintcorr}) requires
a gauge link ${\cal U}$ along a path connecting the quark operator positions
$-z/2$ and $z/2$. Two important choices of path are a direct
straight line and a staple-shaped connection characterized by an additional
vector $v$, cf.~Fig.~\ref{fig:staple}. These different choices will give rise to different genuine twist three contributions to the correlators.

The appropriate choice of gauge link path depends on the physical context. In the TMD limit, the staple-shaped gauge
link is most relevant, since it encodes final/initial state interactions
in SIDIS/DY processes. On the other hand, GPDs are defined with a
straight gauge link; as discussed in more detail below and displayed in Fig.~\ref{fig:staple}, only under certain circumstances do GTMDs with a staple-shaped gauge link have a proper GPD limit, with the staple link collapsing into a straight gauge link. In general, GTMDs defined from the outset with a straight gauge link play a separate role, and both the straight and staple-shaped gauge link choices will be treated in this work. Two specific motivations for doing so are the following:
\begin{itemize}
\item
In the context of quark orbital angular momentum, as accessed via the GTMD $F_{14} $ discussed in detail further below, both the straight and the staple-shaped gauge connections have a definite, distinct physical meaning \cite{Burkardt:2012sd}. A straight gauge link enters the definition of Ji quark orbital angular momentum \cite{Ji:2012sj}, whereas a staple-shaped gauge link generates Jaffe-Manohar quark orbital angular momentum \cite{Hatta:2011ku}. Note that $F_{14} $ is a genuine GTMD quantity, i.e., a quantity which does not have a TMD or GPD limit.
\item
A central aspect of the following treatment are Lorentz invariance relations (LIRs). In the staple link case, these contain twist-three contributions (frequently referred to as ``LIR violating terms'', though their role is to maintain Lorentz invariance) which do not reduce to GTMDs. To ascertain their concrete physical content in terms of quark-gluon-quark correlations, it is useful to combine the staple-link LIR with the straight-link LIR (in which these contributions are absent) as well as the straight and staple-link equations of motion. The resulting information is not directly available considering the staple link case alone.
\end{itemize}

In the most basic definition of GTMDs, a staple-shaped gauge link with
a staple direction vector $v$ on the light cone is chosen \cite{Meissner:2009ww},
such that $v$ has only a minus component, $v=(0,v^{-},0,0)$. On the other
hand, the quark operator separation $z$ is of the form
$z=(0,z^{-},z_T )$, with a two-dimensional transverse vector $z_T $. Note that $z$ is Fourier conjugate to the quark
momentum $k$, and GTMDs are defined in terms of $k^- $-integrated
correlators, setting $z^{+} =0$. Thus, when one forms the GPD limit
of GTMDs by integration over the transverse momentum $k_T $, one sets
$z_T =0$, and $v$ and $z$ then lie along one common axis.
In that case, the staple legs collapse onto that one common axis, the
parts of the staple legs extending beyond the region in between the
quark operators cancel, and one is left with a straight gauge link
connecting those operators, as is appropriate for GPDs.

However, such a light-cone choice of the staple direction $v$ meets with
rapidity divergences, which, in the application to TMDs, are commonly
regulated by taking $v$ off the light cone into the space-like region
\cite{Collins:2011zzd}. Then, $v$ is of the form $v=(v^{+} ,v^{-},0,0)$, and the GPD
limit ceases to be straightforward; even after integration over $k_T $,
i.e., setting $z_T =0$, $v$ and $z$ do not lie on a common axis and
the staple-shaped gauge link does not collapse onto a simple straight
link connecting the quark operators. The $k_T $-integrated quantities
formed in this way are not directly GPDs, but differ from GPDs by
contributions which formally vanish in the $v^{+} \rightarrow 0$
light-cone limit. An alternative possibility of treating this issue
arising with staple links is to modify the GTMD definition such that
correlators are not rigidly defined with $z^{+} =0$, but instead such
that the longitudinal part of $z$ is parallel to $v$ for any chosen $v$,
i.e., $z_L =(z^{+} ,z^{-} )$ is parallel to $v=(v^{+} ,v^{-} )$. In that
case, integration over $k_T $ does indeed lead to collapse of the staple
link into a straight gauge link, but this straight gauge link now does
not lie on the light cone anymore. In effect, in this way one generates
quasi-GPDs in the sense discussed by Ji \cite{Ji:2013dva}.

In the present treatment, both GTMDs defined from the beginning with
straight gauge links, as well as GTMDs defined with staple-shaped gauge
links will be discussed. For the latter case, the discussion will be
confined to the $v^{+} =0$ limit; $v^{+} \neq 0$ corrections will not be
worked out explicitly. However, it should be kept in mind that these
corrections may be important in future applications, and places where
they arise will be pointed out as appropriate below.

\subsection{Parametrization of unintegrated correlation function}
\label{sec2B}
We consider the parametrization of the completely unintegrated off-forward correlator, $W^{\Gamma}_{\Lambda \Lambda'}$ above, in terms of Generalized Parton Correlation Functions (GPCFs) for the vector, $\gamma^\mu$, and axial vector, $\gamma^\mu \gamma^5$, operators. As motivated above, we are also interested in the case of a straight gauge link; the parametrization given in \cite{Meissner:2009ww}, by contrast, is constructed for a staple-shaped gauge link, and its form was chosen such that it is not straightforwardly related to the straight-link case.

In this respect, it should be noted that there is considerable freedom in constructing GPCF parametrizations. This is due to the fact that not all Lorentz structures one can write down are independent of one another; they are related by Gordon identities and other relations, as laid out in detail in \cite{Meissner:2009ww}. After exhausting these relations, 16 GPCFs $A^F_i $ remain to parametrize the staple-link vector correlator, and also 16 GPCFs $A^G_i $ remain to parametrize the staple-link axial vector correlator. In the straight-link case, to be discussed in more detail below, 8 GPCFs remain in each case. The staple-link parametrizations given in \cite{Meissner:2009ww} in neither case contain 8 GPCFs relevant for the straight-link case; some of these were instead chosen to be eliminated in favor of terms intrinsically related to a staple-link structure. The vector correlator parametrization of \cite{Meissner:2009ww} contains only 7 GPCFs relevant for the straight-link case; one additional one therefore has to be reinstated. The axial vector correlator parametrization of \cite{Meissner:2009ww} contains only 3 GPCFs relevant for the straight-link case, and therefore 5 have to be reinstated. Thus, one cannot simply delete the Lorentz structures containing the staple direction vector $v$ (denoted $N$ in \cite{Meissner:2009ww}, up to a rescaling) from the parametrizations given in \cite{Meissner:2009ww} and already arrive at a valid straight-link parametrization. Additional terms are needed, as given below. It would be possible to construct staple-link parametrizations differing from the ones in \cite{Meissner:2009ww}, each containing a full set of 8 structures relevant for the straight-link case, and each an additional 8 structures containing the staple direction $v$, such that deletion of the latter 8 immediately leads to a valid straight-link parametrization. We do not pursue this here to the full extent, but only give the straight-link parametrizations.

In the case of the vector correlator, this is rather simple. The construction of the staple-link parametrization in \cite{Meissner:2009ww} can be followed verbatim even in the straight-link case, merely omitting all structures containing the staple direction vector $v$, except for the very last step. In that very last step, the single missing straight-link structure, namely, $i\sigma^{k\Delta } \Delta^{\mu } $, is eliminated in favor of a staple-link related structure. In the straight-link case, the staple-link related structure is not available, and therefore the aforementioned straight-link structure must be kept instead. Thus, one has the straight-link vector correlator parametrization\footnote{We use the notation $\sigma^{\mu a}= \sigma^{\mu \nu} a_\nu$, and  $\sigma^{a b}= \sigma^{\mu \nu} a_\mu b_\nu$. }
\begin{eqnarray}
\label{vector_GPCF} 
W^{\gamma^\mu}_{\Lambda' \Lambda} &=&  \overline{U}(p',\Lambda') \left[\frac{P^\mu}{M} A_1^F + \frac{k^\mu}{M} A_2^F + \frac{\Delta^\mu}{M} A_3^F + \frac{ i\sigma^{\mu k}}{M} A_5^F + \frac{ i\sigma^{\mu\Delta}}{M} A_6^F + \frac{i\sigma^{k\Delta}}{M^2}\left(\frac{P^\mu}{M} A_8^F + \frac{k^\mu}{M} A_9^F +\frac{\Delta^\mu}{M} A_{17}^F\right) \right]U(p,\Lambda)\nonumber \\
&&
\end{eqnarray}
where the first 7 terms are identical to the ones given in \cite{Meissner:2009ww}, and the last one, containing the additional invariant amplitude $A^F_{17}$, is associated with the aforementioned missing Lorentz structure.

The case of the axial vector correlator is more involved, and the complete construction of the straight-link parametrization is given in Appendix \ref{app:axialGPCFs}. We arrive at the form 
\begin{eqnarray}
\label{axialvector_GPCF}
W^{\gamma^\mu \gamma^5}_{\Lambda' \Lambda}  &=& \overline{U}(p',\Lambda')\left[\frac{i\epsilon^{\mu P k \Delta}}{M^3} A_1^G +
\frac{i\sigma^{P\mu}\gamma^5}{M} A_{17}^G + \frac{i\sigma^{Pk}\gamma^5}{M^2}\left(\frac{P^\mu}{M} A_{18}^G + \frac{k^\mu}{M} A_{19}^G + \frac{\Delta^\mu}{M} A_{20}^G \right) \right. \\
& & \hspace{8cm} + \left. \frac{i\sigma^{P\Delta}\gamma^5}{M^2}\left(\frac{P^\mu}{M} A_{21}^G + \frac{k^\mu}{M} A_{22}^G
+\frac{\Delta^\mu}{M} A_{23}^G \right)\right]U(p,\Lambda) \nonumber
\end{eqnarray}
which in fact has only one term in common with the staple-link parametrization given in \cite{Meissner:2009ww}, namely, the one associated with the invariant amplitude $A^G_1 $; we make choices differing from the ones in \cite{Meissner:2009ww} even within the straight-link sector.
All GPCFs in these straight-link parametrizations are functions of $k^2, k\cdot P, k\cdot \Delta, \Delta^2 , P\cdot \Delta $.
In staple-link parametrizations, such as the ones given in \cite{Meissner:2009ww}, the GPCFs additionally depend on all scalar products involving the additional vector $v$ characterizing the staple link.

It is interesting to note that, for both the vector and axial vector operators,  8 GPCFs enter the parametrization for the straight gauge link case. This is the same as the total number of GPDs (including twist 2, twist 3 and twist 4). This is expected because the GPDs are defined with quarks separated only along the light cone. The number of GTMDs on the other hand is 16. Because the underlying structure functions, the GPCFs, are fewer in number, we expect the GTMDs to be connected to one another. These relations between the GTMDs are known as the Lorentz Invariance Relations and we discuss them in Sec.~\ref{sec4}. 

\subsection{Generalized Transverse Momentum-Dependent Parton Distributions}
\label{sec2C}
The unintegrated correlator definining the Generalized Transverse Momentum-Dependent Parton Distributions (GTMDs) is given by,
\begin{eqnarray}
\label{eq:GTMDcorr}
W_{\Lambda' \Lambda }^\Gamma(P,x,k_T,\xi, \Delta_{T} ; {\cal U} ) & = & \int d k^- W_{\Lambda' \Lambda}^\Gamma(P,k,\Delta ; {\cal U} ) \nonumber\\
&=& \displaystyle \frac{1}{2}\int \frac{d z^- \, d^2 z_T}{(2 \pi)^3} e^{ixP^+ z^- - i k_T\cdot z_T} \left. \langle p', \Lambda' \mid 
\left. \bar{\psi} \left(-\frac{z}{2}\right) \Gamma {\cal U} \psi\left(\frac{z}{2}\right)   \mid p, \Lambda \rangle \right. \right|_{z^+=0}. 
\end{eqnarray} 
Its parametrization in terms of GTMDs, as defined in Ref.\cite{Meissner:2009ww}, reads as follows.\footnote{Note that the form of this GTMD parametrization, as well as the GPD parametrization exhibited further below, is independent of the choice of gauge link, contrary to the GPCF parametrization discussed above. Thus, the relations between GTMDs and GPDs given for staple-shaped gauge links in \cite{Meissner:2009ww} remain true for straight gauge links.}
For $\Gamma= \gamma^+ $, $\gamma^+ \gamma^5$, $i\sigma^{i+} \gamma^{5} $, one has,
\begin{eqnarray}
\label{vector}
W^{\gamma^+}_{\Lambda' \Lambda} & = & 
\frac{1}{2M}\overline{U}(p',\Lambda') \left[F_{11} + \frac{ i\sigma^{i +} k^i}{P^+} F_{12} + \frac{ i\sigma^{i +}\Delta^i}{P^+} F_{13} + \frac{i\sigma^{ij} k^i \Delta^{j} }{M^2}  F_{14}  \right]U(p,\Lambda) \\
& = & 
\left[F_{11} +\frac{i\Lambda\epsilon^{ij}k^i\Delta^j}{M^2}F_{14}\right]\delta_{\Lambda'\Lambda} + \left[\frac{\Lambda\Delta^1 +i\Delta^2}{2M}(2F_{13} - F_{11}) + \frac{\Lambda k^1 + i k^2}{M} F_{12}\right]\delta_{-\Lambda'\Lambda}
\label{vectorimf}
\\
\label{axial}
W^{\gamma^+\gamma^5}_{\Lambda' \Lambda} 
& = & \frac{1}{2M}\overline{U}(p',\Lambda') \left[-\frac{i\epsilon^{ij} k^i \Delta^j}{M^2}G_{11} + \frac{ i\sigma^{i +}\gamma^5 k^i}{P^+} G_{12} + \frac{ i\sigma^{i +}\gamma^5\Delta^i}{P^+} G_{13} + i\sigma^{+ -}\gamma^5  G_{14}  \right]U(p,\Lambda) \\
&=&  \left[ -\frac{i(k^1\Delta^2 -k^2\Delta^1)}{M^2} G_{11} + \Lambda G_{14}\right]\delta_{\Lambda'\Lambda} \nonumber \\
& & \hspace{3cm} +\left[\frac{\Delta^1 +i\Lambda\Delta^2}{M}\left(G_{13}  + \frac{i\Lambda(k^1\Delta^2-k^2\Delta^1)}{2M^2}G_{11}\right) + \frac{k^1 + i \Lambda k^2}{M} G_{12}\right]\delta_{-\Lambda' \Lambda}\label{axialimf} \\
W^{i\sigma^{i+}\gamma^5}_{\Lambda'\Lambda} &=& \frac{1}{2M}\overline{U}(p',\Lambda')\left[i\epsilon^{ij}\left(\frac{k^j}{M}H_{11} +\frac{\Delta^j}{M}H_{12}\right) + \frac{Mi\sigma^{i+}\gamma^5}{P^+}H_{13} + \frac{k^i i\sigma^{k+}\gamma^5 k^k}{MP^+} H_{14} \right.\nonumber\\ 
& & +\left.\frac{\Delta^i i\sigma^{k+}\gamma^5 k^k}{MP^+} H_{15} +
\frac{\Delta^i \, i\sigma^{k+}\gamma^5\Delta^k}{MP^+} H_{16} + \frac{k^ii\sigma^{+-}\gamma^5}{M} H_{17} + \frac{\Delta^i \, i\sigma^{+-}\gamma^5}{M}H_{18} \right] U(p,\Lambda)\\
&=& \left[i\epsilon^{ij}\left(\frac{k^j}{M} H_{11} + \frac{\Delta^j}{M}H_{12}\right) +\Lambda \left(\frac{k^i}{M} H_{17} + \frac{\Delta^i}{M}H_{18}\right)\right]\delta_{\Lambda'\Lambda} + \left[- i\epsilon^{ij}\frac{\Lambda\Delta^1 + i\Delta^2}{2M}\left(\frac{k^j}{M}H_{11} + \frac{\Delta^j}{M}H_{12}\right) \right. \nonumber\\
& & +\left.  (\delta_{i1} + i\Lambda\delta_{i2})H_{13} + \frac{k^1 + i\Lambda k^2}{M}\left(\frac{k^i}{M} H_{14} + \frac{\Delta^i}{M}H_{15}\right) +\frac{(\Delta^1 +i\Lambda\Delta^2)\Delta^i}{M^2} H_{16} \right]\delta_{-\Lambda'\Lambda}
\end{eqnarray}
For each correlator listed, the second equality follows once $P^{+} $ is taken to be much larger than all other mass scales. On the other hand, for $\gamma^i, \gamma^i \gamma^5$, one has
\begin{eqnarray}
\label{eq:tw3vec_metz}
W^{\gamma^i}_{\Lambda' \Lambda} &=&
\frac{1}{2P^{+} }\overline{U}(p',\Lambda') \left[ \frac{k^i }{M} F_{21} +\frac{\Delta^{i} }{M} F_{22} + \frac{Mi\sigma^{i+} }{P^{+} } F_{23} +\frac{k^i i\sigma^{k+} k^k }{MP^{+} } F_{24} \right. \nonumber \\
& & \left. + \frac{\Delta^{i} i\sigma^{k+} k^k }{MP^{+} } F_{25} + \frac{\Delta^{i} i\sigma^{k+} \Delta^{k} }{MP^{+} } F_{26} + \frac{i\sigma^{ji} k^j }{M} F_{27} + \frac{i\sigma^{ji} \Delta^{j} }{M} F_{28} \right] U(p,\Lambda ) \\
& = & \left[ \frac{k^i}{P^+} F_{21} +    \frac{\Delta^{i} }{P^+} F_{22}  - i \Lambda \epsilon^{ij} \left(\frac{k^j}{P^+}   F_{27} + \frac{\Delta^j}{P^+} F_{28} \right) \right] \, \delta_{\Lambda' \Lambda} + 
\frac{1}{2P^+}\left[ - \frac{i\Delta^2+\Lambda\Delta^1}{M}(k^i F_{21} + \Delta^i F_{22}) \right.\nonumber \\
& + & \left. 2M(\Lambda\delta_{i1} +i\delta_{i2})F_{23} 
+ \frac{2}{M}(\Lambda k^1 + i k^2)(k^i F_{24} + \Delta^i F_{25})+\frac{2}{M}(\Lambda \Delta^1 + i \Delta^2)\Delta^i F_{26}\right ]\delta_{-\Lambda' \Lambda} 
\label{eq:tw3vec_imf}
\\
\label{eq:tw3axvec_metz}
W^{\gamma^i \gamma^5}_{\Lambda' \Lambda} &=&
\frac{1}{2P^{+} } \overline{U} (p',\Lambda' ) \left[ -\frac{i\epsilon^{ji} k^j }{M} G_{21} - \frac{i\epsilon^{ji} \Delta^{j} }{M} G_{22} + \frac{Mi\sigma^{i+} \gamma^{5} }{P^{+} } G_{23} + \frac{k^i i\sigma^{k+} \gamma^{5} k^k }{MP^{+} } G_{24} \right. \nonumber \\
& & \left. + \frac{\Delta^{i} i\sigma^{k+} \gamma^{5} k^k }{MP^{+} } G_{25} + \frac{\Delta^{i} i\sigma^{k+} \gamma^{5} \Delta^{k} }{MP^{+} } G_{26} + \frac{k^i i\sigma^{+-} \gamma^{5} }{M} G_{27} + \frac{\Delta^{i} i\sigma^{+-} \gamma^{5} }{M} G_{28} \right] U(p,\Lambda ) \\
& = &\left[ i\epsilon^{ij}(\frac{k^j}{P^+} G_{21} + \frac{\Delta^j}{P^+} G_{22}) + \Lambda (\frac{k^i}{P^+} G_{27} + \frac{\Delta^i}{P^+} G_{28}) \right] \delta_{\Lambda' \Lambda} +  \frac{1}{2P^+}\left[-i\epsilon^{ij} \frac{i\Delta^2+\Lambda\Delta^1}{M}(k^j G_{21} + \Delta^j G_{22}) \right.  \nonumber 
\\ &+& \left. 2 M(\delta_{i1} +i\Lambda \delta_{i2})G_{23}+ \frac{2}{M}(k^1 + i\Lambda k^2)(k^i G_{24} + \Delta^i G_{25})+\frac{2}{M}(\Delta^1 + i \Lambda\Delta^2)\Delta^i G_{26}\right] \delta_{-\Lambda' \Lambda}
\label{eq:tw3axvec_imf}
 \end{eqnarray}

The GTMDs considered here are complex functions of the set of kinematical variables $x,\xi, k_T^2, k_T \cdot \Delta_T, t$; in the case of a staple-shaped gauge link, they furthermore depend on the vector $v$ characterizing the staple,
\begin{equation}
\label{eq:GTMD}
X(x,\xi, k_T^2, k_T \cdot \Delta_T, t, v)= X^e(x,\xi, k_T^2, k_T \cdot \Delta_T, t, v)+ i X^o(x,\xi, k_T^2, k_T \cdot \Delta_T, t, v)
\end{equation}
with $X=F_{1j}, G_{1j}, H_{1j}$, at twist two, and $X=F_{2j}, G_{2j}$, at twist three. $X^e$ is symmetric under $v\rightarrow -v$ ($T$-even), while $X^o$ reverses its sign for $v\rightarrow -v$ (T-odd).
Due to Hermiticity and time reversal invariance, we have that the following GTMD components are odd for $\xi \rightarrow - \xi$, $k_T \cdot \Delta_T \rightarrow -k_T \cdot \Delta_T$,
\begin{subequations}
\label{hermiticity}
\begin{eqnarray}
&& F_{12}^e, F_{22}^e, F_{23}^e, F_{24}^e, F_{26}^e, F_{27}^e, 
G_{13}^e, G_{21}^e, G_{25}^e, G_{28}^e, H_{11}^e, H_{15}^e, H_{18}^e \\
&& F_{11}^o, F_{13}^o, F_{14}^o, F_{21}^o, F_{25}^o, F_{28}^o, G_{11}^o, G_{12}^o, G_{14}^o, G_{22}^o, G_{23}^o, G_{24}^o, G_{26}^o, G_{27}^o, H_{12}^o, H_{13}^o, H_{14}^o, H_{16}^o, H_{17}^o
\end{eqnarray}
\end{subequations}
This influences which $k_T $-moments of these GTMDs can appear in the $\xi =0$ case.

\subsection{Generalized Parton Distributions}
\label{sec2D}
The Generalized Parton Distributions (GPDs) are obtained by formally integrating Eq.(\ref{eq:GTMDcorr}) over the transverse parton momentum, $k_T$, provided that the gauge link has the appropriate form, cf.~the discussion in Sec.~\ref{sec:upath},
\begin{eqnarray}
\label{eq:GPDcorr}
F_{\Lambda' \Lambda}^\Gamma(x,\xi,t) & = & \displaystyle\frac{1}{2}\int \frac{d z^-}{2 \pi} e^{ixP^+ z^-} \left. \langle p', \Lambda' \mid 
 \bar{\psi} \left(-\frac{z}{2}\right) \Gamma {\cal U} \psi\left(\frac{z}{2}\right)   \mid p, \Lambda \rangle \right|_{z^+=0,z_T=0} , 
\end{eqnarray} 
For $\gamma^+, \gamma^+ \gamma^5, i\sigma^{i+}\gamma^5$ one has,
\begin{eqnarray}
\label{GPDvec}
F^{\gamma^+}_{\Lambda' \Lambda } &=&\frac{1}{2P^+}\overline{U}(p',\Lambda')\left[\gamma^+ H + \frac{i\sigma^{+\Delta}}{2M}E\right]U(p,\Lambda)
= H \delta_{\Lambda,\Lambda'} +\frac{ (\Lambda\Delta^1 +i\Delta^2)}{2M}E\delta_{-\Lambda,\Lambda'} \\
\label{GPDaxvec}
F^{\gamma^+\gamma^5}_{\Lambda' \Lambda } &=& \frac{1}{2P^+}\overline{U}(p',\Lambda')\left[\gamma^+\gamma^5 \widetilde{H} + \frac{\Delta^+\gamma^5}{2M}\widetilde{E}\right]U(p,\Lambda)
= \Lambda\widetilde{H} \delta_{\Lambda,\Lambda'} +\frac{ (\Delta^1 +i\Lambda\Delta^2)}{2M}\xi\widetilde{E}\delta_{-\Lambda,\Lambda'} 
\\
\label{GPDtensor}
F^{i\sigma^{i+}\gamma^5}_{\Lambda' \Lambda } &=& \frac{i\epsilon^{ij}}{2P^+}\overline{U}(p',\Lambda')\left[i\sigma^{+j}H_T +\frac{\gamma^+\Delta^j - \Delta^{+} \gamma^{j} }{2M}E_T +\frac{P^+\Delta^j }{M^2}\widetilde{H}_T -\frac{P^+\gamma^j}{M}\widetilde{E}_T\right]U(p,\Lambda)
\\
&=& \left[\frac{i\epsilon^{ij}\Delta^j}{2M}(E_T +2\widetilde{H}_T) +\frac{\Lambda\Delta^i}{2M}\left(\widetilde{E}_T -\xi E_T \right) \right]\delta_{\Lambda\Lambda'} +\left[(\delta_{i1} + i\Lambda \delta_{i2})H_{T} - \frac{i\epsilon^{ij}\Delta^j(\Lambda\Delta^1 + i\Delta^2)}{2M^2}\widetilde{H}_T\right]\delta_{-\Lambda\Lambda'} \nonumber \\
& &
\end{eqnarray}
whereas for $\gamma^i, \gamma^i \gamma^5$, 
\begin{eqnarray}
F^{\gamma^i}_{\Lambda' \Lambda } &=&
\frac{M}{2(P^+)^2}\overline{U}(p',\Lambda')\left[i\sigma^{+i}H_{2T} +\frac{\gamma^+\Delta^i - \Delta^{+} \gamma^{i} }{2M}E_{2T} + \frac{P^+\Delta^i}{M^2}\widetilde{H}_{2T}-\frac{P^+\gamma^i}{M}\widetilde{E}_{2T}\right]U(p,\Lambda)
\\
&=& \left[\frac{\Delta^i}{2P^+}E_{2T} +\frac{\Delta^i}{P^+}\widetilde{H}_{2T} +\frac{i\Lambda\epsilon^{ij}\Delta^j}{2P^+}\left( \widetilde{E}_{2T} -\xi E_{2T} \right) \right]\delta_{\Lambda \Lambda'} \\
& & \hspace{6cm}
+ \left[\frac{-M(\Lambda\delta_{i1} + i\delta_{i2})}{P^+} H_{2T} -\frac{(\Lambda\Delta^1 +i\Delta^2)\Delta^i}{2MP^+}\widetilde{H}_{2T} \right]\delta_{\Lambda-\Lambda'} \nonumber \\
F^{\gamma^i\gamma^5}_{\Lambda' \Lambda } &=& \frac{i\epsilon^{ij}M}{2(P^+)^2}\overline{U}(p',\Lambda')\left[i\sigma^{+j}H_{2T}' +\frac{\gamma^+\Delta^j -\Delta^{+} \gamma^{j} }{2M}E_{2T}' + \frac{P^+\Delta^j}{M^2}\widetilde{H}_{2T}'-\frac{P^+\gamma^j}{M}\widetilde{E}_{2T}'\right]U(p,\Lambda) 
\\
&=&\left[\frac{i\epsilon^{ij}\Delta^j}{2P^+}E_{2T}' +\frac{i\epsilon^{ij}\Delta^j}{P^+}\widetilde{H}_{2T}' -\frac{\Lambda\Delta^i}{2P^+}\left( \widetilde{E}_{2T}' -\xi E_{2T}' \right) \right]\delta_{\Lambda\Lambda'} \\
& & \hspace{6cm} + \left[\frac{ M(\delta_{i1} + i\Lambda\delta_{i2})}{P^+} H_{2T}' -\frac{i\epsilon^{ij}(\Lambda\Delta^1 +i\Delta^2)\Delta^j}{2MP^+}\widetilde{H}_{2T}' \right]\delta_{\Lambda-\Lambda'} \nonumber
\end{eqnarray}
The gauge connection for GPDs is a straight link, implying that all GPDs are naive T-even. We use the GPD parametrization from Ref.\cite{Meissner:2009ww}. 
As in the first parametrization introduced by Ji \cite{Ji:1996ek}, the letter $H$ signifies that in the forward limit these GPDs correspond to a PDF, while the ones denoted by $E$ are completely new functions; $H$, $E$, $\widetilde{H}$, $\widetilde{E}$ parametrize the chiral-even quark operators. In the chiral-odd sector, $H_T$, $E_T$, $\widetilde{H}_T$, $\widetilde{E}_T$ describe the tensor quark operators, the subscript $T$ signifying that the quarks flip helicity or are transversely polarized \cite{Diehl:2001pm}.  
The matrix structures that enter the twist three vector $(\gamma^i)$ and axial vector $(\gamma^i\gamma^5)$ cases are identical to the ones occurring at the twist two level in the chiral-odd tensor sector. 
Hence, the GPDs have similar names: 
the corresponding twist three GPD, occurring with the same matrix coefficient, is named $F_{2T}$ if parametrizing the vector case $\gamma^i$ and $F_{2T}'$ if parametrizing the axial vector case $\gamma^i\gamma^5$, with $F=H,E,\widetilde{H}, \widetilde{E}$.

\subsection{Helicity Structure}
\label{sec2e}
To elucidate the helicity structure, which is needed to connect to phenomenological applications and which also serves as a heuristic tool in the construction of LIR and EoM relations below, we introduce the quark-proton helicity amplitudes, \cite{Diehl:2003ny}, 
\begin{eqnarray}
A_{\Lambda' \lambda', \Lambda \lambda} = \int \frac{d z^- \, d^2 z_T}{(2 \pi)^3} e^{ixP^+ z^- - i k_T\cdot z_T} \left. \langle p', \Lambda' \mid {\cal O}_{\lambda' \lambda}(z) \mid p, \Lambda \rangle \right|_{z^+=0},
\end{eqnarray} 
where at twist two the bilocal quark field operators, 
\begin{eqnarray}
\label{tw2_operator}
{\cal O}_{\pm \pm}(z) & = & \frac{1}{4}\bar{\psi}\left(-\frac{z}{2}\right) \gamma^+(1 \pm \gamma^5)  \psi\left(\frac{z}{2}\right) \equiv \phi_\pm^\dagger \phi_\pm
\end{eqnarray}
define (non flip) transitions between quark $\pm, \pm$ helicity states. Note that, in this section only, for the purpose of discussing helicity structure, we drop the gauge link in the bilocal operators to simplify notation.

The various LIRs and EoM relations that we derive in subsequent sections correspond to different helicity combinations obtained varying the initial and final proton helicity states. We obtain 8 distinct relations from the following combinations, $(+,+) \pm (-,-)$, and $(+,-) \pm (-,+)$, in the vector and axial vector sector, respectively. 
In what follows we derive all four spin correlations.  

%
The correlation functions in Eqs.(\ref{vectorimf},\ref{axialimf}) can be written in terms of the quark-proton helicity amplitudes as,
\begin{eqnarray}
W^{\gamma^+}_{\Lambda' \Lambda } & = & A_{\Lambda' +,\Lambda +} +  A_{\Lambda' -,\Lambda -} \\
W^{\gamma^+\gamma^5}_{\Lambda' \Lambda } & = & A_{\Lambda' +,\Lambda +} -  A_{\Lambda' -,\Lambda -}.
\end{eqnarray}
One finds the following expressions for the proton non flip terms, 
\begin{subequations}
\label{eq:FGnonflip}
\begin{eqnarray}
\label{F11}
 F_{11} & = & \frac{1}{2} (W^{\gamma^+}_{+ +} + W^{\gamma^+}_{- -} )= \frac{1}{2}  (A_{++,++} + A_{+-,+-} + A_{-+,-+} + A_{--,--} )\\
\label{F14}
 i  \frac{({\bf k}_T \times {\bf \Delta}_T )_3 }{M^2} F_{14} & = & \frac{1}{2}  (W^{\gamma^+}_{+ +} - W^{\gamma^+}_{- -} )= \frac{1}{2}  (A_{++,++} + A_{+-,+-} - A_{-+,-+} - A_{--,--} ) \\
\label{G14}
 G_{14} & = & \frac{1}{2}  (W^{\gamma^+\gamma^5}_{+ +} - W^{\gamma^+\gamma^5}_{- -} )= \frac{1}{2}  (A_{++,++} - A_{+-,+-} - A_{-+,-+} + A_{--,--} )\\
\label{G11}
-i  \frac{ ({\bf k}_T \times {\bf \Delta}_T )_3 }{M^2} G_{11} & = & \frac{1}{2}  (W^{\gamma^+\gamma^5}_{+ +} + W^{\gamma^+\gamma^5}_{- -} ) = \frac{1}{2}  (A_{++,++} - A_{+-,+-}  + A_{-+,-+} - A_{--,--}), 
 \end{eqnarray}
\end{subequations}
where, because of the constraints in Eqs.(\ref{hermiticity}),  the combinations on the {\it rhs} of Eqs.(\ref{F11}, \ref{G14}) and Eqs.(\ref{F14},\ref{G11}) are purely real and imaginary, respectively.

The distributions in both transverse coordinate and momentum space corresponding to these GTMDs were analyzed in detail in Refs.\cite{Lorce:2011ni,Lorce:2011kd}. 
$F_{11}$ describes an unpolarized quark and proton state, and it reduces to the PDF $f_1$ in the forward, $k_T$ integrated, limit; $G_{14}$ describes the quark helicity distribution, or $g_1$ in the forward, $k_T$ integrated, limit. $F_{14}$ and $G_{11}$ do not have GPD or TMD limits. 
However, in the forward limit, their average over $k_T$ weighted by $k_T^2 $ gives \cite{Lorce:2011ni},
\begin{eqnarray}
\label{eq:F14OAM}
(L_q)_3  &=& \int dx \int d^2 k_T  \frac{1}{2} \left({\bf k}_T \times i \frac{\partial}{ \partial {\bf \Delta}_T}\right)_3 \left( W_{++}^{\gamma^{+} } - W_{--}^{\gamma^{+} } \right) = -\int dx \int d^2 k_T  \frac{k_T^2}{M^2} F_{14}  \\
\label{eq:G11LdotS}
2(L_q)_3 (S_q)_3 &=&  \int dx \int d^2 k_T  \frac{1}{2} \left({\bf k}_T \times i\frac{ \partial}{ \partial {\bf \Delta}_T}\right)_3 \left( W_{++}^{\gamma^{+} \gamma^{5} } + W_{--}^{\gamma^{+} \gamma^{5} } \right) = \int dx  \int d^2 k_T \frac{k_T^2}{M^2} G_{11} \quad,
\end{eqnarray}
where 
Eq.~(\ref{eq:F14OAM}) represents the quark OAM along the $z$ axis in a longitudinally polarized proton, while Eq.(\ref{eq:G11LdotS}) gives the quark OAM along the $z$ axis for a longitudinally polarized quark, or a spin-orbit term. 

The proton spin flip terms read,
\begin{subequations}
\label{eq:FGflip}
\begin{eqnarray}
\label{F12F13}
- \frac{i( {\bf k}_T\times {\bf \Delta}_T )_3 }{M}F_{12}&=& \frac{1}{2}((\Delta^1 -i\Delta^2)W^{\gamma^+}_{- +} +(\Delta^1 + i\Delta^2)W^{\gamma^+}_{+ -}) \nonumber  \\
&=& \frac{1}{2}\left((\Delta^1 - i\Delta^2)(A_{-+, ++} + A_{--,+-})+(\Delta^1 + i\Delta^2)(A_{++, -+} + A_{+-,--}) \right)\nonumber\\
&&\\
 \frac{k_T\cdot\Delta_T}{M}F_{12} + \frac{\Delta_T^2}{2M}(2F_{13} - F_{11}) &=& \frac{1}{2}((\Delta^1 -i\Delta^2)W^{\gamma^+}_{- +} - (\Delta^1 + i\Delta^2)W^{\gamma^+}_{+ -})
 \nonumber \\
&=& \frac{1}{2}\left((\Delta^1 - i\Delta^2)(A_{-+, ++} + A_{--,+-}) - (\Delta^1 +i\Delta^2)(A_{++, -+} + A_{+-,--}) \right)
\end{eqnarray}
\end{subequations}
and,
\begin{subequations}
\label{eq:FGflip2}
\begin{eqnarray}
\label{G12G13}
\frac{\Delta_T^2}{M}G_{13} + \frac{k_T\cdot\Delta_T}{M}G_{12}&=&\frac{1}{2}((\Delta^1 -i\Delta^2)W^{\gamma^+\gamma^5}_{-+} + (\Delta^1 +i\Delta^2)W^{\gamma^+\gamma^5}_{+-}) \nonumber \\
&=&\frac{1}{2}\left((\Delta^1 - i\Delta^2)(A_{-+, ++} - A_{--,+-}) + (\Delta^1 + i\Delta^2)(A_{++, -+} - A_{+-,--}) \right) \nonumber\\
&&\\
\frac{i({\bf k}_T\times {\bf \Delta}_T )_3 }{M}\left(\frac{\Delta_T^2}{2M^2} G_{11} - G_{12}\right)&=&\frac{1}{2}((\Delta^1 -i\Delta^2)W^{\gamma^+\gamma^5}_{-+} - (\Delta^1 +i\Delta^2)W^{\gamma^+\gamma^5}_{+-}) \nonumber \\
&=&\frac{1}{2}\left((\Delta^1 + i\Delta^2)(A_{-+, ++} - A_{--,+-}) - (\Delta^1 +i\Delta^2)(A_{++, -+} - A_{+-,--}) \right)\nonumber\\
&&
\end{eqnarray}
\end{subequations}
At twist three, the bilocal operators can be written as the overlap of a dynamically independent quark field, $\phi$ (good component), and a dynamically dependent  quark-gluon composite field, $\chi$ (bad component) \cite{Jaffe:1996zw}, 
\begin{subequations}
\begin{eqnarray}
\label{tw3_operator}
{\cal O}^q_{-^* +}(z) & = & \frac{1}{8} \bar{\psi} \left(-\frac{z}{2}\right) (\gamma^1 - i\gamma^2) (1 + \gamma^5)  \psi  \left(\frac{z}{2}\right) = \chi^\dagger_{+} \phi_{+}  \\
{\cal O}^q_{+ -^*}(z) & = & \frac{1}{8} \bar{\psi}\left(-\frac{z}{2}\right) (\gamma^1 + i\gamma^2) (1 + \gamma^5)  \psi  \left(\frac{z}{2}\right)  =  \phi^\dagger_{+} \chi_{+}   \\
{\cal O}^q_{+^* -}(z) & = &  -\frac{1}{8} \bar{\psi} \left(-\frac{z}{2}\right) (\gamma^1 + i\gamma^2) (1 - \gamma^5)  \psi  \left(\frac{z}{2}\right) = - \chi^\dagger_{-} \phi_{-} \\
{\cal O}^q_{- +^*}(z) & = & - \frac{1}{8} \bar{\psi} \left(-\frac{z}{2}\right) (\gamma^1 - i\gamma^2) (1 - \gamma^5) \psi \left( \frac{z}{2} \right)  =  - \phi^\dagger_{-}  \chi_{-}  
\end{eqnarray}
\end{subequations}
Notice that the $*$ on the {\it lhs}  symbolizes the helicity of the quark within the quark-gluon composite field, $\chi$ (on the {\it rhs}), whose helicity is always opposite  
so that angular momentum is conserved \cite{Kogut:1969xa}. 
As a result, one can form twice as many helicity amplitudes as compared to the twist two case \cite{Jaffe:1996zw}, 
\begin{subequations}
\begin{eqnarray}
\label{helamp_1}
A^{tw 3}_{\Lambda' \lambda'^*, \Lambda \lambda} & = & \displaystyle\int \frac{d z^- \, d^2 z_T}{(2 \pi)^3} e^{ixP^+ z^- - i k_T\cdot z_T} \left. \langle p', \Lambda' \mid {\cal O}_{\lambda'^* \lambda}(z) \mid p, \Lambda \rangle \right|_{z^+=0}, \\
A^{tw 3}_{\Lambda' \lambda'^, \Lambda \lambda^*} & = & \displaystyle\int \frac{d z^- \, d^2 z_T}{(2 \pi)^3} e^{ixP^+ z^- - i k_T\cdot z_T} \left. \langle p', \Lambda' \mid {\cal O}_{\lambda'  \lambda^*}(z) \mid p, \Lambda \rangle \right|_{z^+=0} .
\end{eqnarray} 
\end{subequations}
Therefore,
\begin{subequations}
\begin{eqnarray}
\label{helamp_2}
A^{tw 3}_{\Lambda' -^*, \Lambda +} & = & W^{\gamma^1}_{\Lambda' \Lambda } + W^{\gamma^1 \gamma^5}_{\Lambda' \Lambda } - i W^{\gamma^2}_{\Lambda' \Lambda } - i W^{\gamma^2 \gamma^5}_{\Lambda' \Lambda } \\
A^{tw 3}_{\Lambda' +, \Lambda -^*} & = & W^{\gamma^1}_{\Lambda' \Lambda } + W^{\gamma^1 \gamma^5}_{\Lambda' \Lambda }  + i W^{\gamma^2}_{\Lambda' \Lambda } + i W^{\gamma^2 \gamma^5}_{\Lambda' \Lambda } \\
A^{tw 3}_{\Lambda' +^*, \Lambda -} & = & - W^{\gamma^1}_{\Lambda' \Lambda } + W^{\gamma^1 \gamma^5}_{\Lambda' \Lambda }  - i W^{\gamma^2}_{\Lambda' \Lambda } + i W^{\gamma^2 \gamma^5}_{\Lambda' \Lambda } \\
A^{tw 3}_{\Lambda' -, \Lambda +^*} & = & - W^{\gamma^1}_{\Lambda' \Lambda } + W^{\gamma^1 \gamma^5}_{\Lambda' \Lambda }  + i W^{\gamma^2}_{\Lambda' \Lambda } - i W^{\gamma^2 \gamma^5}_{\Lambda' \Lambda }
\end{eqnarray} 
\end{subequations}

\noindent At the twist-three level, the following are the expressions for the proton helicity non-flip terms, 

\begin{subequations}
\begin{eqnarray}
\label{eq:F14tw3} -\frac{i \epsilon^{ij}k^j}{P^+} F_{27}  -\frac{i \epsilon^{ij}\Delta^j}{P^+}F_{28} &=& \frac{1}{2}\left(W^{\gamma^i}_{++} - W^{\gamma^i}_{--} \right)  \\
\frac{k^i}{P^+} F_{21} + \frac{\Delta^i}{P^+} F_{22} &=& \frac{1}{2}\left(W^{\gamma^i}_{++} + W^{\gamma^i}_{--}\right)  \\
\frac{k^i}{P^+} G_{27} + \frac{\Delta^i}{P^+} G_{28}&=& \frac{1}{2}\left(W^{\gamma^i\gamma^5}_{++} - W^{\gamma^i\gamma^5}_{--}\right)  \\
\label{eq:G11tw3} \frac{i \epsilon^{ij}k^j}{P^+} G_{21} + \frac{i \epsilon^{ij}\Delta^j}{P^+} G_{22}&=& \frac{1}{2}\left(W^{\gamma^i\gamma^5}_{++} + W^{\gamma^i\gamma^5}_{--}\right)  \\
\nonumber
\end{eqnarray}
\end{subequations}

\noindent As we show in subsequent sections, Eqs.(\ref{eq:F14tw3}) and (\ref{eq:G11tw3}) allow us to identify the twist-three GTMDs that enter the EoM relations for $F_{14}$ and $G_{11}$ respectively.\\ 

\noindent Writing the GTMDs that enter the proton helicity flip case one has,

\begin{subequations}
\begin{eqnarray}
-\frac{i\epsilon^{ij} M\Delta^j}{P^+} F_{23} -i\frac{({\bf k}_T\times{\bf \Delta}_T )_3 }{MP^+}(k^i F_{24} +\Delta^i F_{25}) &=& \frac{1}{2}\left((\Delta^1 -i\Delta^2)W^{\gamma^i}_{- +} +(\Delta^1 + i\Delta^2)W^{\gamma^i}_{+ -}\right)  \\
 -\frac{\Delta_T^2}{2MP^+}(k^i F_{21} + \Delta^i F_{22}) +\frac{M\Delta^i}{P^+}F_{23} +\frac{ k_T\cdot\Delta_T}{MP^+}(k^i F_{24} &+& \Delta^i F_{25}) +\frac{\Delta_T^2\Delta^i}{MP^+}F_{26} \nonumber\\ 
 &=& \frac{1}{2}\left((\Delta^1 -i\Delta^2)W^{\gamma^i}_{- +} - (\Delta^1 + i\Delta^2)W^{\gamma^i}_{+ -}\right) 
\end{eqnarray}
\end{subequations}
and,
\begin{subequations}
\begin{eqnarray}
\frac{M\Delta^i}{P^+}G_{23} + \frac{k_T\cdot\Delta_T}{MP^+}(k^i G_{24} + \Delta^i G_{25}) + \frac{\Delta^i\Delta_T^2}{M}G_{26}&=&\frac{1}{2}\left((\Delta^1 -i\Delta^2)W^{\gamma^i\gamma^5}_{-+} + (\Delta^1 +i\Delta^2)W^{\gamma^i\gamma^5}_{+-}\right) \nonumber \\
&&\\
-\frac{i\epsilon^{ij}\Delta_T^2}{2MP^+}\left(k^j G_{21} - \Delta^j G_{22}\right) -\frac{i\epsilon^{ij}\Delta^j}{P^+}G_{23} -\frac{i({\bf k}_T\times{\bf \Delta}_T )_3 }{MP^+}\left(k^i G_{24} \right. &+& \left.\Delta^i G_{25}\right)  \nonumber\\
&=&\frac{1}{2}\left((\Delta^1 -i\Delta^2)W^{\gamma^i\gamma^5}_{-+} - (\Delta^1 +i\Delta^2)W^{\gamma^i\gamma^5}_{+-}\right) \nonumber \\ 
\end{eqnarray}
\end{subequations}

The helicity amplitude structure is preserved when going to either the GPD or the TMD limit. It plays an important role in defining the observables for the various quantities.     
The GTMDs defined so far are related to GPDs by integrating them over $k_T$ and to TMDs by taking the forward limit ($\Delta \rightarrow 0$).

\section{Equation of Motion Relations}
\label{sec3}

\subsection{Construction of Equation of Motion Relations}
\label{sec3a}
Equation of motion relations connect different GTMD correlators of the type defined in Eq.~(\ref{eq:GTMDcorr}), in which the quark creation and annihilation operators are located at positions $-z/2$ and $z/2$. To construct them, it is useful to consider initially a somewhat more general correlator in which the quark creation and annihilation operators are located at more freely variable positions $z_{in} $ and $z_{out} $, respectively.
Central to the construction is the observation that, taken between physical particle states, matrix elements of operators that vanish according to the classical field equations of motion vanish in the quantum theory\footnote{Note that the argument given in \cite{Politzer:1980me} is formulated for local operators; its extension to nonlocal operators such as considered here calls for further justification, as noted in \cite{Collins:2011zzd}.} \cite{Politzer:1980me}. Thus, in view of the classical quark field equations of motion
\begin{subequations}
\label{eq:Eom1}
\begin{eqnarray}
\label{eq:Eom1a}
(i\slashed{D} -m)\psi & = & (i\slashed{\partial} + g\slashed{A} -m)\psi=0, \\
\label{eq:Eom1b}
\bar{\psi}(i\overleftarrow{\slashed{D}}+m) & = & \bar{\psi}(i\overleftarrow{\slashed{\partial}} - g\slashed{A} +m) =0
\end{eqnarray}
\end{subequations}
one has the vanishing correlation function
\begin{eqnarray}
0 &=&
\int \frac{dz_{in}^{-} \, d^2 z_{in,T}}{(2 \pi)^3}
\int \frac{dz_{out}^{-} \, d^2 z_{out,T}}{(2 \pi)^3}
e^{ik (z_{out} -z_{in} ) +i\Delta (z_{out} +z_{in} )/2} \nonumber \\
& & \hspace{3cm}
\cdot \left. \langle p', \Lambda' \mid 
\bar{\psi} (z_{in} ) \left[ (i\overleftarrow{\slashed{D}}+m) \Gamma {\cal U}
\pm \Gamma {\cal U} (i\slashed{D} -m) \right]
\psi (z_{out} )
\mid p, \Lambda \rangle \right|_{z_{in}^{+} = z_{out}^{+} =0}
\label{correom}
\end{eqnarray}
where, specifically, $\Gamma = i \sigma^{i+} \gamma^5 =
\gamma^+\gamma^i \gamma^5 - \gamma^i\gamma^+ \gamma^5$ with a transverse
vector index $i=1,2$, cf.~Sec.~\ref{sec2C}.
Note that the $\slashed{D} $ and $\overleftarrow{\slashed{D}}$
operators act on the $z_{out} $ and $z_{in} $ arguments, respectively.
Furthermore, no derivatives with respect to $z_{in}^{+} $ or $z_{out}^{+} $
appear in the square bracket; these derivatives are accompanied in the
Dirac operator by a factor $\gamma^{+} $, implying that the terms in
question vanish once multiplied by the structure $\Gamma $, which
contains an additional factor $\gamma^{+} $. Thus, introducing the
equations of motion as in (\ref{correom}) is consistent with an
a priori specification $z_{in}^{+} = z_{out}^{+} =0$.

Performing an integration by parts with respect to both $z_{out} $ and
$z_{in} $ yields
\begin{eqnarray}
0 &=& \left.
\int \frac{dz_{in}^{-} \, d^2 z_{in,T}}{(2 \pi)^3}
\int \frac{dz_{out}^{-} \, d^2 z_{out,T}}{(2 \pi)^3}
e^{ik (z_{out} -z_{in} ) +i\Delta (z_{out} +z_{in} )/2}
\right\{ 
\label {partint} \\
& & \hspace{2.4cm}
\langle p', \Lambda' \mid 
\bar{\psi} (z_{in} )
\left[ -i\slashed{\partial}_{in} {\cal U} \Gamma
\mp i\Gamma \slashed{\partial}_{out} {\cal U}
-g\slashed{A} (z_{in} ) {\cal U} \Gamma
\pm \Gamma {\cal U} g\slashed{A} (z_{out} )
\right]
\psi (z_{out} )
\mid p, \Lambda \rangle \nonumber \\
& & \hspace{2cm}
+ \left. \left.
\langle p', \Lambda' \mid 
\bar{\psi} (z_{in} )
\left[
\left( -\slashed{k} + \frac{\slashed{\Delta } }{2} \right) {\cal U} \Gamma
\mp \Gamma {\cal U} \left( -\slashed{k} - \frac{\slashed{\Delta } }{2} \right)
+ (m\mp m)\Gamma {\cal U} \right]
\psi (z_{out} )
\mid p, \Lambda \rangle
\right\} \right|_{z_{in}^{+} = z_{out}^{+} =0} \nonumber
\end{eqnarray}
Two types of contributions are generated. The second line of (\ref{partint}) contains the terms in which the derivatives act on the gauge links; these terms will ultimately result in quark-gluon-quark correlators. The third line of (\ref{partint}) contains the standard terms in which the derivatives act on the exponential in the Fourier transformation; these terms result in quark-quark correlators. Proceeding by changing integration variables,
\begin{equation}
b = \frac{z_{in} + z_{out}}{2},  \qquad z = z_{out} - z_{in},
\label{eq:bz}
\end{equation}
and translating the matrix elements by $-b$, one obtains
\begin{eqnarray}
0 &=& \left. \left(
\int \frac{db^- \, d^2 b_T }{(2 \pi)^3} e^{ib\Delta }
e^{ibp} e^{-ibp^{\prime } } \right)
\int \frac{dz^- \, d^2 z_T }{(2 \pi)^3}
e^{ikz}
\right\{ 
\label{bztransform} \\
& & \hspace{2.4cm}
\langle p', \Lambda' \mid 
\bar{\psi} (-z/2)
\left[ \left. (-i\overrightarrow{\slashed{\partial} }
-g\slashed{A} ) {\cal U} \Gamma \right|_{-z/2}
\pm \left. \Gamma {\cal U} (-i\overleftarrow{\slashed{\partial} }
+g\slashed{A} ) \right|_{z/2}
\right]
\psi (z/2)
\mid p, \Lambda \rangle \nonumber \\
& & \hspace{2cm}
+ \left. \left.
\langle p', \Lambda' \mid 
\bar{\psi} (-z/2)
\left[
\left( -\slashed{k} + \frac{\slashed{\Delta } }{2} \right) {\cal U} \Gamma
\mp \Gamma {\cal U} \left( -\slashed{k} - \frac{\slashed{\Delta } }{2} \right)
+ (m\mp m)\Gamma {\cal U} \right]
\psi (z/2)
\mid p, \Lambda \rangle
\right\} \right|_{z^+ = b^+ =0} \nonumber
\end{eqnarray}
having taken into account the phases generated in the proton states by
the translation. Thus, a $\delta $-function which enforces momentum
conservation as expected, $\delta^{3} (p^{\prime } - p -\Delta )$,
is factored out; it follows that the rest of the expression by itself must already vanish. Proceeding to simplify the Dirac structures
(employing, e.g., the identity $\gamma^\mu \gamma^\rho \gamma^\nu =
g^{\mu \rho} \gamma^\nu + g^{\nu \rho} \gamma^\mu - g^{\mu \nu} \gamma^\rho
- i \epsilon^{\sigma \mu \nu \rho} \gamma_\sigma \gamma^5$), one can
finally identify from the third line of (\ref{bztransform}) the GTMD correlators defined in Eq.~(\ref{eq:GTMDcorr}),
and one thus arrives at the equation of motion relations
\begin{subequations}
\begin{eqnarray}
\label{eq:EoMfinal1}
&& -\frac{\Delta^+}{2} W^{\gamma^i\gamma^5}_{\Lambda' \Lambda } + ik^+\epsilon^{ij}W^{\gamma^j}_{\Lambda' \Lambda } +\frac{\Delta^{i} }{2} W^{\gamma^+\gamma^5}_{\Lambda' \Lambda } -i \epsilon^{ij}k^j W^{\gamma^+}_{\Lambda' \Lambda } + \mathcal{M}^{i, S}_{\Lambda' \Lambda } = 0 \\
\label{eq:EoMfinal2}
&& -k^+ W^{\gamma^i\gamma^5}_{\Lambda' \Lambda } +\frac{i\Delta^+}{2}\epsilon^{ij} W^{\gamma^j}_{\Lambda' \Lambda } +k^i W^{\gamma^+\gamma^5}_{\Lambda' \Lambda } -i \epsilon^{ij} \frac{\Delta^{j} }{2} W^{\gamma^+}_{\Lambda' \Lambda } + m W^{i\sigma^{i+}\gamma^5}_{\Lambda' \Lambda } + \mathcal{M}^{i, A}_{\Lambda' \Lambda } = 0,
\end{eqnarray}
\end{subequations}
which relate the correlation functions for different Dirac structures,
$\gamma^i \gamma^5, \gamma^i, \gamma^+ \gamma^5, \gamma^+$, $i\sigma^{i+} \gamma^{5} $, and in which
the genuine/dynamic \cite{Kanazawa:2015ajw} twist-three terms, copied from the second line of (\ref{bztransform}), are given
by\footnote{Note that the expression for ${\cal M}^{i,S}_{\Lambda^{\prime } \Lambda } $ quoted in \cite{Rajan:2016tlg} is missing an overall factor $i$.}
\begin{subequations}
\begin{eqnarray}
{\cal M}^{i, S}_{\Lambda^{\prime} \Lambda } \! &=& \! \frac{i}{4}
\int \frac{d z^- d^2 z_T}{(2 \pi)^3} e^{ixP^+ z^- - i k_T\cdot z_T} 
\langle p^{\prime } ,\Lambda^{\prime } \mid \overline{\psi} \left(-\frac{z}{2} \right) \left[
\left. (\overrightarrow{\slashed{\partial } } -ig\slashed{A} )
{\cal U} \Gamma \right|_{-z/2} + \left. \Gamma {\cal U}
(\overleftarrow{\slashed{\partial } } +ig\slashed{A} ) \right|_{z/2}
\right] \psi \left(\frac{z}{2} \right) \mid p, \Lambda \rangle_{z^+=0} \nonumber \\ \label{qgqterms1} \\
{\cal M}^{i, A}_{\Lambda^{\prime } \Lambda } \! &=& \! \frac{i}{4}
\int \frac{d z^- d^2 z_T}{(2 \pi)^3} e^{ixP^+ z^- - i k_T\cdot z_T} 
\langle p^{\prime } ,\Lambda^{\prime } \mid \overline{\psi} \left(-\frac{z}{2} \right) \left[
\left. -(\overrightarrow{\slashed{\partial } } -ig\slashed{A} )
{\cal U} \Gamma \right|_{-z/2} + \left. \Gamma {\cal U}
(\overleftarrow{\slashed{\partial } } +ig\slashed{A} ) \right|_{z/2}
\right] \psi \left( \frac{z}{2} \right)  \mid p, \Lambda \rangle_{z^+=0} \nonumber \\ \label{qgqterms}
\end{eqnarray}
\end{subequations}
with $\Gamma = i\sigma^{i+}\gamma^5$. In the following, only the case
of vanishing skewness, $\Delta^{+} =0$, will be considered further.

Relations (\ref{eq:EoMfinal1}) and (\ref{eq:EoMfinal2}) are generalizations to the off-forward case of the EoM relations involving the $k_T $-unintegrated correlator first introduced in 
\cite{Mulders:1995dh,Tangerman:1994bb,Goeke:2003az}. In particular, Eq.(\ref{eq:EoMfinal2}) leads to the relation between the polarized structure functions $g_1$ and $g_2$ first obtained in the forward limit using the same method in Refs.\cite{Mulders:1995dh,Tangerman:1994bb}. However, notice that, at variance with \cite{Mulders:1995dh,Tangerman:1994bb}, because of the symmetrization introduced in Eqs.~(\ref{correom})-(\ref{bztransform}), the imaginary parts in Eq.(\ref{eq:EoMfinal2}) appear only for the non forward terms (terms multiplied by $\Delta$).  
As will be discussed further below, these relations represent a first step towards deriving a connection between twist-two GTMDs and twist-three GPDs using a procedure alternative to OPE that highlights the sensitivity to the quark intrinsic transverse momentum. In our case, they attain additional significance in that they provide a framework for describing partonic OAM in the proton in terms of specific distributions, thus helping to clarify possible mechanisms that generate it. A prerequisite for understanding what produces OAM in the proton is that one examines the dynamics encoded in the correlator components at the unintegrated level.


\begin{figure}
\vspace{-0.5cm}
\includegraphics[width=9cm]{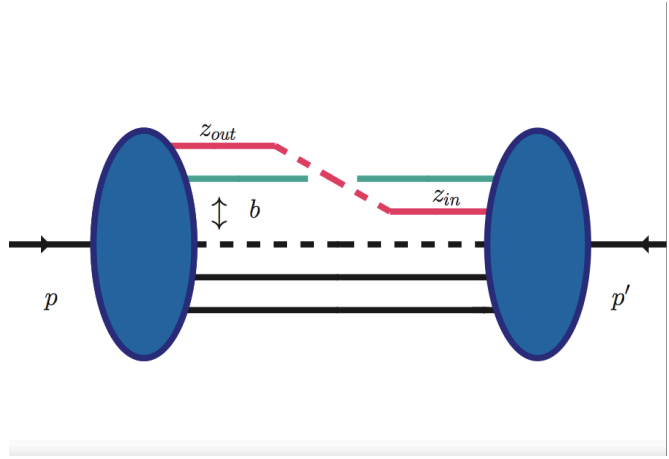}
\vspace{-2cm}
\caption{Kinematical variables for the correlation function describing a GTMD. The matrix element in the correlator is a function of $z=z_{out}-z_{in}$ where $z_{out}=z/2$ ($z_{in}=-z/2$) is the argument of $\psi$ ($\bar{\psi}$); $b=(z_{in}+z_{out})/2$ is the Fourier conjugate of $\Delta=p'-p$.}
\label{fig:kinem}
\end{figure}

\subsection{Gauge Link Structure and Intrinsic Twist Three Term}
\label{sec3c}
The form of the intrinsic twist-three terms given in Section \ref{sec3a} is valid for an arbitrary choice of gauge link ${\cal U} $. The gauge link depends parametrically on the locations of its endpoints; the derivative operators quantify those dependences. More concrete forms are obtained by considering particular gauge link paths. An important choice is the staple-shaped gauge link path, the geometry of which was already discussed in detail in Sec.~\ref{sec:upath}, with the legs of the staple described by a four-vector $v$; this contains also the straight gauge link path in the limit $v=0$.
Given this concrete choice, a more explicit form of the intrinsic twist-three
contributions can be derived.

To establish notation,
consider a staple-shaped gauge link ${\cal U} $ connecting the space-time
points $y$ and $y^{\prime } $ via three straight segments,
\begin{eqnarray}
{\cal U} &=&
{\cal P} \exp \left( -ig \int_{y}^{y+v} dx^{\mu } A_{\mu } (x) \right)
{\cal P} \exp \left( -ig \int_{y+v}^{y^{\prime } +v} dx^{\mu } A_{\mu } (x)\right)
{\cal P} \exp \left( -ig \int_{y^{\prime } +v}^{y^{\prime } }
dx^{\mu } A_{\mu } (x) \right) \\
& \equiv & U_1 (0,1) U_2 (0,1) U_3 (0,1) \ ,
\end{eqnarray}
which each can be parametrized in terms of a real parameter $t$ as
\begin{eqnarray}
U_1 (a,b) &=& {\cal P} \exp \left( -ig \int_{a}^{b} dt\, v^{\mu }
A_{\mu } (y+tv) \right) \\
U_2 (a,b) &=& {\cal P} \exp \left( -ig \int_{a}^{b} dt\,
(y^{\prime } -y)^{\mu } A_{\mu } (y+v+t(y^{\prime } -y)) \right) \\
U_3 (a,b) &=& {\cal P} \exp \left( -ig \int_{a}^{b} dt\, (-v^{\mu } )
A_{\mu } (y^{\prime } +v-tv) \right)
\end{eqnarray}
As noted above, the four-vector $v$ describes the legs of the staple-shaped
path. The parametrization includes the special case $v=0$, in which the
staple degenerates to a straight link between $y$ and $y^{\prime } $ given
by $U_2 (0,1)$, whereas $U_1 =U_3 =1$. In the following, $U_i $ given
without an argument means $U_i \equiv U_i (0,1)$.

As shown in Appendix \ref{Appendix:qgq}, with this parametrization, one
arrives at the explicit expression
\begin{eqnarray}
\left( \frac{\partial }{\partial y^{\nu } } -igA_{\nu } (y) \right) {\cal U}
&=& ig U_1 \int_{0}^{1} ds\, U_2 (0,s)
(y^{\prime } -y)^{\mu }
F_{\mu \nu } (y+v+s(y^{\prime } -y)) (1-s) U_2 (s,1) U_3
\nonumber \\
& & +ig\int_{0}^{1} ds\, U_1 (0,s) v^{\mu } F_{\mu \nu } (y+sv)
U_1 (s,1) U_2 U_3
\end{eqnarray}
in which only field strength terms remain.
In complete analogy, one also obtains for the adjoint term,
\begin{eqnarray}
{\cal U} \left( \frac{\overleftarrow{\partial}}{\partial y^{\prime \nu } } 
+iA_{\nu } (y^{\prime } ) \right) &=&
ig U_1 \int_{0}^{1} ds\, U_2 (0,s)
(y^{\prime } -y)^{\mu }
F_{\mu \nu } (y+v+s(y^{\prime } -y)) s U_2 (s,1) U_3
\nonumber \\
& & -U_1 U_2 ig\int_{0}^{1} ds\, U_3 (0,s) v^{\mu }
F_{\mu \nu } (y^{\prime } +v-sv) U_3 (s,1) \quad,
\end{eqnarray}
where in each integral, $s$ parametrizes the position of the color field strength insertion along the gauge link connecting the quark positions.
These forms are still completely general.
In the following, in particular the $k_T $-integral of the genuine twist-three terms will be of interest, in which case, cf.~(\ref{qgqterms1},\ref{qgqterms}), the transverse separation $z_T $ is set to zero and $z$ has only a minus component, $z=(0,z^{-} ,0,0)$. Specializing furthermore to the case where also $v$ has only a minus component, $v=(0,v^{-} ,0,0)$, cf.~the discussion in Sec.\ref{sec:upath}, the staple legs collapse onto a common axis. In this case we define $U(x,x^{\prime})$ to denote a straight Wilson line connecting the locations $x$ and $x^{\prime} $, and obtain, upon identifying the endpoints $y=-z/2$ and $y^{\prime } =z/2$,
\begin{eqnarray}
\left. (\overrightarrow{\slashed{\partial } } -ig\slashed{A} ) {\cal U} \right|_{-z/2} &=&
igz^{-} \int_{0}^{1} ds\, (1-s) U(-z/2,-z/2+v+sz)
\gamma_{\mu } F^{+\mu } (-z/2+v+sz) U(-z/2+v+sz,z/2) \nonumber \\
& & + igv^{-} \int_{0}^{1} ds\, U(-z/2,-z/2+sv) \gamma_{\mu } F^{+\mu } (-z/2+sv) U(-z/2+sv,z/2) \label{qgqright} \\
\left. {\cal U} (\overleftarrow{\slashed{\partial } } +ig\slashed{A} ) \right|_{z/2} &=&
igz^{-} \int_{0}^{1} ds\, s\, U(-z/2,-z/2+v+sz) \gamma_{\mu } F^{+\mu } (-z/2+v+sz) U(-z/2+v+sz,z/2) \nonumber \\
& & -igv^{-} \int_{0}^{1} ds\, U(-z/2,z/2+sv) \gamma_{\mu } F^{+\mu } (z/2+sv) U(z/2+sv,z/2) \label{qgqleft}
\end{eqnarray}

Note that, in both expressions, the first line stems from the variation of the Wilson line which connects the ends of the staple legs, whereas the second line stems from the variation of the staple leg attached to the endpoint with respect to which the derivative is taken. The straight gauge link case is obtained by setting $v=0$, i.e., only the first lines in (\ref{qgqright}) and (\ref{qgqleft}) remain. This limit was already given in \cite{Rajan:2016tlg}.

Particularly compact expressions are obtained if one further integrates over the longitudinal momentum fraction $x$, in which case $z=0$ altogether, cf.~(\ref{qgqterms1},\ref{qgqterms}). For $z=0$, the first lines in (\ref{qgqright}) and (\ref{qgqleft}) vanish, {\it i.e.}, the genuine twist-three terms integrate to zero for a straight gauge link. On the other hand, in the general staple link case, the second lines remain, and give identical contributions up to a relative minus sign. Combining with the Dirac structure $\Gamma $ and assembling the complete genuine twist-three expressions, one has in the completely integrated limit,
\begin{eqnarray}
\int dx \int d^2k_T \, {\cal M}^{i, S}_{\Lambda^{\prime} \Lambda } &=& i\epsilon^{ij} gv^{-} \frac{1}{2P^{+} }
\int_{0}^{1} ds\, \langle p^{\prime } ,\Lambda^{\prime } | \bar{\psi } (0) \gamma^{+}  U(0,sv) F^{+j} (sv) U(sv,0)
\psi (0) | p,\Lambda \rangle \label{xintqgq} \\
\int dx \int d^2 k_T \, {\cal M}^{i, A}_{\Lambda^{\prime } \Lambda } &=& -gv^{-} \frac{1}{2P^{+} }
\int_{0}^{1} ds\, \langle p^{\prime } ,\Lambda^{\prime } | \bar{\psi } (0) \gamma^{+} \gamma^{5} U(0,sv) F^{+i} (sv) U(sv,0) \psi (0) | p,\Lambda \rangle \label{xintqgqaxial}
\end{eqnarray}
Note that the $\epsilon^{ij} $ in (\ref{xintqgq}) can be absorbed into the dual field strength $\widetilde{F}^{+i} = -\epsilon^{ij} F^{+j}$, useful for the analysis within instanton models \cite{Kiptily:2002nx}, in which $\widetilde{F} =\pm F$.
On the other hand, compact expressions also for the second Mellin moments result if one specializes to the straight link case. A weighting by a factor $x$ can be generated by taking a derivative with respect to $z^{-} $, cf.~(\ref{qgqterms1},\ref{qgqterms}); in the limit $z=0$, only the contributions from the derivative acting on either of the $z^{-} $ prefactors in the first lines of (\ref{qgqright}) and (\ref{qgqleft}) remain. Thus, one arrives at
\begin{eqnarray}
\int dx\, x \int d^2 k_T \, {\cal M}^{i,S}_{\Lambda^{\prime } \Lambda } &=&
\frac{ig}{4(P^{+} )^2 } \langle p^{\prime } ,\Lambda^{\prime } | \bar{\psi } (0) \gamma^{+} \gamma^{5} F^{+i} (0) \psi (0) | p,\Lambda \rangle
\\
\int dx\, x \int d^2 k_T \, {\cal M}^{i,A}_{\Lambda^{\prime } \Lambda } &=&
\frac{g}{4(P^{+} )^2 } \epsilon^{ij} \langle p^{\prime } ,\Lambda^{\prime } | \bar{\psi } (0) \gamma^{+} F^{+j} (0) \psi (0) | p,\Lambda \rangle
\label{mellin_2A}
\end{eqnarray}
for straight gauge links. 
Note that one can obtain, {\it e.g.}, the right-hand side of (\ref{mellin_2A}) by evaluating the $v^{-} $-derivative of (\ref{xintqgq}) at $v^{-} =0$, and multiplying by a factor $-i/(2P^{+})$. In other words, we uncover a connection between straight-link quark-gluon-quark correlators such as (\ref{mellin_2A}) and $v^{-} $-derivatives of Qiu-Sterman type terms such as (\ref{xintqgq}), where the latter can be accessed using Lattice QCD TMD data \cite{in_prepar}, such as given in \cite{Musch:2011er,Engelhardt:2015xja,Yoon:2017qzo}.
\subsection{EoM Relations involving Orbital Angular Momentum}
\label{sec3e}
Altogether, Eqs.~(\ref{eq:EoMfinal1}) and (\ref{eq:EoMfinal2}) generate 32 individual
relations between GTMDs, obtained by inserting the parametrizations (\ref{vectorimf}),(\ref{axialimf}),(\ref{eq:tw3vec_imf}),(\ref{eq:tw3axvec_imf}): each of the two relations is a two-component equation in the transverse plane;
furthermore, the resulting 4 individual component relations are complex, {\it i.e.}, each comprises a relation for the real (T-even) and the imaginary (T-odd) parts of GTMDs. 
The resulting 8 relations finally each contain 4 possible helicity combinations, as discussed in Section \ref{sec2e}, for the proton helicity conserving, Eqs.(\ref{eq:FGnonflip}), and the helicity flip, Eqs.(\ref{eq:FGflip}) combinations, respectively.
We refrain from quoting all 32 of these relations. They can be specialized to the $\Delta =0$ TMD limit and to the $k_T $-integrated GPD limit. In the TMD limit, a number of known TMD relations \cite{Bacchetta:2006tn} is reproduced, including explicit expressions for the genuine twist-3 parts in terms of quark-gluon-quark correlators, encoded in the ${\cal M}^{i,S}_{\Lambda^{\prime } \Lambda } $ and ${\cal M}^{i,A}_{\Lambda^{\prime } \Lambda} $ terms. 
For the $k_T $-integrated case, we focus on purely transverse momentum transfer, {\it i.e.}, vanishing skewness, $\xi =0$. In this case, there are potentially 8 relations: Of the original 32, 16 are $\xi$-odd, and of course only the T-even relations are relevant for the GPD limit. Among these 8 relations, we discuss in detail
three which involve exclusively $k_T^2 $ moments of GTMDs and GPDs. 
These three are moreover singled out by the fact that they are also accompanied by three corresponding LIRs. 

In this section, we present, in particular, the EoM relations describing the quark OAM and spin-orbit contributions. 
These involve $F_{14}$, which is obtained for the helicity configuration (\ref{F14}) describing an unpolarized quark in a longitudinally polarized proton, and a relation for $G_{11}$, obtained for the helicity configuration (\ref{G11}), describing a longitudinally polarized quark in an unpolarized proton. 
These configurations are obtained by taking the helicity combinations $(\Lambda' \Lambda )=(++) \pm (--)$, in Eqs.~(\ref{eq:EoMfinal1}) and (\ref{eq:EoMfinal2}), respectively.
The relations we obtain constitute $x$-dependent identities tying the definitions, respectively, of partonic OAM, $L_z$, and the longitudinal contribution to the spin-orbit coupling $L \cdot S$, to directly observable twist-three distributions. 
We present the third EoM relation, which instead involves transverse polarization, in Section \ref{sec5}. As we show below, after taking $\Delta^+ = 0$ (without loss of generality in the angular momentum sum rule), 
we obtain the following EoM relations from Eqs.~(\ref{eq:EoMfinal1}) and (\ref{eq:EoMfinal2}), respectively,
\vspace{0.5cm}
\begin{eqnarray}
\label{eq:EoMF14}
x \widetilde{E}_{2T}(x)  &=& - \widetilde{H}(x) + F_{14}^{(1)}(x) - {\cal M}_{F_{14} }  \\
\label{eq:EoMG11}
x\left[2\widetilde{H}_{2T}'(x) + E_{2T}'(x) \right] & = &- H(x)  + \frac{m}{M}( 2 \widetilde{H}_T(x)+ E_T(x))  - G_{11}^{(1)}(x) - {\cal M}_{G_{11} }
\end{eqnarray}
where we defined
\begin{equation}
\label{ktmoment}
X^{(1)} = 2\int d^2 k_T  \frac{k_T^2}{M^2}
\frac{k_T^2 \Delta_{T}^{2} - (k_T \cdot \Delta_{T} )^2 }{k_T^2 \Delta_{T}^{2} } X (x, 0, k_T^2, k_T\cdot\Delta_T, \Delta_T^2)
\end{equation}
Note that, in the forward limit, this reduces to the standard $k_T^2 $-moment,
\begin{equation}
\left. X^{(1)} \right|_{\Delta_{T} =0} =
\int d^2 k_T \frac{k_T^2 }{M^2 } X (x,0,k_T^2,0,0)
\end{equation}
The genuine twist three contributions are defined as\footnote{Note that in \cite{Rajan:2016tlg}, the first of these relations was quoted with an erroneous additional normalization factor $2M$.}
\begin{eqnarray}
&& \mathcal{M}_{F_{14}} (x) = \int d^2 k_T \frac{\Delta^{i} }{\Delta_T^2} \left( \mathcal{M}^{i,S}_{++} - \mathcal{M}^{i,S}_{--} \right) \\
&& \mathcal{M}_{G_{11}} (x) = \int d^2 k_T i\epsilon^{ij} \frac{\Delta^{j} }{\Delta_T^2} \left( {\mathcal{M}}^{i,A}_{++} + {\mathcal{M}}^{i,A}_{--} \right) \quad ,
\end{eqnarray}
where the expressions for ${\mathcal{M}}^{i,S(A)}_{\Lambda \Lambda'}$  given in Eqs.(\ref{qgqterms1},\ref{qgqterms},\ref{xintqgq},\ref{xintqgqaxial}), 
 can be interpreted as quantifying the quark-gluon-quark interaction experienced by a quark of specific $x$, in the given helicity configuration.

Eqs.(\ref{eq:EoMF14}, \ref{eq:EoMG11}) are the equation of motion relations involving the OAM and the longitudinal part of the spin-orbit $L \cdot S$ distributions, defined through $F_{14}$ and $G_{11}$, in Eqs.~(\ref{eq:F14OAM}) and (\ref{eq:G11LdotS}), respectively. They are particularly  important among the various GTMD EoM relations that we can write because they allow us to define observables other than the GTMDs to measure the OAM distribution in the proton.  

All of the distributions in the EoM relations are defined according to the scheme of Ref.~\cite{Meissner:2009ww} (see Section \ref{sec3}): $H$ and $\widetilde{H}$ are twist two GPDs, in the vector and axial vector sector respectively; $\widetilde{E}_{2T}$ is a twist three GPD in the vector sector, $\widetilde{H}_{2T}'$ and $E_{2T}'$ are  axial vector twist three GPDs.

Eq.~(\ref{eq:EoMF14}) relates an {\it intrinsic} \cite{Kanazawa:2015ajw} twist three GPD, $\widetilde{E}_{2T}$, on the {\it lhs} \cite{Rajan:2016tlg}, to a twist two GPD, $\widetilde{H}$, the $k_T$-moment of the GTMD, $F_{14}$, Eq.(\ref{ktmoment}), and a {\it genuine} twist three term, ${\cal M}_{F_{14} } $. It is obtained by contracting Eq.~(\ref{eq:EoMfinal1}) with $\Delta^{i} /\Delta_{T}^{2} $, forming the $(\Lambda^{\prime } \Lambda )=(++)-(--)$ combination of helicity components, and inserting the GTMD parametrizations of the correlators, yielding
\begin{equation}
0 = -2x\left( \frac{k_T \cdot \Delta_{T} }{\Delta_{T}^{2} } F_{27} + F_{28} \right) +G_{14} -2 \frac{k_T^2 \Delta_{T}^{2} - (k_T \cdot \Delta_{T} )^{2} }{M^2 \Delta_{T}^{2} } F_{14} +\frac{\Delta^{i} }{\Delta_{T}^{2} } \left( {\cal M}_{++}^{i,S} - {\cal M}_{--}^{i,S} \right)
\end{equation}
Integrating over $k_T $ and identifying the resulting GPDs \cite{Meissner:2009ww} gives


\begin{equation}
0 = x \widetilde{E}_{2T} + \widetilde{H} - F_{14}^{(1)} + \int d^2 k_T \frac{\Delta^{i} }{\Delta_T^2}\left( \mathcal{M}^{i,S}_{++} - \mathcal{M}^{i,S}_{--} \right),
\end{equation}
i.e., one obtains Eq.~(\ref{eq:EoMF14}).
Recalling the discussion in Sec.~\ref{sec:upath}, in the case of a staple-shaped gauge link, this requires that the legs of the staple properly collapse upon $k_T $-integration such as to produce GPDs with straight gauge link.

Eq.~(\ref{eq:EoMG11}) was derived in a similar way. It relates the twist-three GPD combination, $2\widetilde{H}_{2T}'(x) + E_{2T}'(x)$, to the GPD $H$, the $k_T $-moment of the GTMD, $G_{11}$, which describes the longitudinal part of the parton spin-orbit distribution, and a {\it genuine} twist three term. Notice the appearance of a quark mass term proportional to the GPD $2\widetilde{H}_{T} + E_{T}$ in the chiral odd sector \cite{Goldstein:2012az}. Contracting Eq.~(\ref{eq:EoMfinal2}) with $i\epsilon^{ij} \Delta^{j} /\Delta_{T}^{2} $, forming the $(\Lambda' \Lambda ) = (++)+(--)$ combination of helicity components, cf.~Eq.~(\ref{G11}), and inserting the GTMD parametrizations of the correlators yields
\begin{eqnarray}
0 &=& 2x \left( \frac{k_T \cdot \Delta_{T} }{\Delta_{T}^{2} } G_{21} +G_{22} \right) +F_{11} +2\frac{k_T^2 \Delta_{T}^{2} - (k_T \cdot \Delta_{T} )^{2} }{M^2 \Delta_{T}^{2} } G_{11} -2\frac{m}{M} \left( \frac{k_T \cdot \Delta_{T} }{\Delta_{T}^{2} } H_{11} + H_{12} \right) \\
& & \hspace{10cm} +i\epsilon^{ij} \frac{\Delta^{j} }{\Delta_{T}^{2} } \left( {\cal M}_{++}^{i,A} + {\cal M}_{--}^{i,A} \right)
\nonumber
\end{eqnarray}
Integrating over $k_T $ and identifying the resulting GPDs gives
\begin{eqnarray}
0 &=& x \left( 2\widetilde{H}_{2T}^{\prime } +E_{2T}^{\prime } \right) + H + G_{11}^{(1)} -\frac{m}{M} \left( 2\widetilde{H}_{T} +E_T \right) +\int d^2 k_T i\epsilon^{ij} \frac{\Delta^{j} }{\Delta_{T}^{2} } \left( {\cal M}_{++}^{i,A} + {\cal M}_{--}^{i,A } \right)
\end{eqnarray}
i.e., Eq.~(\ref{eq:EoMG11}).
\\

\section{Generalized Lorentz Invariance Relations}
\label{sec4}
The underlying Lorentz structure of the unintegrated correlator, Eqs. (\ref{vector_GPCF},\ref{axialvector_GPCF}) allows one to find relations between the $x$-dependent $k_T$-moments of GTMDs and GPDs. As stated before, this is due to the fact that, for the straight gauge link case, the total number of GPCFs is less than the number of GTMDs. Similar relations connecting the various TMDs, in the forward limit, were derived in Refs.\cite{Goeke:2005hb,Mulders:1995dh}. These equations are a consequence of the covariant definition of the correlation function, and they are therefore referred to as Lorentz Invariance Relations (LIRs). 

The following LIRs, which we derive further below, involve the $k_T$-moments of the GTMDs respectively describing the OAM and longitudinal spin-orbit terms which also enter the EoMs derived in Section \ref{sec3}, Eqs.(\ref{eq:EoMF14},\ref{eq:EoMG11}),
\begin{eqnarray}
\frac{d F_{14}^{(1)}}{ d x}  &=& \widetilde{E}_{2T} +H +E \quad  \Rightarrow \quad  F_{14}^{(1)} = - \int_x^1 dy \left[\widetilde{E}_{2T} +H +E  \right]
\label{oamlir} \\
 \frac{d G_{11}^{(1)}}{d x} &=& -\left( E_{2T}' + 2\widetilde{H}_{2T}' + \widetilde{H}\right) \quad 
 \Rightarrow \quad G_{11}^{(1)} = \int_x^1 dy \left[  2\widetilde{H}_{2T}' + E_{2T}' + \widetilde{H}  \right]
\label{spinorbitlir}
\end{eqnarray}
On the left hand side, we have $k_T^2$-moments of twist two GTMDs. These GTMDs are unique in that, in the limit $t=0$,  they carry the physical meaning of parton longitudinal OAM distribution, $F_{14}^{(1)}$, and longitudinal parton spin-orbit distribution, $G_{11}^{(1)}$. On the right hand side, the integral expressions for the intrinsic twist three GPDs $\widetilde{E}_{2T} +H +E$, and $ 2\widetilde{H}_{2T}' + E_{2T}' + \widetilde{H}$, allow us to access both OAM and the longitudinal spin-orbit term directly from deeply virtual exclusive measurements as these GPDs enter as coefficients of specific azimuthal angular modulations of the cross section. Note that these $x$-dependent relations are valid also for $\Delta_{T} \neq 0$.

Note also that, at variance with previous work \cite{Penttinen:2000dg,Kiptily:2002nx,Hatta:2012cs,Ji:2013dva}, Eq~(\ref{oamlir}) allows us to obtain directly information on the OAM  distribution because its form is not integrated in $x$ (it occurs at the $k_T$-integrated level).     
Eq.(\ref{spinorbitlir}) is new: it allows us to connect the longitudinal spin-orbit $x$-distribution, $G_{11}^{(1)}$, to a specific twist three GPD combination, $ 2\widetilde{H}_{2T}' + E_{2T}'$ that uniquely appears in off-forward processes.

If we were to work with a staple gauge link, the number of GPCFs
would increase to 16 for both the vector and the axial vector case. In this
scenario, since the number of GTMDs is the same as the number of GPCFs,
we do not expect there to be any LIRs connecting exclusively GTMDs (or their GPD limits). Indeed, if we do try to write these
relations, we find that extra terms appear that consist of GPCFs that cannot be
combined to form either GPDs or GTMDs. These extra terms, which are required in order to properly encode Lorentz invariance in the relations, have been termed LIR breaking terms \cite{Accardi:2009au}. For example, (\ref{oamlir}) is modified to read
\begin{eqnarray}
\frac{d F_{14}^{(1)}}{dx} &=& \widetilde{E}_{2T} +H+E + {\cal A}_{F_{14} } \ ,
\label{eq:LIRviol}
\end{eqnarray}
with,
\begin{eqnarray}
{\cal A}_{F_{14} } = v^{-} \frac{(2P^{+} )^2 }{M^2 } \int d^2 k_T \int dk^{-} \left[
\frac{k_{T} \cdot \Delta_{T} }{\Delta_{T}^{2} } (A_{11}^F +xA_{12}^F )
+A_{14}^F + \frac{k_T^2 \Delta_{T}^{2} - (k_T \cdot \Delta_{T} )^{2} }{\Delta_{T}^{2} } \left(
\frac{\partial A_8^F }{\partial (k\cdot v)} +
x\frac{\partial A_9^F }{\partial (k\cdot v)} \right) \right]
\label{lir_staple}
\end{eqnarray}
where the 4-vector
$v=(0,v^{-} ,0,0)$ describes the direction of the staple, which here is taken to
extend along the light cone. The amplitudes $A_i^F$ are the ones appearing in the parametrization given in \cite{Meissner:2009ww}, appropriate for a staple link structure, up to a rescaling stemming from the fact that the staple vector $v$ used here and the analogous vector $N$ used in \cite{Meissner:2009ww} are related by a rescaling. Note that, if one were to take $v$ off the light cone, $v=(v^{+} ,v^{-} ,0,0)$, cf.~the discussion in Sec.~\ref{sec:upath}, additional terms would appear in (\ref{lir_staple}) that formally vanish as $v^{+} \rightarrow 0$; examples of such terms in the case of TMD LIRs have been given in \cite{Accardi:2009au}. Of course, the GPCFs themselves then also depend on $v^{+} $.

In what follows, we work with straight gauge links, where terms such as (\ref{lir_staple}) are absent; in Sec.~\ref{seclirqgq}, we return to the staple link case and obtain a concrete expression for (\ref{lir_staple}) in terms of quark-gluon-quark correlators, by combining the LIR with the corresponding EoM.

\subsection{Construction of Lorentz invariance relations}
\label{sec4A}
The general structure of the unintegrated correlation function was written in terms of all the independent Lorentz structures multiplied by scalar functions,  $A_i^{F}, A_i^G$ in Section \ref{sec2B}.  
The correlation function integrated in $k^-$ (and $k_T$ dependent) was parametrized in terms of GTMDs in Ref.\cite{Meissner:2009ww} (Section \ref{sec2C}). GTMDs can, therefore, be expressed through $k^-$ integrals of the scalar functions $A_i^{F}, A_i^G$. These expressions are given in Appendix \ref{gtmdfroma}. 
As was shown in Section \ref{sec2B}, the total number of independent functions in the unintegrated corrrelator is 8 for the vector and 8 for the axial vector sectors. The total number of twist two plus twist three GTMDs is 12 vector and 12 axial vector \cite{Meissner:2009ww}. Since this number exceeds the number of $A_i^F, A_i^G$ functions, the GTMDs will be related to one another. This type of relation that is just originating from the parametrization in terms of Lorentz covariant structures is called a LIR.         

In the following, we describe the procedure used to derive LIRs between the $k_T$-moments of the twist two GTMDs listed in Section \ref{sec2C}, Eqs.(\ref{F14}, \ref{G11}), and the twist three GPDs listed in Section \ref{sec2D}. It is based on the following integral relation for amplitudes $A$ depending on the integration variable $k$ via Lorentz invariants as $A\equiv A(k\cdot P,k^2,k\cdot \Delta )$,
\begin{eqnarray}
\frac{d}{dx} \int d^2 k_T \int dk^{-} \frac{k_T^2 \Delta_{T}^{2} - (k_T \cdot \Delta_{T} )^2 }{\Delta_{T}^{2} } \ {\cal X} [A;x] &=&
\int d^2 k_T \int dk^{-} (k\cdot P -xP^2 ) {\cal X} [A;x] \label{intrel} \\
& & + \int d^2 k_T \int dk^{-} \frac{k_T^2 \Delta_{T}^{2} - (k_T \cdot \Delta_{T} )^2 }{\Delta_{T}^{2} } \frac{\partial {\cal X} }{\partial x} [A;x] \nonumber
\end{eqnarray}
where ${\cal X} [A;x]$ is a linear combination of amplitudes $A$ in which
the coefficients, aside from containing the invariants $k\cdot P $, $k^2 $
and $k\cdot \Delta $, may have an explicit $x$-dependence. This is an
off-forward extension of relations used previously in the analysis of TMD LIRs
\cite{Accardi:2009au,Mulders:1995dh,Bacchetta:2003vn,Tangerman:1994bb};
here, the presence of the additional invariant $k\cdot \Delta $ must be properly accounted for.
In view of this complication, it is worth laying out the elements of the
derivation of (\ref{intrel}); this is presented in Appendix \ref{app:lirintrel}.

\subsection{Relating $k_T^2$ moments of GTMDs to GPDs}
\label{sec4B}
Since a generic GTMD $X$ can be expressed in the form $X=\int dk^{-} {\cal X} [A;x]$, as given in Appendix \ref{gtmdfroma}, one can use (\ref{intrel}) to cast the $x$-derivative of its $k_T $-moment, $X^{(1)} $, cf.~(\ref{ktmoment}), in terms of $A$ amplitudes. In particular,
\begin{eqnarray}
\frac{d}{dx} F_{14}^{(1)} &=& \frac{4P^{+} }{M^2 } \int d^2 k_T \int dk^{-} \left[ (k\cdot P -xP^2 ) (A^F_8 +xA^F_9 )
+ \frac{k_T^2 \Delta_{T}^{2} - (k_T \cdot \Delta_{T} )^2 }{\Delta_{T}^{2} } A^F_9 \right] \label{f14xdev} \\
\frac{d}{dx} G_{11}^{(1)} &=& \frac{4P^{+} }{M^2 } \int d^2 k_T \int dk^{-} \left[ (k\cdot P -xP^2 ) \left( A^G_1 +\frac{A^G_{18} +xA^G_{19} }{2} \right) + \frac{k_T^2 \Delta_{T}^{2} - (k_T \cdot \Delta_{T} )^2 }{\Delta_{T}^{2} } \frac{A^G_{19} }{2} \right] \label{g11xdev}
\end{eqnarray}
To complete the LIRs, one constructs the appropriate combinations of
GPDs which yield the right-hand sides. The relevant combinations,
cf.~Appendix \ref{gtmdfroma}, are\footnote{Note that one could equally quote the left-hand sides
of (\ref{hpe})-(\ref{h2tpme2tp}) in terms of $k_T $-integrals of GTMDs instead of quoting directly their GPD limits, cf.~Appendix \ref{gtmdfroma}. This would facilitate a consistent regularization of the obtained LIRs at the level of $k_T$-integrals of GTMDs.}
\begin{eqnarray}
H+E 
&=& 2P^{+} \int d^2 k_T \int dk^{-} \ 2 \left( \frac{k_T \cdot \Delta_{T} }{\Delta_{T}^{2} } A^F_5 +A^F_6 + \frac{P\cdot k -xP^2 }{M^2 } (A^F_8 +xA^F_9 ) \right)
\label{hpe} \\
\widetilde{E}_{2T} 
&=& 2P^{+} \int d^2 k_T \int dk^{-} \ (-2) \left( \frac{k_T \cdot \Delta_{T} }{\Delta_{T}^{2} } A^F_5 + A^F_6 + \frac{(k_T \cdot \Delta_{T} )^2 -k_T^2 \Delta_{T}^{2} }{M^2 \Delta_{T}^{2} } A^F_9
\right) \\
\widetilde{H} &=& 2P^{+} \int d^2 k_T \int dk^{-} \left( -A^G_{17} + \frac{xP^2 -k\cdot P }{M^2 } (A^G_{18} +xA^G_{19} ) \right) \\
E^{\prime }_{2T} + 2\widetilde{H}^{\prime }_{2T}
&=& 2P^{+} \int d^2 k_T \int dk^{-} \left( 2 \ \frac{xP^2 -k\cdot P }{M^2 } A^G_1 + A^G_{17} + \frac{(k_T \cdot \Delta_{T} )^2 - k_T^2 \Delta_{T}^{2} }{M^2 \Delta_{T}^{2} } A^G_{19} \right)
\label{h2tpme2tp}
\end{eqnarray}
To construct the appropriate combinations completing
the LIRs, we examine the expression for the proton helicity combination
associated with the GTMD appearing on the left hand side of (\ref{f14xdev}),
(\ref{g11xdev}), and find the twist-three GPDs
corresponding to that same helicity structure. The GPCF substructure of
the twist-three GPDs need not, in general, completely match the GPCF
combination of the $x$-derivative of the $k_T^2$ moment of the GTMD.
One may need to add a twist-two GPD with the appropriate GPCF substructure.

In particular, $F_{14}$ describes an unpolarized quark in a longitudinally
polarized proton, Eq.(\ref{F14}) for $\Gamma=\gamma^+$; the twist-three
GPD with a similar proton helicity combination is $\widetilde{E}_{2T}$.
Comparing their GPCF decompositions, we see that if we add $H+E$, we
arrive at the LIR,
\begin{equation}
\label{eq:LIRF14}
\frac{d F_{14}^{(1)}}{dx}  = \widetilde{E}_{2T} +H +E.
\end{equation}

Similarly, $G_{11}$  describes a longitudinally polarized quark in an
unpolarized proton, Eq.(\ref{G11}) with $\Gamma=\gamma^+\gamma^5$.
The corresponding twist-three combination with the same proton helicity
combination is $2 \widetilde{H}_{2T}' + {E}_{2T}'$, with their GPCF
substructure given above. By adding the GPD $\widetilde{H}$, this gives us,
\begin{equation}
\label{eq:LIRG11}
\frac{d G_{11}^{(1)}}{dx}  = -\left(2\widetilde{H}_{2T}' + E_{2T}' \right) - \widetilde{H}.
\end{equation}

As already noted further above, in the case of a staple link, these Lorentz Invariance Relations acquire LIR violating terms that we introduce as, cf.~(\ref{eq:LIRviol}), 
\begin{eqnarray}
\label{eq:LIRviolf14}
\frac{d F_{14}^{(1)}}{dx}  & = & \widetilde{E}_{2T} +H +E + {\cal A}_{F_{14} } \\
\label{eq:LIRviolg11}
\frac{d G_{11}^{(1)}}{dx}  & = & -\left(2\widetilde{H}_{2T}' + E_{2T}'
 \right) - \widetilde{H}+ {\cal A}_{G_{11} } 
 \end{eqnarray}
These relations are a central result of our paper: they give a connection valid point by point in the kinematical variables $x$ and $t=-\Delta_{T}^{2} $ among the $k_T$ moments of GTMDs that define dynamically OAM and longitudinal spin-orbit coupling, specific twist three GPDs, and LIR violating terms that can be expressed in terms of genuine twist three contributions; the latter connection will be elucidated using the example of ${\cal A}_{F_{14} } $, cf.~(\ref{eq:LIRviolf14}), in the next section.

\subsection{Intrinsic twist three contributions}
\label{seclirqgq}
Lorentz invariance relations (LIRs) derived in the presence of a
staple-shaped gauge link generally include additional terms beyond
those found for straight gauge links, as exemplified by (\ref{eq:LIRviol}),(\ref{lir_staple}) in
comparison to (\ref{oamlir}). 
Whereas the staple LIR by itself does not
yield the concrete physical content of these terms, considering it in the
context of the straight-link LIR as well as staple and straight link EoMs provides more detailed insight into their
meaning. To illustrate this, it is useful to pursue the case of the
LIRs (\ref{oamlir}) and (\ref{eq:LIRviol}), relevant for the description of quark orbital
angular momentum in the nucleon, further.
Subtracting the former LIR from the latter yields
\begin{eqnarray}
{\cal A}_{F_{14} } (x) & \equiv &
v^{-} \frac{(2P^{+} )^2 }{M^2 } \int d^2 k_T \int dk^{-} \left[
\frac{k_{T} \cdot \Delta_{T} }{\Delta_{T}^{2} } (A_{11} +xA_{12} )
+A_{14} + \frac{k_T^2 \Delta_{T}^{2} - (k_T \cdot \Delta_{T} )^2 }{\Delta_{T}^{2} } \left(
\frac{\partial A_8 }{\partial (k\cdot v)} +
x\frac{\partial A_9 }{\partial (k\cdot v)} \right) \right] \nonumber \\
&=& \frac{d F_{14}^{(1)}}{dx} -
\left. \frac{d F_{14}^{(1)}}{dx} \right|_{v=0}
\label{af14diff}
\end{eqnarray}
giving a concrete expression for ${\cal A}_{F_{14} } $ in terms of the GTMD $F_{14} $.
Note that, here, the discussion given in Sec.~\ref{sec:upath} should be kept in mind: Formulating the LIRs (as well as the EoMs below) in terms of GPDs assumes that, in the staple-link case, the legs of the staples have properly collapsed upon $k_T $-integration such as to produce GPDs with their straight gauge link structures. This requires the staple link vector $v$ to lie on the light cone, $v=(0,v^{-} ,0,0)$. Corrections to the above relation would arise from several sources if one were to take the staple vector $v$ off the light cone, $v=(v^{+} ,v^{-} ,0,0)$.
On the one hand, the cancellation between the straight and staple link GTMD precursors of the GPDs in (\ref{oamlir}) and (\ref{eq:LIRviol}) would be incomplete; there would be residual terms corresponding to the difference between the two cases (unless one opts for the alternative quasi-GPD scheme also mentioned in Sec.~\ref{sec:upath}). On the other hand, as already noted in connection with eq.~(\ref{lir_staple}), additional amplitudes would enter the GPCF expression.

Now, the difference of GTMD $k_T$-moments in (\ref{af14diff}) can also be extracted from the EoMs: subject again to the above caveats,
the GPD terms in the EoM (\ref{eq:EoMF14}) are identical for a straight
link and a staple link, and subtracting an instance of (\ref{eq:EoMF14}) with a
straight link from an instance with a staple link yields
\begin{equation}
F_{14}^{(1)} -
\left. F_{14}^{(1)} \right|_{v=0} =
{\cal M}_{F_{14} } - \left. {\cal M}_{F_{14} } \right|_{v=0}
\end{equation}
Thus, the additional terms in the staple LIR (\ref{eq:LIRviol}) are associated with quark-gluon-quark correlations,
\begin{equation}
{\cal A}_{F_{14} } (x) = \frac{d}{dx} \left(
{\cal M}_{F_{14} } - \left. {\cal M}_{F_{14} } \right|_{v=0} \right)
\label{amrel}
\end{equation}
Therefore, we see that, comparing the genuine twist-three terms entering the staple link LIR and the staple link EoM, these encode independent information: the EoM contains ${\cal M}_{F_{14} } $ alone, whereas the LIR contains the difference of ${\cal M}_{F_{14} } $ and ${\cal M}_{F_{14} } |_{v=0} $.

As was shown in Refs.~\cite{Hatta:2011ku,Burkardt:2012sd}, in the forward limit, and integrated over momentum fraction $x$, the quantity $-F_{14}^{(1)}$
corresponds to Jaffe-Manohar quark orbital angular momentum in the staple link case, whereas it corresponds to Ji quark orbital angular momentum in the $v=0$ straight link case. Using (\ref{xintqgq}), we obtain a concrete expression for the difference between the two,
\begin{eqnarray}
& & -\left. \int dx\, \left( F_{14}^{(1)} - \left. F_{14}^{(1)} \right|_{v=0} \right) \right|_{\Delta_{T} =0} = \label{torqcomp} \\
& & \hspace{4.5cm}
-\left. \frac{\partial }{\partial \Delta^{i} }
i\epsilon^{ij} gv^{-} \frac{1}{2P^{+} } \int_{0}^{1} ds\, \langle p^{\prime } ,+|\bar{\psi } (0) \gamma^{+} U(0,sv) F^{+j} (sv) U(sv,0) \psi (0) | p,+\rangle \right|_{\Delta_{T} =0} , \nonumber
\end{eqnarray}
where it has been used that $2(\Delta^{i} /\Delta_{T}^{2} ) f^i = (\partial /\partial \Delta^{i} ) f^i $ in the limit $\Delta_{T} \rightarrow 0$ for a vector function $f$ which vanishes at least linearly in that limit (this is clear if one decomposes $f$ using $\Delta_{T} $,
$f^i = \Delta^{i} f^{||} + \epsilon^{ij} \Delta^{j} f^{\perp } $); note that the function on which the $\Delta_{T} $-derivative acts in (\ref{torqcomp}) satisfies this requirement since the left-hand side is regular at $\Delta_{T} =0$.
In deriving (\ref{torqcomp}), it has furthermore been used that, once one is considering $\Delta_{T} $-derivatives, the $(++)$ and $(--)$ helicity combinations contribute equally to quark orbital angular momentum. Eq.~(\ref{torqcomp}) can be interpreted in terms of the accumulated torque experienced by the struck
quark in a deep inelastic scattering process as a result of final state
interactions \cite{Burkardt:2012sd}. The genuine twist-three term ${\cal A}_{F_{14} } (x)$ entering the staple link LIR thus rather directly encodes information about this torque, via repeated integration in $x$. 
Eq.~(\ref{torqcomp}) reproduces\footnote{To see the correspondence, it is useful to reinstate into the expression given in \cite{Burkardt:2012sd} a small momentum transfer,
and to translate the matrix element such that the quark operators are located at the origin, as they are in (\ref{torqcomp}). Taking into account the resulting phases stemming from the proton states, one can then identify
$\int d^3 r\, r^i \exp (i\vec{\Delta} \vec{r} ) = i (2\pi )^3 \delta^{3} (\Delta ) \partial /\partial \Delta^{i} $. In view of the standard normalization of states $\langle p^{\prime } +|p +\rangle = 2P^{+} (2\pi )^3 \delta^{3} (p^{\prime } -p)$, the correspondence becomes evident.} the expression for the torque given in \cite{Burkardt:2012sd}. 

Analogous considerations apply to the staple link version of the other LIR derived in section \ref{sec4B}. For the spin-orbit sum rule, one has
\begin{equation}
{\cal A}_{G_{11} } = \frac{d}{dx} \left( G_{11}^{(1)} - \left. G_{11}^{(1)} \right|_{v=0} \right) = -\frac{d}{dx} \left( {\cal M}_{G_{11} } - \left. {\cal M}_{G_{11} } \right|_{v=0} \right)
\end{equation}
and in the completely integrated, forward limit,
 \begin{eqnarray}
&& \int dx \left. \left( G_{11}^{(1)} - \left. G_{11}^{(1)} \right|_{v=0} \right) \right|_{\Delta_{T} = 0} = \\
&& \hspace{4.5cm}
  -\left. \frac{\partial }{\partial \Delta^{i} } i\epsilon^{ij}
 gv^{-} \frac{1}{2P^{+} } \int_{0}^{1} ds\, \langle p^{\prime } ,+|\bar{\psi } (0) \gamma^{+} \gamma^{5} U(0,sv) F^{+j} (sv) U(sv,0) \psi (0) | p,+\rangle \right|_{\Delta_{T} =0} \nonumber
\end{eqnarray}
This term is analogous to Eq.~(\ref{torqcomp}), the only difference being in $\gamma^+ \rightarrow \gamma^+ \gamma^5$. 
\subsection{Eliminating GTMD moments from LIR and EoM relations}
\label{secWW}
We now merge the information from the LIR, Eqs.(\ref{eq:LIRviolf14}, \ref{eq:LIRviolg11}), and EoM relations Eqs.~(\ref{eq:EoMF14},\ref{eq:EoMG11}) such as to eliminate the GTMD moments. By eliminating $F_{14}^{(1)}$ between Eqs.(\ref{eq:LIRviolf14}) and (\ref{eq:EoMF14}), and $G_{11}^{(1)}$ between Eqs.~(\ref{eq:LIRviolg11}) and (\ref{eq:EoMG11}), respectively, we obtain relations involving only twist two and twist three GPDs including their corresponding genuine twist terms. 
Considering again separately the vector and axial vector cases one has,   
\begin{eqnarray}
\label{F14_WW1}
\widetilde{E}_{2T} &=&
- \int_x^1 \frac{dy}{y}(H + E) - \left[ \frac{\widetilde{H}}{x} -\int_x^1 \frac{dy}{y^2} \widetilde{H}\right]  - \left[ \frac{1}{x}\mathcal{M}_{F_{14}} - \int_x^1 \frac{dy}{y^2} \mathcal{M}_{F_{14}}  \right]
- \int_x^1 \frac{dy}{y} {\cal A}_{F_{14} }
\\
\label{G11_WW1}
2\widetilde{H}_{2T}' + E_{2T}' &=& 
-\int_x^1 \frac{dy}{y} \widetilde{H} - \left[  \frac{H}{x} - \int_x^1 \frac{dy}{y^2} H \right] + \frac{m}{M}\left[ \frac{1}{x}(2 \widetilde{H}_T + E_T ) - \int_x^1 \frac{dy}{y^2}(2 \widetilde{H}_T + E_T )\right] 
\nonumber \\
&& - \left[\frac{1}{x} \mathcal{M}_{G_{11}} -\int_x^1 \frac{dy}{y^2} \mathcal{M}_{G_{11}} \right] +\int_x^1 \frac{dy}{y} {\cal A}_{G_{11} }
\end{eqnarray}
These relations are valid for either a staple or a straight gauge link structure (with staple vector $v$ on the light cone in the former case), keeping in mind that ${\cal A}_{F_{14} } \equiv 0$ and ${\cal A}_{G_{11} } \equiv 0$ in the straight-link case. Since the GPDs in these relations by definition are identical in the staple and straight link cases, subtracting a straight-link instance of (\ref{F14_WW1}) from a staple-link instance again yields the relation (\ref{amrel}) between quark-gluon-quark terms (upon differentiation with respect to $x$), and one likewise obtains the analogous relation for ${\cal A}_{G_{11} } $. A converse way of stating this is that the terms containing ${\cal M}_{F_{14} } $ and ${\cal A}_{F_{14} } $ always conspire such that only a straight-link quark-gluon-quark contribution remains, even if (\ref{F14_WW1}) is formally written for the staple-link case; the same is true for (\ref{G11_WW1}).

If one disregards the quark-gluon-quark contributions, and the quark mass term in Eq.~(\ref{G11_WW1}), one obtains generalizations of the relation derived by Wandzura and Wilczek (WW) in Ref.~\cite{Wandzura:1977qf},
\begin{eqnarray}
\widetilde{E}_{2T}^{WW} =
- \int_x^1 \frac{dy}{y}(H + E) - \left[ \frac{\widetilde{H}}{x} -\int_x^1 \frac{dy}{y^2} \widetilde{H}\right] \\
(2\widetilde{H}_{2T}' + E_{2T}')^{WW}  = 
 -\int_x^1 \frac{dy}{y} \widetilde{H} - \left[  \frac{H}{x} - \int_x^1 \frac{dy}{y^2} H \right]
\end{eqnarray}
isolating the twist-two components of $\widetilde{E}_{2T}$ and $(2\widetilde{H}_{2T}' + E_{2T}')$.
We can then re-express Eqs.(\ref{F14_WW1}, \ref{G11_WW1}) as,
\begin{eqnarray}
\label{F14_WW2}
&& \widetilde{E}_{2T} = \widetilde{E}_{2T}^{WW}
 + \widetilde{E}_{2T}^{(3)} +  \widetilde{E}_{2T}^{LIR} 
\\
\label{G11_WW2}
&& \overline{E}_{2T}'  = \overline{E}_{2T}'^{WW} + \overline{E}_{2T}'^{(3)} +\overline{E}_{2T}'^{LIR}  + \overline{E}_{2T}'^{m}
\end{eqnarray}
where we defined
\[ \overline{E}_{2T}'= 2\widetilde{H}_{2T}' + E_{2T}' \ . \]
Here, $\widetilde{E}_{2T}^{LIR} $ and $\overline{E}_{2T}'^{LIR} $ are the LIR violating terms containing $ {\cal A}_{F_{14} } $ and ${\cal A}_{G_{11} } $, respectively, $\widetilde{E}_{2T}^{(3)}$ and $\overline{E}_{2T}'^{(3)}$ are the genuine twist three terms containing $ {\cal M}_{F_{14} } $ and ${\cal M}_{G_{11} } $, and $\overline{E}_{2T}'^{m}$ is the quark mass dependent term.

\subsection{$x^0$, $x$ and $x^2$ Moments}
\label{sec:moments}
We now consider the $x$ moments for the twist three GPDs entering Eqs.(\ref{F14_WW1},\ref{G11_WW1}). Integral relations for twist three GPDs were first obtained in Ref.\cite{Kiptily:2002nx,Penttinen:2000dg} directly from the OPE while in this paper 
we derive them by integrating the $x$-dependent expressions found from the LIR and EoM.\footnote{Notice the notation difference between Refs.\cite{Penttinen:2000dg,Kiptily:2002nx,Hatta:2012cs} and the classification scheme followed in this paper \cite{Meissner:2009ww}: $\int dx\, x G_2=-\int dx\, x(\widetilde{E}_{2T} + H + E)$.}
It is therefore important to check how the two approaches correspond to one another. 
For the vector case we have, 
\begin{subequations}
\begin{eqnarray}
\label{vec_sumrule0}
\int dx \widetilde{E}_{2T} &=& -\int dx (H+E) \quad \;\;\; \Rightarrow \int dx \, \left(  \widetilde{E}_{2T} + H+ E  \right) =0 \\
\label{vec_sumrule1}
\int dx x \widetilde{E}_{2T} &=& -\frac{1}{2}\int dx x(H+E) - \frac{1}{2} \int dx \widetilde{H}    \\
\label{vec_sumrule2}
\int dx x^2 \widetilde{E}_{2T} &=& -\frac{1}{3}\int dx x^2(H+E) -\frac{2}{3}\int dx x\widetilde{H} - \left. \frac{2}{3}\int dx x \, \mathcal{M}_{F_{14}} \right|_{v=0} \ .
\end{eqnarray}
\end{subequations}
where one can see that the contributions from ${\cal A}_{F_{14}}$ and ${\cal M}_{F_{14}}$ cancel in the first two expressions integrating by parts; it is assumed that the integrands are sufficiently well behaved at the boundaries for all such integrations.
Notice that Eq.(\ref{vec_sumrule0}) is an extension of the Burkhardt-Cottingham sum rule to the off-forward case. 
Eq.(\ref{vec_sumrule1}), taken in the forward limit, is a
sum rule for Ji quark angular momentum,
\begin{eqnarray}
&& J_q^{Ji} = \frac{1}{2} \Delta \Sigma_q + L_q^{Ji}
\end{eqnarray}
as can be seen by identifying the terms,
\begin{eqnarray}
\label{JiSR}
&& J_q^{Ji} = \frac{1}{2}\int dx\, x (H+E),  \quad \Delta \Sigma_q = \int dx \widetilde{H} , \quad  L_q^{Ji} = \int dx\, x (\widetilde{E}_{2T} +H+E) \ . 
\end{eqnarray}
Finally, Eq.(\ref{vec_sumrule2}) is the only one containing a genuine twist three contribution. It should be noticed that this contribution was surmised to be the same for all helicity configurations in Ref.\cite{Kiptily:2002nx}, while here we see that they are distinct terms.

In order to gauge the size of the OAM component, one can use data on the twist two GPDs contributing to the WW definition, and simultaneously extract the twist three GPDs. Detailed comparisons between the two sets of measurements will allow us to constrain this quantity. 

The axial vector moments are given by,
\begin{subequations}
\begin{eqnarray}
\label{ax_sumrule0}
\int dx \left(E_{2T}' +2\widetilde{H}_{2T}'\right) &=& -\int dx  \widetilde{H} \quad \;\;\; \Rightarrow  \int dx \left(E_{2T}' +2\widetilde{H}_{2T}' +   \widetilde{H}\right)  =0 \\
\label{ax_sumrule1}
\int dx x\left(E_{2T}' +2\widetilde{H}_{2T}'\right) &=& -\frac{1}{2}\int dx x \widetilde{H} -\frac{1}{2}\int dx H +\frac{m}{2M}\int dx (E_T +2 \widetilde{H}_T) \\
\label{ax_sumrule2}
\int dx \, x^2\left(E_{2T}' +2\widetilde{H}_{2T}'\right) &=& -\frac{1}{3}\int dx x^2 \widetilde{H} -\frac{2}{3}\int dx x H +\frac{2m}{3M}\int dx x (E_T +2 \widetilde{H}_T) -\left. \frac{2}{3}\int dx x \mathcal{M}_{G_{11}} \right|_{v=0}
\end{eqnarray}
\end{subequations}
Eqs.(\ref{ax_sumrule0},\ref{ax_sumrule1},\ref{ax_sumrule2}) are also consistent with those found in Ref.\cite{Kiptily:2002nx} and revisited in Ref.\cite{Bhoonah:2017olu}. In particular, Eq.(\ref{ax_sumrule0}) is an extension of the Burkhardt-Cottingham sum rule to the off-forward case. Similarly to the vector case, the various terms in Eq.(\ref{ax_sumrule1})  can be rearranged so as to single out the second moment of a twist-three GPD, namely the combination $2 \widetilde{H}_{2T}' + E_{2T}' + \widetilde{H}$, which in the forward limit can be interpreted through the LIR in Eq.~(\ref{spinorbitlir}) as the longitudinal contribution to the parton spin-orbit interaction $(L_z S_z)_q$, cf.~(\ref{eq:G11LdotS}),
\begin{equation}
2 (L_z S_z)_q = \int dx x\left(E_{2T}' +2\widetilde{H}_{2T}' + \widetilde{H} \right).
\end{equation}
One then has, in the forward limit,
\begin{equation}
\label{eq:spinorbit1}
\frac{1}{2} \int dx x \widetilde{H} + \frac{m_q}{2M} \kappa_{T}^{q}  = \int dx x (2 \widetilde{H}_{2T}' + E_{2T}' + \widetilde{H}) + \frac{1}{2} \, e_q
\end{equation}
corresponding to the sum rule
\begin{equation}
2(J_z S_z)_q  = 2(L_z S_z )_q + 2(S_z S_z)_q 
\end{equation}
where the transverse anomalous magnetic moment, $\kappa_{T}^{q} $, and the quark number, $e_q $,
\begin{equation}
\kappa_{T}^{q} = \int dx \, (E_T + 2\widetilde{H}_{T} ) \ , \quad \quad \: e_q = \int dx\, H
\end{equation}
have been defined.

The quark mass-dependent term which appears in Eq.(\ref{ax_sumrule1}), technically through the equations of motion,  is due to transverse angular momentum components that are present for non-zero quark mass. 
Note that this term is chiral even, being given by the product of two chiral-odd quantities. 
We thus find the following partitioning of the terms representing total angular momentum,
\begin{equation}
2(J_z S_z)_q \equiv 2[(J \cdot S)_q - (J_T  \cdot S_T)_q] = \frac{1}{2} \int dx x \widetilde{H} + \frac{m_q}{2M} \kappa_T^q \ .
\end{equation}
In the chiral limit, only the longitudinal polarization component is available to the quarks, and the correlation $(J_z S_z )$ is then quantified correctly by helicity-weighting the correlator yielding $J_z $, cf.~(\ref{JiSR}), which converts $H+E$ into $\widetilde{H}$. No contribution from $\widetilde{E}$ appears due to time reversal invariance. In the presence of a non-zero quark mass, this is modified by the transverse anomalous magnetic moment term, which accounts for the fact that also transverse polarization components are available to massive quarks. 
%
%
Note that one does not have to polarize the proton to observe these correlations between quark spin and angular momentum.
\section{LIR and EoM relations involving transverse spin configurations}
\label{sec5}
The main results of this paper are given by the EoM relations in Eqs.(\ref{eq:EoMF14},\ref{eq:EoMG11}), the LIR relations in Eqs.(\ref{oamlir},\ref{spinorbitlir}), and the WW relations in Section \ref{sec4}, which were obtained for 
longitudinal proton polarization at $\xi=0$.
%
Most of the  LIRs \cite{Kanazawa:2015ajw}, however, including the original ones \cite{Mulders:1995dh,Tangerman:1994bb}, were originally derived for the proton helicity flip case, or for transversely polarized proton configurations. 
It is therefore interesting to study the extension to the off-forward case for these helicity configurations. We obtain the following EoM result for the axial-vector GTMD, 
\begin{equation}
\frac{1}{2} G_{12}^{(1)} = x\left[H_{2T}' -\frac{\Delta_T^2}{4M^2} E_{2T}'\right] -\frac{\Delta_T^2}{4M^2}(H+E) -\frac{m}{M}\left[H_T -\frac{\Delta_T^2}{4M^2}E_T\right] - {\cal M}_{G_{12} },
\label{g12eom}
\end{equation}
where the genuine twist-three term
\begin{equation}
{\cal M}_{G_{12} } = -\int d^2 k_T i\epsilon^{ij} \frac{\Delta^{j} }{\Delta_{T}^{2} } \left[ \frac{\Delta^{1} +i\Delta^{2} }{2M} {\cal M}_{+-}^{i,A} +\frac{-\Delta^{1} +i\Delta^{2} }{2M} {\cal M}_{-+}^{i,A} -\frac{\Delta_{T}^{2} }{4M^2 } {\cal M}_{++}^{i,A} - \frac{\Delta_{T}^{2} }{4M^2 } {\cal M}_{--}^{i,A} \right]
\label{MG12}
\end{equation}
has been defined.

Our derivation proceeds in analogy to the steps used in the longitudinally polarized case, with a few important differences, as follows. Multiplying the $(\Lambda' \Lambda )=(+-)$ component of (\ref{eq:EoMfinal2}) with $(\Delta^{1} +i\Delta^{2} )$ and the $(\Lambda' \Lambda )=(-+)$ component of (\ref{eq:EoMfinal2}) with $(\Delta^{1} - i\Delta^{2} )$, subtracting these two component equations and contracting with $i\epsilon^{ij} \Delta^{j} /(2M\Delta_{T}^{2} )$ yields, upon inserting the parametrizations in terms of GTMDs,
\begin{eqnarray}
0 &=& \frac{k_T \cdot \Delta_{T} }{2M^2 } F_{12} +\frac{\Delta_{T}^{2} }{2M^2 } \left( F_{13} -\frac{F_{11} }{2} \right) +\frac{k_T^2 \Delta_{T}^{2} -(k_T \cdot \Delta_{T} )^{2} }{M^2 \Delta_{T}^{2} } \left(G_{12} -\frac{\Delta_{T}^{2} }{2M^2 } G_{11} \right) \\
& & -x\left( \frac{k_T \cdot \Delta_{T} }{2M^2 } G_{21} +\frac{\Delta_{T}^{2} }{2M^2 } G_{22} +G_{23} + \frac{k_T^2 \Delta_{T}^{2} - (k_T \cdot \Delta_{T} )^{2} }{M^2 \Delta_{T}^{2} } G_{24} \right) \\
& & +\frac{m}{M} \left( \frac{k_T \cdot \Delta_{T} }{2M^2 } H_{11} +\frac{\Delta_{T}^{2} }{2M^2 } H_{12} + H_{13} + \frac{k_T^2 \Delta_{T}^{2} - (k_T \cdot \Delta_{T} )^{2} }{M^2 \Delta_{T}^{2} } H_{14} \right) \\
& & -\frac{i\epsilon^{ij} \Delta^{j} }{2M\Delta_{T}^{2} } \left( (\Delta^{1} +i\Delta^{2} ){\cal M}_{+-}^{i,A} +(-\Delta^{1} +i\Delta^{2} ) {\cal M}_{-+}^{i,A} \right)
\end{eqnarray}
and, upon integration with respect to $k_T $ and identifying the corresponding GPDs,
\begin{eqnarray}
\label{eq:EoMG12}
0 &=& \frac{\Delta_{T}^{2} }{4M^2 } E +\frac{1}{2} G_{12}^{(1)} -\frac{\Delta_{T}^{2} }{4M^2 } G_{11}^{(1)} -x\left( H_{2T}^{\prime } +\frac{\Delta_{T}^{2} }{2M^2 } \widetilde{H}_{2T}^{\prime } \right) +\frac{m}{M} \left( H_T +\frac{\Delta_{T}^{2} }{2M^2 } \widetilde{H}_{T} \right) \\
& & -\frac{i\epsilon^{ij} \Delta^{j} }{2M\Delta_{T}^{2} } \int d^2 k_T \left( (\Delta^{1} +i\Delta^{2} ) {\cal M}_{+-}^{i,A} +(-\Delta^{1} +i\Delta^{2} ) {\cal M}_{-+}^{i,A} \right)
\end{eqnarray}
Finally, eliminating $G_{11}^{(1)} $ using Eq.~(\ref{eq:EoMG11}), we obtain Eq.~(\ref{g12eom}).

Eq.~(\ref{g12eom}) is a direct, off-forward GPD extension of the well-known relation involving the polarized twist three PDF $g_T$, the $k_T$-moment of the TMD $g_{1T}$ and transversity, $h_1$ \cite{Mulders:1995dh},
as can be seen by identifying, in the forward limit,
$H_{2T}^{\prime } \rightarrow g_T $, $H_T \rightarrow h_1 $, and $G_{12} \rightarrow g_{1T} $,
\begin{equation}
0 = \frac{1}{2} g_{1T}^{(1)} (x) -xg_T (x) +\frac{m}{M} h_1 (x) + x\widetilde{g}_{T} (x)
\end{equation}
Note that our definition of the $k_T $-moment $X^{(1)} $ differs from the one in Refs.~\cite{Accardi:2009au,Mulders:1995dh} by a factor of 2.
The intrinsic twist-three contribution can be given explicitly in terms of quark-gluon-quark correlators as,
\begin{equation}
x\widetilde{g}_{T} (x) = \left. \frac{1}{4M} \int d^2 k_T \left( {\cal M}_{+-}^{1,A} +i{\cal M}_{+-}^{2,A} +{\cal M}_{-+}^{1,A} -i{\cal M}_{-+}^{2,A} \right) \right|_{\Delta_{T} =0} = \left. {\cal M}_{G_{12} } \right|_{\Delta_{T} =0}
\label{gttildedef}
\end{equation}
This simplified form for ${\cal M}_{G_{12} } $ in the $\Delta_{T} =0$ limit is obtained from (\ref{MG12}) by considering approaches to the $\Delta_{T} =0$ limit along both the $\Delta^{1} $ and the $\Delta^{2} $ axes.


The EoM relation (\ref{g12eom}) is accompanied by a corresponding LIR,
which one obtains in complete analogy to the treatment in Sec.~\ref{sec4B}
by considering the appropriate decompositions into amplitudes $A^G $,
cf.~Appendix \ref{gtmdfroma}. For straight gauge links, one has
\begin{eqnarray}
\frac{d}{dx} G_{12}^{(1)} &=& \frac{4P^{+} }{M^2 } \int d^2 k_T \int dk^{-} \frac{P^2 }{M^2 } \left[ (k\cdot P -xP^2 ) (A^G_{18} +xA^G_{19} ) + \frac{k_T^2 \Delta_{T}^{2} - (k_T \cdot \Delta_{T} )^2 }{\Delta_{T}^{2} } A^G_{19} \right]
\label{g12xdev}
\end{eqnarray}
as well as
\begin{eqnarray}
\widetilde{H}
&=& 2P^{+} \int d^2 k_T \int dk^{-} \left( -A^G_{17} + \frac{xP^2 -k\cdot P }{M^2 } (A^G_{18} +xA^G_{19} ) \right) \\
H^{\prime }_{2T} -\frac{\Delta_{T}^{2} }{4M^2 } E^{\prime }_{2T}
&=& 2P^{+} \int d^2 k_T \int dk^{-} \frac{P^2 }{M^2 } \left( -A^G_{17} - \frac{(k_T \cdot \Delta_{T} )^2 -k_T^2 \Delta_{T}^{2} }{M^2 \Delta_{T}^{2} }
A^G_{19} \right)
\end{eqnarray}
leading to the LIR
\begin{eqnarray}
\frac{1}{2} \frac{d G_{12}^{(1)}}{dx}   &=& H_{2T}' -\frac{\Delta_T^2}{4M^2}E_{2T}'-
\left(1+\frac{\Delta_T^2}{4M^2}\right)
\widetilde{H} + {\cal A}_{G_{12} }
\label{g12lir_bis}
\end{eqnarray}
where ${\cal A}_{G_{12} } \equiv 0$ in the straight-link case, but in the presence of a staple link with staple direction $v$ on the light cone, cf.~the analogous discussion in Sec.~\ref{seclirqgq}, one has the genuine twist-three contribution
\begin{equation}
{\cal A}_{G_{12} } = -\frac{d}{dx} \left(
{\cal M}_{G_{12} } - \left. {\cal M}_{G_{12} } \right|_{v=0} \right)
\label{ag12def}
\end{equation}
This LIR is likewise an off-forward generalization of a well-known structure function relation; setting $\Delta_{T} =0$ in (\ref{g12lir_bis}) directly yields
\begin{equation}
\frac{1}{2} \frac{d}{dx} g_{1T}^{(1)} (x) = g_T (x) -g_1 (x) - \widehat{g}_{T} (x)
\end{equation}
in which the genuine twist-three contribution is given in terms of quark-gluon-quark correlators as
\begin{equation}
\widehat{g}_{T} (x) = \left. \frac{d}{dx} \left( {\cal M}_{G_{12} } -\left. {\cal M}_{G_{12} } \right|_{v=0} \right) \right|_{\Delta_{T} =0}
\label{gthatdef}
\end{equation}
By using the same techniques as in Section \ref{sec4}, eliminating the term containing $G_{12} $, we obtain the following relation,
\begin{eqnarray}
H_{2T}' -\frac{\Delta_T^2}{4M^2}E_{2T}' &=& 
%
\left(1+ \frac{\Delta_T^2}{4M^2} \right)\int_x^1\frac{dy}{y}\widetilde{H} + 
\frac{m}{M} \left[ \frac{1}{x}\left(H_T -\frac{\Delta_T^2}{4M^2}E_{T}\right)   
-  \int_x^1\frac{dy}{y^2} \left(H_T -\frac{\Delta_T^2}{4M^2}E_{T}\right)\right] \nonumber\\
&& +\frac{\Delta_T^2}{4M^2}\left[ \frac{1}{x}(H+E) - \int_x^1\frac{dy}{y^2}(H+E) \right] + \left[ \frac{{\mathcal{M}}_{G_{12}}}{x} -\int_x^1 \frac{dy}{y^2}{\mathcal{M}}_{G_{12}} \right]
-\int_x^1 \frac{dy}{y} {\cal A}_{G_{12} } \quad .
\label{g12_eliminated}
\end{eqnarray}
Notice that this relation reduces in the forward limit to the one for the polarized structure functions, $g_1$ and $g_T$ \cite{Accardi:2009au}, namely, taking $H_{2T}' \rightarrow g_T=g_1+g_2$, $\widetilde{H} \rightarrow g_1$, and $H_T \rightarrow h_1 $, as well as taking into account (\ref{gttildedef}), (\ref{ag12def}) and (\ref{gthatdef}),
\begin{equation}
g_T = \int_x^1 \frac{dy}{y} g_1  + \frac{m}{M} \left( \frac{1}{x} h_1 -\int_x^1 \frac{dy}{y^2 } h_1 \right) + \left( \widetilde{g}_{T} -\int_x^1 \frac{dy}{y} \widetilde{g}_{T} \right) +\int_x^1 \frac{dy}{y} \widehat{g}_{T}
\end{equation}
 
Taking moments of (\ref{g12_eliminated}) in $x$, one obtains,
\begin{subequations}
\begin{eqnarray}
\label{eq:g12_x0}
\int dx \left(H_{2T}' -\frac{\Delta_T^2}{4M^2}E_{2T}'\right)
&=& \left(1+ \frac{\Delta_T^2}{4M^2} \right) \int dx \widetilde{H} 
\quad \;\;\;  \stackrel{\Delta_T \rightarrow 0}{\Rightarrow} \int dx \left(H_{2T}'   - \widetilde{H} \right) \equiv \int dx \, g_2 =0 
\\
\label{eq:g12_x1}
\int dx \, x \left(H_{2T}' -\frac{\Delta_T^2}{4M^2}E_{2T}'\right)
&=& \frac{1}{2} \left(1+ \frac{\Delta_T^2}{4M^2} \right)\int dx x \widetilde{H}+\frac{\Delta_T^2}{8M^2}\int dx (H+E) +\frac{m}{2M}\int dx \left(H_T -\frac{\Delta_T^2}{4M^2} E_{T}\right) \nonumber 
\\ 
&& \\
\label{eq:g12_x2}
\int dx \, x^2 \left(H_{2T}' -\frac{\Delta_T^2}{4M^2}E_{2T}'\right)
& = & \frac{1}{3} \left(1+ \frac{\Delta_T^2}{4M^2} \right) \int dx x^2 \widetilde{H}+\frac{\Delta_T^2}{6M^2}\int dx x (H+E) +\frac{2m}{3M} \int dx x \left(H_T -\frac{\Delta_T^2}{4M^2} E_{T}\right) \nonumber \\
& & \hspace{7cm} +\left. \frac{2}{3}\int dx\, x\, \mathcal{M}_{G_{12}} \right|_{v=0} \ .
\end{eqnarray}
\end{subequations}
Eq.(\ref{eq:g12_x0}) is the off-forward generalization of the original Burkhardt-Cottingham sum rule; similarly, Eq.(\ref{eq:g12_x1}) is the generalization of the Efremov-Leader-Teryaev sum rule \cite{Efremov:1996hd},
\begin{equation}
\int dx \, x \left[ g_T(x) - \frac{1}{2} g_1(x) \right] = 0 \quad , 
\end{equation}
which is valid in the chiral limit, $m \rightarrow 0$, 
whereas Eq.(\ref{eq:g12_x2}) in the forward and chiral limits reduces to\footnote{Note that definitions of $d_2 $ in the literature vary by a factor of 2.},
\begin{equation}
 \int dx \, x^2 \left[ g_T(x) - \frac{1}{3} g_1(x) \right] = \left.
\frac{2}{3}  \int dx\, x\, \mathcal{M}_{G_{12}} \right|_{v=0,\Delta_{T} =0} = \frac{1}{3} d_2 \quad . 
\end{equation}
$d_2$, which incorporates  quark-gluon-quark correlations, is one of the few quantities where these effects can be obtained unambiguously from inclusive polarized scattering experiments (\cite{Flay:2016wie} and references therein).
One can write explicitly the helicity structure of ${\cal M}_{G_{12}}$ as, 
\begin{eqnarray}
d_2 = 
\left. \frac{1}{2M} \int dx \, x \, \int d^2 k_T \, \left( {\cal M}_{+-}^{1,A} +i{\cal M}_{+-}^{2,A} +{\cal M}_{-+}^{1,A} -i{\cal M}_{-+}^{2,A} \right) \right|_{v=0,\Delta_{T} =0}  
\end{eqnarray}
A relation involving the GTMD $F_{12}$, the imaginary part of which in the forward limit is (minus) the Sivers function $f_{1T}^\perp$ \cite{Meissner:2009ww},  is obtained by considering the combination $(\Delta^1 - i\Delta^2)$ times the $(\Lambda' \Lambda)= (-+)$ helicity component added to $(\Delta^1 + i\Delta^2)$ times the $(\Lambda' \Lambda)= (+-)$ helicity component of Eq.~(\ref{eq:EoMfinal1}) and multiplying by $\Delta^i/M\Delta_{T}^{2} $,

\begin{eqnarray}
-x\left(F_{23} +\frac{k_T^2\Delta_T^2 - (k_T\cdot\Delta_T)^2}{M^2 \Delta_{T}^{2} } F_{24}\right) &+&\frac{1}{2M^2}\left(\Delta_T^2 G_{13} + k_T\cdot\Delta_T G_{12}\right) + \frac{k_T^2\Delta_T^2 - (k_T\cdot\Delta_T)^2}{M^2 \Delta_{T}^{2} } F_{12}\nonumber \\
&+&\frac{\Delta^i}{2M\Delta_{T}^{2} } \left((\Delta^1 - i\Delta^2){\cal M}_{-+}^{i,S} + (\Delta^1 + i\Delta^2){\cal M}_{+-}^{i,S }\right) = 0.
\label{f12gtmdeq}
\end{eqnarray}


\noindent In the forward limit, Eq.~(\ref{f12gtmdeq}) is a relation between purely imaginary quantities. This follows from the fact that, for $\Delta =0$, the real parts of all GTMDs entering Eq.~(\ref{f12gtmdeq}) except for $G_{12} $ vanish \cite{Meissner:2009ww}; on the other hand, $G_{12} $ is multiplied by $\Delta_{T} $. As a consequence, also the real part of the genuine twist three term vanishes for $\Delta =0$. Turning therefore to the imaginary part, in the forward limit, one has the TMD identifications $F_{23}^{o} =f_T^{\prime } $ and $F_{24}^{o} = f_T^{\perp } $ \cite{Meissner:2009ww}. Integrated over $k_T $, the term in the first parenthesis thus combines to $\int d^2 k_T f_T =0$ \cite{Bacchetta:2006tn}. One is therefore left with the following $k_T $-integrated relation involving the Sivers function $f_{1T}^{\perp } $ in the forward limit,
\begin{equation}
f_{1T}^{\perp (1)} = -F_{12}^{o(1)} = \left. {\cal M}_{F_{12}} \right|_{\Delta_{T} =0}
\label{sivers_qgq}
\end{equation}
where the quark-gluon-quark term
\begin{equation}
{\cal M}_{F_{12}} = -2i \frac{\Delta^{i} }{2M\Delta_{T}^{2} } \int d^2 k_T \left( (\Delta^{1} -i\Delta^{2} ) {\cal M}_{-+}^{i,S} + (\Delta^{1} +i\Delta^{2} ) {\cal M}_{+-}^{i,S} \right)
\end{equation}
has been introduced. Eq.~(\ref{sivers_qgq}) indicates a correspondence of ${\cal M}_{F_{12} } $ to the well-known Qiu-Sterman term $T_q (x,x)$ in the forward limit. 
Indeed, in analogy to the discussion surrounding Eq.~(\ref{torqcomp}), a compact expression for ${\cal M}_{F_{12} } $ can be obtained in the fully integrated case. Approaching the $\Delta_{T} =0 $ limit either along the $\Delta^{2} =0$ axis or the $\Delta^{1} =0$ axis, one has
\begin{equation}
\left. {\cal M}_{F_{12} } \right|_{\Delta_{T} =0} =
-i \frac{1}{M} \int d^{2} k_T ({\cal M}_{+-}^{1,S} + {\cal M}_{-+}^{1,S} ) = \frac{1}{M} \int d^2 k_T ({\cal M}_{+-}^{2,S} - {\cal M}_{-+}^{2,S} )
\end{equation}
Considering, for example, the form given in terms of ${\cal M}_{\Lambda^{\prime } \Lambda }^{1,S} $, and integrating with respect to $x$, one can insert (\ref{xintqgq}) and obtain the Sivers shift
\begin{equation}
\langle k_2 \rangle = M \frac{1}{2} \int dx f_{1T}^{\perp (1)} = gv^{-} \frac{1}{2P^{+} } \int_{0}^{1} ds \langle P, S_1 | \bar{\psi } (0) \gamma^{+} U(0,sv) F^{+2} (sv) U(sv,0) \psi (0) | P, S_1 \rangle
\end{equation}
after having converted the states from the helicity basis to a spin quantization axis in 1-direction, and having used a rotation by $\pi $ in the transverse plane to combine terms associated with spin in the $\pm 1 $-directions. The case of spin in the 2-direction can be treated analogously.
One thus obtains the standard Qiu-Sterman form \cite{Burkardt:2012sd} in the forward limit.
For $\Delta_T \neq 0$, ${\cal M}_{12}$ is an off-forward/generalized analogue of the Qiu-Sterman $T_q(x,x)$ term.

%
The EoM relations presented so far in either the longitudinal or transverse proton polarization cases allow us to decompose specific twist-three GPDs into a linear combination of a twist-two GPD, a quark-gluon-quark correlation, the $k_T^2$ moment of a twist-two GTMD and a mass term in the axial-vector case. 
The $k_T^2$ moment of the GTMD can be eliminated using the LIRs. The resulting relations, when integrated over $x$, are analogous to the relations provided by Kiptily and Polyakov in \cite{Kiptily:2002nx}.    
Note, however, that not all EoM relations are of this form; in general, also other GTMD moments besides $k_T^2 $ moments appear in the EoM relations that we have not discussed in detail in this work. For instance, one finds a relation in which the GTMD $G_{13}$ contributes to the EoM weighted by $(k_T\cdot \Delta_T)$. Moreover, that EoM relation cannot have a LIR counterpart within the twist two and twist three sectors, since $G_{13}$ is the only GTMD in those sectors which contains the invariant amplitude $A^G_{21}$, cf.~Appendix~\ref{gtmdfroma}.

\section{Conclusions and Outlook}
\label{sec6}
We presented the derivation of a set of relations connecting $k_T^2 $-moments of GTMDs and twist-two as well as twist-three GPDs, known as Lorentz Invariance Relations (LIRs) and Equation of Motion (EoM) relations. LIRs stem from the Lorentz structure of the off-forward correlation function. By examining their gauge link structure, we find that two different types of relations exist: one obtained by considering a staple-shaped gauge link, where an explicit quark-gluon-quark contribution appears, and one for the straight gauge link, where this term is instead absent. On the other hand, the QCD equations of motion yield complementary relations containing explicit quark-gluon-quark contributions that have a different structure than the ones in the LIRs. By inserting the LIRs in the equations of motion we can eliminate the $k_T^2$-moments of GTMDs, and obtain relations directly between twist-two and twist-three GPDs. In the absence of genuine twist-three terms, these relations represent off-forward generalizations of the original Wandzura-Wilzcek relations connecting twist-two and twist-three PDFs.  

Within our general scheme of constructing LIRs, we focus particularly on ones involving the $k_T^2$-moments of the GTMDs $F_{14}$ and $G_{11}$, which describe the $x$-density distributions of the quark OAM, $L_z$, and longitudinal spin-orbit interaction, $L_z S_z$.
Our detailed study of the $k_T$-dependence of these OAM-related observables provides physical insight that buttresses previous suggestions in the literature, stemming from OPE-based integral relations, that partonic OAM is described by twist-three GPDs. 

Our results, therefore, represent a step forward in comprehending parton OAM in the proton, on two accounts. On the one hand, the obtained relations are key to accessing information from experiment on the missing piece in the proton's angular momentum budget: we obtain the $x$-dependent distribution of OAM  through the GPDs $\widetilde{E}_{2T}$, $H$ and $E$, which can be readily measured from various azimuthal angular modulations in DVCS and related processes. 
The new $x$-dependent expressions written in terms of twist-three GPDs including the genuine quark-gluon-quark terms bring, for the first time, partonic OAM within experimental grasp.
On the other hand, taking integrals in $x$, and using the QCD equations of motion, one recovers the sum rule relating the second Mellin-Barnes moment of a specific twist-three GPD combination, here called $\widetilde{E}_{2T} + H+ E$,
to the moments of twist-two GPDs yielding the combination $J_q-S_q$. Our result is therefore not only consistent with previous findings hinting at a twist-three nature of OAM \cite{Penttinen:2000dg,Kiptily:2002nx,Hatta:2012cs,Ji:2012sj}: it goes beyond these predictions by providing
a physical link, missing from earlier work, which explains how OAM is described at twist-three through its connection with the $k_T^2$-moment of a GTMD. 
The, perhaps, most distinguishing merit of these new relations lies in that they provide a handle on the dynamical underpinnings of the parton correlations through which OAM is generated. OAM is present because of the transverse motion of partons when they are displaced from the origin. This is described in QCD by a twist-three parton correlation;
the correlation is generated by the Lorentz invariant structure of the proton matrix elements appearing in the QCD equations of motion.  

The LIRs will allow us to directly connect, on the one hand, twist-three GPD measurements of OAM and spin-orbit correlations, and on the other hand, Lattice QCD evaluations of GTMDs. The $k_T^2 $-moment of $F_{14}$ has already been accessed in a preliminary Lattice QCD calculation \cite{Engelhardt:2017miy}: GTMD $k_T^2$-moments can be obtained  
by generalizing the proton matrix elements of quark bilocal operators used to study TMDs, namely, by supplementing the transverse momentum information with transverse position information through the introduction of an additional nonzero momentum transfer. The calculation in Ref.~\cite{Engelhardt:2017miy} also includes the gauge connection between the quarks in the quark bilocal operators, enabling the evaluation of both the staple gauge link path used in TMD calculations, characterizing Jaffe-Manohar (JM) OAM, and the straight path yielding Ji OAM. Although this exploration was performed at the pion mass $m_\pi$ = 518 MeV, its results suggest a sizable difference between the two definitions.

Our findings provide a perspective for accessing experimentally all terms appearing in both the JM and the Ji definitions: Ji OAM is given by the Wandzura-Wilczek component of $\widetilde{E}_{2T}$, which is described in terms of twist-two GPDs, while JM OAM is given by the sum of these terms and the genuine/intrinsic twist-three contribution, which we identified as an integral over ${\cal A}_{F_{14} }$, 
technically a Lorentz invariance relation violating term. Such a term may be obtained by a careful analysis of DVCS type experiments (see {\it e.g.} an analogous term in the forward case for the axial vector components $g_1(x)$ and $g_2(x)$, 
\cite{Accardi:2009au}). 

Our findings extend to other GTMDs: here, we have treated specifically $G_{11}$, encoding spin-orbit correlations, and $G_{12}$, the off-forward extension of $g_{1T}$, leading to a direct measurement of the color force between quarks. 

Understanding the role of GTMDs and twist-three GPDs in quark OAM has
initiated a fruitful interaction between phenomenology, theory and
Lattice QCD which we intend to pursue further. In particular, the structure of the underlying
QCD matrix element suggests the study of experimental
processes containing two hadronic reaction planes, one associated
with the hadron momentum transfer, and one associated with the transverse
momentum of the hadronized ejected quark. 
We envisage developing the
description of such two-jet processes to underpin future experimental
efforts to access quark OAM directly from GTMDs.
Investigations of  experimental hard scattering processes/observables that measure OAM have started, and the opportunity to measure OAM using deeply virtual multiple coincidence exclusive processes will be soon within reach at the new Jlab upgrade and, even more promisingly, at an upcoming Electron Ion Collider (EIC).
Having understood the mechanisms that regulate quark OAM in the proton paves the way for future studies of the gluon sector which will be crucial to understand the spin of hadrons.
\acknowledgments
We thank Aurore Courtoy for taking part in the initial stages of this work. We are also grateful to Ted Rogers for many useful discussions, as well as to Matthias Burkardt and Markus Diehl. 
This research is funded by DOE grants DE-SC0016286 (S.L. and A.R.) and DE-FG02-96ER40965 (M.E.), by the Jefferson Science Associates grant (A.R.), and by the DOE Topical Collaboration on TMDs.

\appendix

\section{Explicit form of quark-gluon-quark terms}
\label{Appendix:qgq}
Consider a staple-shaped gauge link ${\cal U} $ connecting the space-time
points $y$ and $y^{\prime } $ via three straight segments,
\begin{eqnarray}
{\cal U} &=&
{\cal P} \exp \left( -ig \int_{y}^{y+v} dx^{\mu } A_{\mu } (x) \right)
{\cal P} \exp \left( -ig \int_{y+v}^{y^{\prime } +v} dx^{\mu } A_{\mu } (x)\right)
{\cal P} \exp \left( -ig \int_{y^{\prime } +v}^{y^{\prime } }
dx^{\mu } A_{\mu } (x) \right) \\
& \equiv & U_1 (0,1) U_2 (0,1) U_3 (0,1) \ ,
\end{eqnarray}
which each can be parametrized in terms of a real parameter $t$ as
\begin{eqnarray}
U_1 (a,b) &=& {\cal P} \exp \left( -ig \int_{a}^{b} dt\, v^{\mu }
A_{\mu } (y+tv) \right) \label{u1link} \\
U_2 (a,b) &=& {\cal P} \exp \left( -ig \int_{a}^{b} dt\,
(y^{\prime } -y)^{\mu } A_{\mu } (y+v+t(y^{\prime } -y)) \right)
\label{u2link} \\
U_3 (a,b) &=& {\cal P} \exp \left( -ig \int_{a}^{b} dt\, (-v^{\mu } )
A_{\mu } (y^{\prime } +v-tv) \right)
\end{eqnarray}
The four-vector $v$ describes the legs of the staple-shaped
path. The parametrization includes the special case $v=0$, in which the
staple degenerates to a straight link between $y$ and $y^{\prime } $ given
by $U_2 (0,1)$, whereas $U_1 =U_3 =1$. In the following, $U_i $ given
without an argument means $U_i \equiv U_i (0,1)$.

The goal of the following treatment is to evaluate
\begin{equation}
\left( \frac{\partial }{\partial y^{\nu } } -igA_{\nu } (y) \right) {\cal U}
=\left( \frac{\partial U_1 }{\partial y^{\nu } }
-igA_{\nu } (y) U_1 \right) U_2 U_3
+U_1 \frac{\partial U_2 }{\partial y^{\nu } } U_3
\label{covaru}
\end{equation}
(note that $U_3 $ is independent of $y$). Consider first
$\partial U_1 /\partial y^{\nu } $. The derivative of the path-ordered
exponential, cf.~(\ref{u1link}), is
\begin{equation}
\frac{\partial U_1 }{\partial y^{\nu } } =
\int_{0}^{1} ds\, U_1 (0,s) \left[ -ig v^{\mu } \partial_{\nu }
A_{\mu } (y+sv) \right] U_1 (s,1)
\label{du1link}
\end{equation}
This can be recast by the following integration by parts. Noting that
\begin{eqnarray}
\frac{d}{ds} U_1 (0,s) &=& U_1 (0,s)
(-ig) v^{\mu } A_{\mu } (y+sv) \\
\frac{d}{ds} U_1 (s,1) &=&
igv^{\mu } A_{\mu } (y+sv) U_1 (s,1)
\end{eqnarray}
it follows that
\begin{eqnarray}
U_1 igA_{\nu } (y+v) - igA_{\nu } (y) U_1
&=& \int_{0}^{1} ds \frac{d}{ds}
\left[ U_1 (0,s) igA_{\nu } (y+sv) U_1 (s,1) \right] \label{partu1} \\
&=& \int_{0}^{1} ds \left[
U_1 (0,s) (-ig) v^{\mu } A_{\mu } (y+sv) igA_{\nu } (y+sv)
U_1 (s,1) \right. \nonumber \\
& & \ \ \ \ \ \ \ \ \ + U_1 (0,s) igA_{\nu } (y+sv) igv^{\mu } A_{\mu } (y+sv) U_1 (s,1)
\nonumber \\
& & \ \ \ \ \ \ \ \ \ \left. + U_1 (0,s) igv^{\mu } \partial_{\mu } A_{\nu } (y+sv)
U_1 (s,1) \right] \nonumber \\
&=& igv^{\mu } \int_{0}^{1} ds\, U_1 (0,s) \left[ 
F_{\mu \nu } (y+sv) +\partial_{\nu } A_{\mu } (y+sv) \right] U_1 (s,1)
\nonumber
\end{eqnarray}
having introduced the field strength
$F_{\mu \nu } = \partial_{\mu } A_{\nu } -\partial_{\nu } A_{\mu }
-ig[A_{\mu } ,A_{\nu } ]$. Adding the left- and right-hand sides of
(\ref{du1link}) to the initial and final expressions in (\ref{partu1}),
respectively, as well as subtracting $U_1 igA_{\nu } (y+v)$ from both sides,
finally yields
\begin{equation}
\left( \frac{\partial }{\partial y^{\nu } } -igA_{\nu } (y) \right) U_1
= igv^{\mu } \int_{0}^{1} ds\, U_1 (0,s) F_{\mu \nu } (y+sv) U_1 (s,1)
-U_1 igA_{\nu } (y+v)
\label{u1term}
\end{equation}
The term $\partial U_2 /\partial y^{\nu } $ can be treated analogously;
the resulting expressions are slightly more involved, since, in this
case, also the line element in $U_2 $ depends explicitly on $y$,
cf.~(\ref{u2link}):
\begin{eqnarray}
\frac{\partial U_2 }{\partial y^{\nu } } &=&
\int_{0}^{1} ds U_2 (0,s) (-ig) \left[
-A_{\nu } (y+v+s(y^{\prime } -y))
+ (y^{\prime } -y)^{\mu } \partial_{\nu }
A_{\mu } (y+v+s(y^{\prime } -y)) (1-s) \right] U_2 (s,1)
\label{du2link}
\end{eqnarray}
Noting, in analogy to above,
\begin{eqnarray}
\frac{d}{ds} U_2 (0,s) &=& U_2 (0,s)
(-ig) (y^{\prime } -y)^{\mu } A_{\mu } (y+v+s(y^{\prime } -y)) \\
\frac{d}{ds} U_2 (s,1) &=&
ig(y^{\prime } -y)^{\mu } A_{\mu } (y+v+s(y^{\prime } -y))
U_2 (s,1)
\end{eqnarray}
one has
\begin{eqnarray}
-igA_{\nu } (y+v) U_2
&=& \int_{0}^{1} ds \frac{d}{ds}
\left[ U_2 (0,s) igA_{\nu } (y+v+s(y^{\prime } -y)) (1-s) U_2 (s,1) \right]
\label{partu2} \\
&=& \int_{0}^{1} ds \left[
U_2 (0,s) (-ig)(y^{\prime } -y)^{\mu } A_{\mu } (y+v+s(y^{\prime } -y))
igA_{\nu } (y+v+s(y^{\prime } -y)) (1-s) U_2 (s,1) \right. \nonumber \\
& & \ \ \ \ \ \ \ \ \ +U_2 (0,s) igA_{\nu } (y+v+s(y^{\prime } -y)) (1-s)
ig(y^{\prime } -y)^{\mu } A_{\mu } (y+v+s(y^{\prime } -y))
U_2 (s,1) \nonumber \\
& & \ \ \ \ \ \ \ \ \ +U_2 (0,s)ig(y^{\prime }-y)^{\mu } \partial_{\mu } A_{\nu } (y+v+s(y^{\prime } -y))
(1-s) U_2 (s,1) \nonumber \\
& & \ \ \ \ \ \ \ \ \ \left. -U_2 (0,s) igA_{\nu } (y+v+s(y^{\prime } -y)) U_2 (s,1) \right]
\nonumber \\
&=& ig \int_{0}^{1} ds\, U_2 (0,s) \left[ (1-s) (y^{\prime }-y)^{\mu }
\left( F_{\mu \nu } (y+v+s(y^{\prime } -y))
+\partial_{\nu } A_{\mu } (y+v+s(y^{\prime } -y)) \right) \right. \nonumber \\
& & \hspace{7cm}
\left. -A_{\nu } (y+v+s(y^{\prime } -y)) \right] U_2 (s,1) \nonumber
\end{eqnarray}
Adding the left- and right-hand sides of (\ref{du2link}) to the initial and
final expressions in (\ref{partu2}), respectively, as well as adding
$igA_{\nu } (y+v) U_2 $ to both sides, finally leaves
\begin{equation}
\frac{\partial U_2 }{\partial y^{\nu } } =
ig \int_{0}^{1} ds\, U_2 (0,s) (1-s) (y^{\prime }-y)^{\mu }
F_{\mu \nu } (y+v+s(y^{\prime } -y)) U_2 (s,1)
+igA_{\nu } (y+v) U_2
\label{u2term}
\end{equation}
Inserting (\ref{u2term}) and (\ref{u1term}) on the right-hand side of
(\ref{covaru}), one finally obtains an expression in which only field
strength terms remain,
\begin{eqnarray}
\left( \frac{\partial }{\partial y^{\nu } } -igA_{\nu } (y) \right) {\cal U}
&=& ig U_1 \int_{0}^{1} ds\, U_2 (0,s)
(y^{\prime } -y)^{\mu }
F_{\mu \nu } (y+v+s(y^{\prime } -y)) (1-s) U_2 (s,1) U_3
\nonumber \\
& & +ig\int_{0}^{1} ds\, U_1 (0,s) v^{\mu } F_{\mu \nu } (y+sv)
U_1 (s,1) U_2 U_3
\end{eqnarray}
In complete analogy, one obtains for the adjoint term,
\begin{eqnarray}
{\cal U} \left( \overleftarrow{\frac{\partial }{\partial y^{\prime \nu } } }
+iA_{\nu } (y^{\prime } ) \right) &=&
ig U_1 \int_{0}^{1} ds\, U_2 (0,s)
(y^{\prime } -y)^{\mu }
F_{\mu \nu } (y+v+s(y^{\prime } -y)) s U_2 (s,1) U_3
\nonumber \\
& & -U_1 U_2 ig\int_{0}^{1} ds\, U_3 (0,s) v^{\mu }
F_{\mu \nu } (y^{\prime } +v-sv) U_3 (s,1)
\end{eqnarray}


\section{Integral relation for the construction of LIRs}
\label{app:lirintrel}
The construction of LIRs is based on the integral relation (\ref{intrel})
for amplitudes $A$ depending on the integration variable $k$ via Lorentz
invariants as $A\equiv A(k\cdot P,k^2,k\cdot \Delta )$,
\begin{eqnarray}
\frac{d}{dx} \int d^2 k_T \int dk^{-} \frac{k_T^2 \Delta_{T}^{2} - (k_T \cdot \Delta_{T} )^2 }{\Delta_{T}^{2} } \ {\cal X} [A;x] &=&
\int d^2 k_T \int dk^{-} (k\cdot P -xP^2 ) {\cal X} [A;x] \label{intrelapp} \\
& & + \int d^2 k_T \int dk^{-} \frac{k_T^2 \Delta_{T}^{2} - (k_T \cdot \Delta_{T} )^2 }{\Delta_{T}^{2} } \frac{\partial {\cal X} }{\partial x} [A;x] \nonumber
\end{eqnarray}
where ${\cal X} [A;x]$ is a linear combination of amplitudes $A$ in which
the coefficients, aside from containing the invariants $k\cdot P $, $k^2 $
and $k\cdot \Delta $, may have an explicit $x$-dependence.

To see this relation, it is useful to handle the dependences of the
amplitudes $A$ on the invariants $k\cdot P $, $k^2 $ and $k\cdot \Delta $
by introducing new variables embodying these invariants,
\begin{eqnarray}
\sigma &\equiv& 2k\cdot P = xP^2 + 2k^-P^+ \Rightarrow
k^- = \frac{1}{2P^+}\left(\sigma - xP^2 \right) \label{sigsub} \\ 
\tau &\equiv& k^2 = x\sigma- x^2 P^2 -k_T^2 \label{tausub} \\
\sigma^{\prime } &\equiv&  k\cdot \Delta
= -k_T \cdot \Delta_T = -|k_T ||\Delta_T | \cos \phi \label{rhosub}
\end{eqnarray}
Note again that the present treatment is for vanishing skewness, in which
case $P^2 = M^2 + \Delta_{T}^{2} /4$. Examining the two terms on the
right-hand side of (\ref{intrelapp}), they take the form
\begin{eqnarray}
I_1 &=& \int d^2 k_T \int dk^{-} (k\cdot P -xP^2 )
{\cal X} [A(2k\cdot P,k^2 ,k\cdot \Delta );x] \\
&=& \frac{1}{8P^{+} } \int d\sigma d\tau d\sigma^{\prime }
\int_{0}^{\infty } dk_T^2 \int_{0}^{2\pi } d\phi \,
\delta (\tau-x\sigma +x^2 P^2 +k_T^2 )
\delta (\sigma^{\prime } +k_T \cdot \Delta_{T} )
(\sigma -2xP^2 ) {\cal X} [A(\sigma ,\tau ,\sigma^{\prime } );x] \nonumber \\
&=& \frac{1}{4P^{+} } \int d\sigma d\tau d\sigma^{\prime } \,
\theta ( x\sigma -\tau -x^2 P^2 )
\theta ( \Delta_{T}^{2} (x\sigma -\tau -x^2 P^2 ) -\sigma^{\prime 2} )
\frac{\sigma -2xP^2 }{\sqrt{\Delta_{T}^{2} (x\sigma -\tau -x^2 P^2 )
-\sigma^{\prime 2} } } {\cal X} [A(\sigma ,\tau ,\sigma^{\prime } );x]
\nonumber \\
I_2 &=& \int d^2 k_T \int dk^{-} \frac{k_T^2 \Delta_{T}^{2} -
(k_T \cdot \Delta_{T} )^2 }{\Delta_{T}^{2} }
\frac{\partial {\cal X} }{\partial x} [A(2k\cdot P,k^2 ,k\cdot \Delta );x]
\label{i2int} \\
&=& \frac{1}{4P^{+} } \int d\sigma d\tau d\sigma^{\prime }
\int_{0}^{\infty } dk_T^2 \int_{0}^{2\pi } d\phi \,
\delta (\tau-x\sigma +x^2 P^2 +k_T^2 )
\delta (\sigma^{\prime } +k_T \cdot \Delta_{T} ) k_T^2 \sin^{2} \phi
\frac{\partial {\cal X} }{\partial x} [A(\sigma ,\tau ,\sigma^{\prime } );x]
\nonumber \\
&=& \frac{1}{2P^{+} } \int d\sigma d\tau d\sigma^{\prime } \,
\theta ( x\sigma -\tau -x^2 P^2 )
\theta ( \Delta_{T}^{2} (x\sigma -\tau -x^2 P^2 ) -\sigma^{\prime 2} )
\sqrt{\frac{x\sigma -\tau -x^2 P^2 }{\Delta_{T}^{2} }
-\frac{\sigma^{\prime 2} }{\Delta_{T}^{4} } }
\frac{\partial {\cal X} }{\partial x} [A(\sigma ,\tau ,\sigma^{\prime } );x]
\nonumber
\end{eqnarray}
where in each case, in the first step, the integration variable $k^{-} $
has been substituted by $\sigma $ according to (\ref{sigsub}), and two
representations of unity have been introduced enforcing the identifications
(\ref{tausub}) and (\ref{rhosub}); also the $k_T$-integration has been
cast in polar coordinates. In the second step, the angular integrations
have been carried out using
\begin{eqnarray}
\int_{0}^{2\pi } d\phi \, \delta (\sigma^{\prime } +k_T \cdot \Delta_{T} ) \sin^2 \phi
&=& \frac{2}{k_T^2 \Delta_{T}^{2} } \sqrt{k_T^2 \Delta_{T}^{2} -\sigma^{\prime 2} }
\ \theta (k_T^2 \Delta_{T}^{2} -\sigma^{\prime 2} ) \label{phisinint} \\
\int_{0}^{2\pi } d\phi \, \delta (\sigma^{\prime } +k_T \cdot \Delta_{T} )
&=& \frac{2}{\sqrt{k_T^2 \Delta_{T}^{2} -\sigma^{\prime 2} } }
\ \theta (k_T^2 \Delta_{T}^{2} -\sigma^{\prime 2} ) \label{phiint}
\end{eqnarray}
followed by the integration over $k_T^2 $. Consider now the left-hand
side of (\ref{intrelapp}). It is of the same form as (\ref{i2int}), except
for containing ${\cal X}$ instead of $\partial{\cal X} /\partial x$,
and for the overall derivative with respect to $x$. Thus, in view of
the last line of (\ref{i2int}), it reads
\begin{eqnarray}
I &=& \frac{d}{dx} \frac{1}{2P^{+} } \int d\sigma d\tau d\sigma^{\prime } \,
\theta ( x\sigma -\tau -x^2 P^2 )
\theta ( \Delta_{T}^{2} (x\sigma -\tau -x^2 P^2 ) -\sigma^{\prime 2} )
\sqrt{\frac{x\sigma -\tau -x^2 P^2 }{\Delta_{T}^{2} }
-\frac{\sigma^{\prime 2} }{\Delta_{T}^{4} } }
{\cal X} [A;x] \\
&=& \frac{1}{2P^{+} } \int d\sigma d\tau d\sigma^{\prime } \,
\delta ( x\sigma -\tau -x^2 P^2 )
\theta ( \Delta_{T}^{2} (x\sigma -\tau -x^2 P^2 ) -\sigma^{\prime 2} )
(\sigma -2x P^2 )
\sqrt{\frac{x\sigma -\tau -x^2 P^2 }{\Delta_{T}^{2} }
-\frac{\sigma^{\prime 2} }{\Delta_{T}^{4} } }
{\cal X} [A;x] \nonumber \\
& & +\frac{1}{2P^{+} } \int d\sigma d\tau d\sigma^{\prime } \,
\theta ( x\sigma -\tau -x^2 P^2 )
\delta ( \Delta_{T}^{2} (x\sigma -\tau -x^2 P^2 ) -\sigma^{\prime 2} )
(\sigma -2x P^2 )
\sqrt{\Delta_{T}^{2} (x\sigma -\tau -x^2 P^2 ) -\sigma^{\prime 2} }
{\cal X} [A;x] \nonumber \\
& & +\frac{1}{2P^{+} } \int d\sigma d\tau d\sigma^{\prime } \,
\theta ( x\sigma -\tau -x^2 P^2 )
\theta ( \Delta_{T}^{2} (x\sigma -\tau -x^2 P^2 ) -\sigma^{\prime 2} )
\cdot \label{iint} \\
& & \hspace{4cm}
\left[
\frac{1}{2} \frac{\sigma -2xP^2 }{\sqrt{\Delta_{T}^{2} (x\sigma -\tau -x^2 P^2 )
-\sigma^{\prime 2} } } {\cal X} [A;x]
+\sqrt{\frac{x\sigma -\tau -x^2 P^2 }{\Delta_{T}^{2} }
-\frac{\sigma^{\prime 2} }{\Delta_{T}^{4} } }
\frac{\partial {\cal X} }{\partial x} [A;x]
\right]
\nonumber
\end{eqnarray}
The last term corresponds to $I_1 +I_2 $; to see (\ref{intrelapp}), it thus
remains to argue that the first two lines in (\ref{iint}) yield no
contribution. In the first term, the $\delta $-function sets
$x\sigma -\tau -x^2 P^2 =0$, and therefore the rest of the integrand is
proportional to $\theta (-\sigma^{\prime 2} ) \sqrt{-\sigma^{\prime 2} } $,
which vanishes for any $\sigma^{\prime } $. In the second line in
(\ref{iint}), the $\delta $-function sets the quantity in the square root
to zero, $\Delta_{T}^{2} (x\sigma -\tau -x^2 P^2 ) -\sigma^{\prime 2} =0$.
It should be emphasized that these properties hinge on the $\sin^{2} \phi $ weighting of the $k_T $-integral, cf.~(\ref{i2int}). Without this weighting,
it is not clear that the two terms do not contribute, and the LIR could
potentially be modified by boundary terms. The possibility of corrections
through boundary terms in LIRs not weighted by $\sin^{2} \phi $ has also
been noted in \cite{Accardi:2009au}. Note furthermore that no pathology
arises in the limit $\Delta_{T} \rightarrow 0$; this limit merely generates
$\delta (\sigma^{\prime } )$ distributions in the integrands, as is clear
from inspecting (\ref{phisinint}) and (\ref{phiint}), which are the source
of the superficially singular dependences on $\Delta_{T}^{2} $. One can
retrace the above derivation analogously in the $\Delta_{T} =0$ limit,
with (\ref{intrelapp}) continuing to hold.

%

\section{GTMDs in terms of A amplitudes}
\label{gtmdfroma}
To relate GTMDs to $A$ amplitudes, one equates the $k^{-} $ integrals of the GPCF parametrizations (\ref{vector_GPCF}) and (\ref{axialvector_GPCF}), for $\mu = +$ and $\mu =i$, a transverse vector index, to the corresponding GTMD parametrizations (\ref{vector}), (\ref{axial}), (\ref{eq:tw3vec_metz}), (\ref{eq:tw3axvec_metz}). Complete correspondence between the structures is achieved by eliminating terms in the GPCF parametrizations containing $\sigma^{i-} $. This can be effected using the Gordon identity
\begin{equation}
0 = \overline{U} (p^{\prime } ,\Lambda^{\prime } ) \left( \frac{\Delta^{\mu } }{2} + i\sigma^{\mu \nu } P_{\nu } \right) U (p,\Lambda ) \ .
\end{equation}
For purely longitudinal $P$ and transverse $\Delta $, this allows one to substitute
\begin{equation}
\overline{U} (p^{\prime } ,\Lambda^{\prime } )
i\sigma^{i-} P^{+} U(p,\Lambda ) = \overline{U} (p^{\prime } ,\Lambda^{\prime } ) \left( -\frac{i\sigma^{i+} P^2 }{2P^{+} } - \frac{\Delta^{i} }{2} \right) U(p,\Lambda )
\end{equation}
and furthermore implies $\overline{U} \sigma^{+-} U=0$; moreover, in combination with
$i\sigma^{\mu \nu } \gamma^{5} = -\frac{1}{2} \epsilon^{\mu \nu \rho \sigma } \sigma_{\rho \sigma } $
it also yields
\begin{equation}
\overline{U} (p^{\prime } ,\Lambda^{\prime } ) i\sigma^{i-} \gamma^{5} P^{+} U(p,\Lambda ) = \overline{U} (p^{\prime } ,\Lambda^{\prime } ) \left( \frac{i\sigma^{i+} \gamma^{5} P^2 }{2P^{+} } -i\epsilon^{ij} \frac{\Delta^{j} }{2} \right) U(p,\Lambda ) \ .
\end{equation}
In addition, it is useful to contract the twist-three equations, which carry a transverse vector index, with the two available transverse vectors $k_T $ and $\Delta_{T} $ in order to extract the full information from the equations. The following relations result: 

For the twist-two vector GTMDs as functions of the $A^F $ amplitudes, one obtains:
\begin{eqnarray}
F_{11} &=& 2P^+\int dk^- \left[A_1^F + x A_2^F -\frac{x\Delta_T^2}{2M^2}(A_8^F +xA_9^F)\right] \\
%
F_{12} &=& 2P^+\int dk^- \left[ A_5^F \right] \\
%
F_{13} &=& 2P^+\int dk^- \left[A_6^F +\frac{P\cdot k -xP^2 }{M^2 } (A_8^F +x A_9^F)\right] \\
%
F_{14} &=& 2P^+\int dk^- \left[A_8^F +xA_9^F\right]
\label{f14amps}
\end{eqnarray}
For the twist-three vector GTMDs as functions of the $A^F $ amplitudes:
\begin{eqnarray}
F_{21} &=& 2P^+\int dk^- \left[A_2^F - \frac{x\Delta_{T}^{2} }{2M^2 } A_9^F
\right] \\
F_{22} &=& 2P^+\int dk^- \left[A_3^F -\frac{x}{2} A_5^F
- \frac{x\Delta_{T}^{2} }{2M^2 } A_{17}^{F} \right] \\
F_{23} &=& 2P^+\int dk^- \frac{P\cdot k -xP^2 }{M^2 } \left[ A_5^F +
\frac{(k_T \cdot \Delta_{T} )^2 - k_T^2 \Delta_{T}^{2} }{M^2
(k_T \cdot \Delta_{T} )} A_9^F \right] \\
F_{24} &=& 2P^+\int dk^- \frac{P\cdot k -xP^2 }{M^2 }
\frac{\Delta_{T}^{2} }{k_T \cdot \Delta_{T} } \left[A_9^F \right] \\
F_{25} &=& 2P^+\int dk^- \frac{xP^2 -P\cdot k}{M^2 } \left[A_9^F \right] \\
F_{26} &=& 2P^+\int dk^- \frac{P\cdot k -xP^2 }{M^2 } \left[
\frac{k_T^2 }{k_T \cdot \Delta_{T} } A_9^F + A_{17}^{F} \right] \\
F_{27} &=& 2P^+\int dk^- \left[A_5^F +
\frac{k_T \cdot \Delta_{T} }{M^2 } A_9^F
+ \frac{\Delta_{T}^{2} }{M^2 } A_{17}^{F} \right] \\
F_{28} &=& 2P^+\int dk^- \left[A_6^F -\frac{k_T^2 }{M^2 } A_9^F
-\frac{k_T \cdot \Delta_{T} }{M^2 } A_{17}^{F} \right]
\end{eqnarray}
For the twist-two axial vector GTMDs as functions of the $A^G $ amplitudes:
%
\begin{eqnarray}
G_{11} &=& 2 P^+ \int dk^- \left[A_1^G + \frac{A_{18}^G + x A_{19}^G}{2}\right] \\
%
G_{12} &=& 2 P^+ \int dk^- \frac{P^2 }{M^2 } [A_
{18}^G + x A_{19}^G] \\
%
G_{13} &=& 2 P^+ \int dk^- \frac{P^2 }{M^2 } \left[
A_{21}^G + x A_{22}^G\right] \\
%
G_{14} &=& 2 P^+ \int dk^- \left[-A_{17}^G + \frac{xP^2 - k\cdot P}{M^2}(A_{18}^G + x A_{19}^G)\right]\\
\end{eqnarray}

For the twist-three axial vector GTMDs as functions of the $A^G $ amplitudes:
\begin{eqnarray}
G_{21} &=& 2P^+\int dk^- \left[ \frac{k_T \cdot \Delta_{T} }{2M^2 } A_{19}^G
+\frac{\Delta_{T}^{2} }{2M^2 } A_{20}^G \right] \\
G_{22} &=& 2P^+\int dk^- \left[ \frac{xP^2 -P\cdot k}{M^2 } A_1^G
+\frac{1}{2} A_{17}^G -\frac{k_T^2 }{2M^2 } A_{19}^G
-\frac{k_T \cdot \Delta_{T} }{2M^2 } A_{20}^G \right] \\
G_{23} &=& 2P^+\int dk^- \frac{P^2 }{M^2 } \left[ -A_{17}^G +
\frac{(k_T \cdot \Delta_{T} )^2 - k_T^2 \Delta_{T}^{2} }{M^2
(k_T \cdot \Delta_{T} )} A_{22}^G \right] \\
G_{24} &=& 2P^+\int dk^- \frac{P^2 }{M^2 } \left[ A_{19}^G +
\frac{\Delta_{T}^{2} }{k_T \cdot \Delta_{T} } A_{22}^G \right] \\
G_{25} &=& 2P^+\int dk^- \frac{P^2 }{M^2 } \left[A_{20}^G -A_{22}^G \right] \\
G_{26} &=& 2P^+\int dk^- \frac{P^2 }{M^2 } \left[A_{23}^G +
\frac{k_T^2 }{k_T \cdot \Delta_{T} } A_{22}^G \right] \\
G_{27} &=& 2P^+\int dk^- \frac{xP^2 -P\cdot k}{M^2 } \left[ A_{19}^G \right] \\
G_{28} &=& 2P^+\int dk^- \frac{xP^2 -P\cdot k}{M^2 } \left[ A_{20}^G \right]
\end{eqnarray}
Combining these relations with ones expressing GPDs in terms of $k_T $-integrals over GTMDs, as given in \cite{Meissner:2009ww}, one can also obtain the GPD combinations relevant for the developments in this work in terms of the $A$ amplitudes. In particular,
\begin{eqnarray}
H+E &=& \int d^2 k_T \ 2\left( \frac{k_T \cdot \Delta_{T} }{\Delta_{T}^{2} } F_{12} + F_{13} \right) \\ 
&=& 2P^{+} \int d^2 k_T \int dk^{-} \ 2 \left( \frac{k_T \cdot \Delta_{T} }{\Delta_{T}^{2} } A^F_5 +A^F_6 + \frac{P\cdot k -xP^2 }{M^2 } (A^F_8 +xA^F_9 ) \right) \\
\widetilde{H} &=& \int d^2 k_T \ G_{14} \\
&=& 2P^{+} \int d^2 k_T \int dk^{-} \left( -A^G_{17} + \frac{xP^2 -k\cdot P }{M^2 } (A^G_{18} +xA^G_{19} ) \right) \\
\widetilde{E}_{2T} &=& \int d^2 k_T \ (-2) \left( \frac{k_T \cdot \Delta_{T} }{\Delta_{T}^{2} } F_{27} +F_{28} \right) \\
&=& 2P^{+} \int d^2 k_T \int dk^{-} \ (-2) \left( \frac{k_T \cdot \Delta_{T} }{\Delta_{T}^{2} } A^F_5 + A^F_6 + \frac{(k_T \cdot \Delta_{T} )^2 -k_T^2 \Delta_{T}^{2} }{M^2 \Delta_{T}^{2} } A^F_9
\right) \\
E^{\prime }_{2T} + 2\widetilde{H}^{\prime }_{2T} &=& \int d^2 k_T \ 2\left( \frac{k_T \cdot \Delta_{T} }{\Delta_{T}^{2} } G_{21} + G_{22} \right) \\
&=& 2P^{+} \int d^2 k_T \int dk^{-} \left( 2 \ \frac{xP^2 -k\cdot P }{M^2 } A^G_1 + A^G_{17} + \frac{(k_T \cdot \Delta_{T} )^2 - k_T^2 \Delta_{T}^{2} }{M^2 \Delta_{T}^{2} } A^G_{19} \right) \\
H^{\prime }_{2T} -\frac{\Delta_{T}^{2} }{4M^2 } E^{\prime }_{2T} &=& \int d^2 k_T \left( G_{23} - \frac{\Delta_{T}^{2} }{M^2 } \frac{(k_T \cdot \Delta_{T} )^2 -k_T^2 \Delta_{T}^{2} }{(\Delta_{T}^{2} )^2 } G_{24} \right) \\
&=& 2P^{+} \int d^2 k_T \int dk^{-} \frac{P^2 }{M^2 } \left( -A^G_{17} - \frac{(k_T \cdot \Delta_{T} )^2 -k_T^2 \Delta_{T}^{2} }{M^2 \Delta_{T}^{2} }
A^G_{19} \right)
\end{eqnarray}

\section{\label{app:axialGPCFs} The axial vector parametrization}

\noindent We outline here the steps used to obtain the GPCFs that parametrize the completely unintegrated quark-quark correlation function for a straight-line gauge link in the axial vector case. They parallel the steps followed in \cite{Meissner:2009ww}. We use the Gordon identities:
\begin{eqnarray}
\overline{U}\gamma^\mu U &=& \overline{U} \left[\frac{P^\mu}{M} + \frac{i\sigma^{\mu\Delta}}{2M}\right]U \label{Gord1}\\
0 &=& \overline{U}\left[\frac{\Delta^\mu}{2M} + \frac{i\sigma^{\mu P}}{M}\right] U \label{Gord2}\\
\overline{U}\gamma^\mu \gamma^5U &=&\overline{U} \left[\frac{\Delta^\mu\gamma^5}{2M} + \frac{i\sigma^{\mu P}\gamma^5}{M}\right]U \label{Gord3}\\
0 &=& \overline{U}\left[\frac{P^\mu\gamma^5}{M} + \frac{i\sigma^{\mu \Delta}\gamma^5}{2M}\right] \label{Gord4}U 
\end{eqnarray}
\noindent The $\epsilon$ identity :
\begin{equation}
g^{\alpha\beta}\epsilon^{\mu\nu\rho\sigma} = g^{\mu\beta}\epsilon^{\alpha\nu\rho\sigma} + g^{\nu\beta}\epsilon^{\mu\alpha\rho\sigma} + g^{\rho\beta}\epsilon^{\mu\nu\alpha\sigma} + g^{\sigma\beta}\epsilon^{\mu\nu\rho\alpha} \label{eps}
\end{equation}
\noindent The $\sigma$ identity :
\begin{equation}
i\sigma^{\mu\nu}\gamma^5 = -\frac{1}{2}\epsilon^{\mu\nu\rho\sigma}\sigma_{\rho\sigma} \label{sig}
\end{equation}

\noindent A complete parametrization of the axial vector Dirac bilinear can be obtained by treating all possible Dirac currents one after another:\\

\noindent 1. Vector current [$\overline{U}(p',\Lambda')\gamma^\mu U(p,\Lambda)$]: Using the Gordon identity in eq.(\ref{Gord1}) all vector currents can be replaced by scalar and tensor currents.\\

\noindent 2. Axial vector current [$\overline{U}(p',\Lambda')\gamma^\mu \gamma^5U(p,\Lambda)$]: Using the Gordon identity in eq.(\ref{Gord3}) all axial vector currents can be replaced by pseudoscalar and pseudotensor currents.\\

\noindent 3. Pseudoscalar current [$\overline{U}(p',\Lambda')\gamma^5 U(p,\Lambda)$]: Using eq.(\ref{Gord4}) and contracting with $P^\mu$ all pseudoscalar currents can be replaced by pseudotensor currents.\\

\noindent 4. Tensor current [$\overline{U}(p',\Lambda')\sigma^{\mu\nu} U(p,\Lambda)$]: Using the $\sigma$ identity in eq.(\ref{sig}) all tensor currents can be replaced by pseudotensor currents.\\

\noindent 5. Pseudotensor current [$\overline{U}(p',\Lambda')\sigma^{\mu\nu} \gamma^5 U(p,\Lambda)$]: All possible pseudotensor currents are of the form
\begin{equation}
\overline{U}(p',\Lambda')\sigma^{\mu a} \gamma^5 U(p,\Lambda), \qquad \overline{U}(p',\Lambda')a^\mu\sigma^{b c} \gamma^5 U(p,\Lambda)
\end{equation}
\noindent where $a, b$ and $c$ can be any of the vectors $P,k$ and $\Delta$.\\

\noindent 6. Scalar current [$\overline{U}(p',\Lambda') U(p,\Lambda)$]:  There is only one possible scalar current 
\begin{equation}
\frac{\overline{U}U}{M^3}i\epsilon^{\mu P k \Delta}
\end{equation}
\noindent A useful relation that can be derived by multiplying the Gordon identity eq.(\ref{Gord2}) by the $\epsilon$ identity eq.(\ref{eps}) and using eq.(\ref{sig})
is
\begin{equation}
0 =\overline{U} \left[\frac{P^\mu}{M} i \sigma^{\nu\rho}\gamma^5 + \frac{P^\nu}{M} i \sigma^{\rho\mu}\gamma^5 + \frac{P^\rho}{M} i \sigma^{\mu\nu}\gamma^5 - i\frac{\epsilon^{\mu\nu\rho\Delta}}{2M}\right]U 
\end{equation}
Contracting with $P^\nu k^\rho$, $P^\nu\Delta^\rho$ and $k^\nu \Delta^\rho$ :
\begin{eqnarray}
0 &=& \overline{U} \left[\frac{P^\mu}{M} i \sigma^{Pk}\gamma^5 + \frac{P^2}{M} i \sigma^{k\mu}\gamma^5 + \frac{k\cdot P}{M} i \sigma^{\mu P}\gamma^5 - i\frac{\epsilon^{\mu Pk\Delta}}{2M}\right]U \\
0 &=& \overline{U} \left[\frac{P^\mu}{M} i \sigma^{P\Delta}\gamma^5 + \frac{P^2}{M} i \sigma^{\Delta\mu}\gamma^5 + \frac{P\cdot \Delta}{M} i \sigma^{\mu P}\gamma^5 \right]U \\
0 &=& \overline{U} \left[\frac{P^\mu}{M} i \sigma^{k\Delta}\gamma^5 + \frac{P\cdot k}{M} i \sigma^{\Delta\mu}\gamma^5 + \frac{P\cdot \Delta}{M} i \sigma^{\mu k}\gamma^5 \right]U
\end{eqnarray}
Using these relations, we can eliminate currents $\sigma^{k\mu}\gamma^5$, $\sigma^{\Delta\mu}\gamma^5$ . On the other hand, contracting with $P^\mu k^\nu \Delta^\rho$ :
\begin{equation}
 0 = \frac{P^2}{M}\overline{U}i\sigma^{k\Delta}\gamma^5 U +\frac{P\cdot k}{M}\overline{U}i\sigma^{\Delta P}\gamma^5 U +\frac{P\cdot \Delta}{M}\overline{U}i\sigma^{Pk}\gamma^5 U
\end{equation}
which allows us to eliminate $\sigma^{k\Delta}\gamma^5 a^\mu$. The parametrization can thus be written as : 
\begin{eqnarray}
W^{\gamma^\mu \gamma^5} & = & \frac{\overline{U}U}{M^3}i\epsilon^{\mu P k \Delta} A_1^G +
\frac{\overline{U}i\sigma^{P\mu}\gamma^5U}{M} A_{17}^G + \frac{\overline{U}i\sigma^{Pk}\gamma^5U}{M^3}\left(P^\mu A_{18}^G + k^\mu A_{19}^G + \Delta^\mu A_{20}^G \right) \nonumber \\ & & 
+ \frac{\overline{U}i\sigma^{P\Delta}\gamma^5U}{M^3}\left(P^\mu A_{21}^G + k^\mu A_{22}^G + \Delta^\mu A_{23}^G \right) 
\end{eqnarray}

\bibliography{OAM_bib}

\end{document}